\documentstyle[12pt]{article}
\def\href#1#2{{#2}}
\begin{document}
\begin{titlepage}
\begin{flushright}
hep-ph/9708379 \\
TS-TH-97-1 \\
August 1997, Revised June 1998
\end{flushright}
\begin{center}
\LARGE
From Sets to Quarks: \\
\normalsize
Deriving the Standard Model plus Gravitation \\
from Simple Operations on Finite Sets \\
Frank D. (Tony) Smith, Jr. \\
\footnotesize
e-mail: fsmith@pegasus.cau.edu \\
and tsmith@innerx.net  \\
P. O. Box 370, Cartersville, GA 30120 USA \\
WWW URLs http://galaxy.cau.edu/tsmith/TShome.html \\
and http://www.innerx.net/personal/tsmith/TShome.html \\
\end{center}
\normalsize
\begin{abstract}

From sets and simple operations on sets, \\
a Feynman Checkerboard physics model is constructed \\
that allows computation of force strength constants and \\
constituent mass ratios of elementary particles,  \\
with a Lagrangian structure that gives a Higgs scalar \\
particle mass of about 146 GeV and a Higgs scalar \\
field vacuum expectation value of about 252 GeV, \\
giving a tree level constituent Truth Quark (top quark) mass \\
of roughly 130 GeV, which is (in my opinion) supported \\
by dileptonic events and some semileptonic events.  \\
See http://galaxy.cau.edu/tsmith/HDFCmodel.html \\
and http://www.innerx.net/personal/tsmith/HDFCmodel.html \\
 Chapter  1 - Introduction. \\
 Chapter  2 - From Sets to Clifford Algebras. \\
 Chapter  3 - Octonions and E8 lattices. \\
 Chapter  4 - E8 spacetime and particles. \\
 Chapter  5 - HyperDiamond Lattices. \\
 Chapter  6 - Internal Symmetry Space. \\
 Chapter  7 - Feynman Checkerboards. \\
 Chapter  8 - Charge = Amplitude to Emit Gauge Boson. \\
 Chapter  9 - Mass = Amplitude to Change Direction. \\
 Chapter 10 - Protons, Pions, and Physical Gravitons. \\

\end{abstract}
\normalsize
\end{titlepage}
\newpage
\setcounter{footnote}{0}
\setcounter{equation}{0}

\tableofcontents

\section{Introduction.}

From finite sets and some simple operations on sets,
a Feynman Checkerboard physics model, whose continuous
version is the  Lie algebra $D_{4}-D_{5}-E_{6}$ model,
is constructed.

\newpage

\subsection{Particle Masses and Force Constants.}

Force strength constants and constituent mass ratios
of elementary particles can be calculated.
If the electron mass is taken as the base mass,
and assumed to be about 0.511 MeV, then the model
gives roughly the following particle masses
(the massless gluons are confined,
the massless Spin(5) gravitons are confined,
but the massless physical gravitons propagate):

\begin{equation}
\begin{array}{|c|c|c|c|c|}
\hline
10^{19} GeV  & & & Planck Mass &   \\
\hline
& & &  &  \\
\hline
10^{15}  & & & Higgs Max &   \\
\hline
& & &  &  \\
\hline
500 &  &  & SSB  &  \\
\hline
250 - 260 & T-\bar{T} &  & Higgs VEV & W^{+} + W^{-} + Z_{0} \\
\hline
130 - 146 & Truth Q & & Higgs Boson & \sqrt{W_{+}^{2} +
W_{-}^{2} + Z_{0}^{2}}\\
\hline
80-92 &  & &  &  W^{+},  W^{-},  Z_{0} \\
\hline
& & &  &  \\
\hline
5-6 & & Beauty Q &  &  \\
\hline
2 & Charm Q & & Tauon & \\
\hline
& & &  &  \\
\hline
1 &  & Proton &  & \\
\hline
0.6 &  & Strange Q &  & \\
\hline
0.3 & Up Q & Down Q &  & \\
\hline
0.24 & & QCD &  &  \\
\hline
0.14 & & Pion &  &  \\
\hline
& & &  &  \\
\hline
0.10 & & & Muon &  \\
\hline
0.0005 & & & Electron &  \\
\hline
 & & & Neutrinos & photon \\
0 & & & ( e , \mu , \tau ) & gluon \\
 & & &  & graviton \\
\hline
\end{array}
\end{equation}

\vspace{12pt}

The fermion masses give the following Kobayashi-Maskawa parameters:
\vspace{12pt}

\begin{equation}
\begin{array}{|c|c|c|c|}
\hline
& d & s & b
  \\
\hline
u & 0.975 & 0.222 & -0.00461 i  \\
c & -0.222 -0.000191 i & 0.974 -0.0000434 i & 0.0423  \\
t & 0.00941 -0.00449 i & -0.0413 -0.00102 i & 0.999  \\
\hline
\end{array}
\end{equation}

\vspace{12pt}

If the geometric gravitational force volume is set at unity,
then the model gives the following force strengths,
at the characteristic energy level of each force:
\vspace{12pt}

\begin{equation}
\begin{array}{|c|c|c|c|c|}
\hline
Gauge \: Group & Force & Characteristic
& Geometric & Total \\
& & Energy & Force & Force \\
& & & Strength & Strength \\
\hline
& & & & \\
Spin(5) & gravity & \approx 10^{19} GeV
& 1 & G_{G}m_{proton}^{2} \\
& & & & \approx 5 \times 10^{-39} \\
\hline
& & & & \\
SU(3) & color & \approx 245 MeV & 0.6286
& 0.6286 \\
\hline
& & & & \\
SU(2) & weak & \approx 100 GeV & 0.2535
& G_{W}m_{proton}^{2} \approx  \\
& & & & \approx 1.05 \times 10^{-5} \\
\hline
& & & & \\
U(1) & e-mag  & \approx 4 KV
& 1/137.03608  & 1/137.03608 \\
\hline
\end{array}
\end{equation}

\vspace{12pt}

The force strengths are given at the characteristic
energy levels of their forces, because the force
strengths run with changing energy levels.
\vspace{12pt}

The effect is particularly pronounced with the color
force.

\vspace{12pt}

The color force strength was calculated
at various energies according to renormalization group
equations, with the following results:
\vspace{12pt}

\begin{equation}
\begin{array}{|c|c|}
\hline
Energy \: Level & Color \: Force \: Strength \\
\hline
&  \\
245 MeV & 0.6286 \\
& \\
5.3 GeV & 0.166 \\
& \\
34 GeV & 0.121  \\
& \\
91 GeV & 0.106 \\
& \\
\hline
\end{array}
\end{equation}

\vspace{12pt}

\subsubsection{Confined Quark Hadrons.}

Since particle masses can only be observed experimentally
for particles that can exist in a free state
("free" means "not strongly bound to other particles,
except for virtual particles of the active vacuum of spacetime"),
\vspace{12pt}

and
\vspace{12pt}

since quarks do not exist in free states,
interpret the calculated quark masses
as constituent masses (not current masses).
\vspace{12pt}

Constituent particles are Pre-Quantum particles
in the sense that their properties are calculated without
using sum-over-histories Many-Worlds quantum theory.
\newline

("Classical" is a commonly-used synonym for "Pre-Quantum".)
\vspace{12pt}

Since experiments are quantum sum-over-histories processes,
experimentally observed particles are Quantum particles.
\vspace{12pt}

Consider the experimentally observed proton.
\vspace{12pt}

A proton is a Quantum particle containing 3 constituent quarks:
\vspace{12pt}

two up quarks and one down quark;
one Red, one Green, and one Blue.
\vspace{12pt}

The 3 Pre-Quantum constituent quarks are called "valence" quarks.
They are bound to each other by SU(3) QCD.
The constituent quarks "feel" the effects of QCD
by "sharing" virtual gluons and virtual quark-antiquark pairs
that come from the vacuum in sum-over-histories quantum theory.
\vspace{12pt}

Since the 3 valence constituent quarks within the proton
are constantly surrounded by the shared virtual gluons
and virtual quark-antiquark pairs,
the 3 valence constituent quarks can be said to
"swim" in a "sea" of virtual gluons and quark-antiquark pairs,
which are called "sea" gluons, quarks, and antiquarks.
\vspace{12pt}

In the model,
the proton is the most stable bound state of 3 quarks,
so that the virtual sea within the proton is at the
lowest energy level that is experimentally observable.
\vspace{12pt}

The virtual sea gluons are massless SU(3) gauge bosons.
Since the lightest quarks are up and down quarks,
the virtual sea quark-antiquark pairs that most often
appear from the vacuum are up or down pairs,
each of which have the same constituent mass, 312.75 MeV.
If you stay below the threshold energy of the strange quark,
whose constituent mass is about 625 MeV,
the low energy sea within the proton contains only
the lightest (up and down) sea quarks and antiquarks,
so that the Quantum proton lowest-energy background sea
has a density of 312.75 MeV.
(In the model, "density" is mass/energy per unit volume,
where the unit volume is Planck-length in size.)
\vspace{12pt}

Experiments that observe the proton as a whole
do not "see" the proton's internal virtual sea, because
the paths of the virtual sea gluon, quarks, and antiquarks
begin and end within the proton itself.
Therefore, the experimentally observed mass
of the proton is the sum of the 3 valence quarks,
3 x 32.75 MeV, or 938.25 MeV
which is very close to the experimental value of
about 938.27 MeV.
\vspace{12pt}

To study the internal structure of hadrons, mesons, etc.,
you should use sum-over-histories quantum theory
of the SU(3) color force SU(3).
Since that is computationally very difficult,
\vspace{12pt}

you can use approximate theories that correspond
to your experimental energy range.
\vspace{12pt}

For instance,
the internal structure of a proton looks like
a nonperturbative QCD soliton.
See WWW URLs
\vspace{12pt}

http://galaxy.cau.edu/tsmith/SolProton.html
\vspace{12pt}

http://www.innerx.net/personal/tsmith/SolProton.html
\vspace{12pt}

For high energy experiments,
such as Deep Inelastic Scattering,
you can use Perturbative QCD.
For low energies,
you can use Chiral Perturbation Theory.
\vspace{12pt}

To do calculations in theories such
as Perturbative QCD and Chiral Perturbation Theory,
you need to use effective quark masses that
are called current masses.  Current quark masses are different
from the Pre-Quantum constituent quark masses of the model.
\vspace{12pt}

The current mass of a quark is defined in the model as
the difference between
the constituent mass of the quark
and
the density of the lowest-energy sea of virtual gluons,
quarks, and antiquarks, or 312.75 MeV.
\vspace{12pt}

Since the virtual sea is a quantum phenomenon,
the current quarks
of Perturbative QCD and Chiral Perturbation Theory are,
in my view, Quantum particles.
\vspace{12pt}

Therefore, the model is unconventional in that:
\vspace{12pt}

the input current quarks of
Perturbative QCD and Chiral Perturbation Theory
are Quantum, and not Pre-Quantum, so that
Perturbative QCD and Chiral Perturbation Theory are
effectively "second-order" Quantum theories
(rather than fundamental theories)
that are most useful in describing phenomena
at high and low energy levels, respectively; and
\vspace{12pt}

a current quark is a composite combination
of a fundamental constituent quark
and Quantum virtual sea gluon, quarks, and antiquarks
(compare the conventional picture of, for example,
hep-ph/9708262, in which current quarks are Pre-Quantum
and constituent quarks are Quantum composites).
\vspace{12pt}

\vspace{12pt}

\vspace{12pt}

\vspace{12pt}

Assuming the accepted values of
the gravitational force strength constant (Newton's constant)
and
the electron mass (0.511 MeV),
then the calculationed ratios give values of all
the other force strength constants and particle masses.
These calculated force strengths and particle masses agree
with conventionally accepted experimental results
within at most about 10 percent in all but one case:
\vspace{12pt}

the mass of the Truth quark (sometimes called the Top quark).
\vspace{12pt}

The tree level constituent mass of the Truth quark
is computed to be roughly 130 GeV, as opposed to
the roughly 175 GeV figure advocated by FermiLab.
\vspace{12pt}

In my opinion, the FermiLab figure is incorrect.
\vspace{12pt}

The Fermilab figure is based on analysis of
semileptonic events.  I think that the Fermilab
semileptonic analysis does not handle background correctly,
and ignores signals in the data that are in rough
agreement with the tree level constituent mass
of about 130 GeV.
\vspace{12pt}

Further, I think that dileptonic events are more
reliable for Truth quark mass determination,
even though there are fewer of them than semileptonic events.
\vspace{12pt}

I think that analysis of dileptonic events
gives a Truth quark mass that is in rough
agreement with the tree level constituent mass
of about 130 GeV.
\vspace{12pt}

More details about these issues,
including gif images of Fermilab data histograms
and other relevant experimental results,
can be found on the World Wide Web at URLs
\vspace{12pt}

http://galaxy.cau.edu/tsmith/TCZ.html
\vspace{12pt}

http://www.innerx.net/personal/tsmith/TCZ.html
\vspace{12pt}

I consider the mass of the Truth quark to be
a good test of the model, as the model can
be falsified if my interpretation of experimental results
turns out to be wrong.

\newpage

\subsection{D4-D5-E6 Lagrangian and 146 GeV Higgs.}

The D4-D5-E6 Lagrangian in 8-dimensional SpaceTime,
  prior to dimensional reduction, is the Integral over 8-dim
                               SpaceTime of

\vspace{12pt}

$$
\int_{V_{8}}
\partial_{8}^{2} \overline{\Phi_{8}} \wedge \star
\partial_{8}^{2} \Phi_{8}
+ F_{8} \wedge \star F_{8} +
\overline{S_{8\pm}} \not \! \partial_{8} S_{8\pm} +
GF + GH
$$

where $F_{8}$ is the 28-dimensional $Spin(8)$ curvature,
$\star$ is the Hodge dual, \\
$\partial_{8}$ is the 8-dimensional covariant derivative, \\
$\Phi_{8}$ is the 8-dimensional scalar field, \\
$\not \!  \partial_{8}$ is the 8-dimensional Dirac operator, \\
$V_{8}$ is 8-dimensional spacetime, \\
$S_{8\pm}$ are the $+$ and $-$ 8-dimensional
half-spinor fermion spaces, and \\
GF and GH are gauge-fixing and ghost terms.

\vspace{12pt}

\subsubsection{Ni-Lou-Lu-Yang R-R Method.}

Before I read quant-ph/9806009,
\vspace{12pt}

         To Enjoy the Morning Flower in the Evening -
   What does the Appearance of Infinity in Physics Imply?
\vspace{12pt}

by Guang-jiong Ni of the Department of Physics,
Fudan University, Shanghai 200433, P. R. China,
\vspace{12pt}

and the related paper hep-ph/9801264 by Guang-jiong Ni,
Sen-yue Lou, Wen-fa Lu, and Ji-feng Yang,
\vspace{12pt}

I did not correctly understand the Higgs mechanism.
\vspace{12pt}

Therefore, in my earlier papers I had wrongly stated that
the D4-D5-E6 model gives a Higgs scalar mass of about 260 GeV
and a Higgs scalar field vacuum expectation value of
about 732 GeV.
\vspace{12pt}

I now see that my earlier values were wrong,
and that the correct values under the D4-D5-E6 model
are a Higgs scalar mass of about 146 GeV and
a Higgs scalar field vacuum expectation value of about 252 GeV.
The fault was not with the D4-D5-E6 model itself,
but with my incorrect understanding of it with
respect to the Higgs mechanism.
\vspace{12pt}

Guang-jiong Ni, Sen-yue Lou, Wen-fa Lu, and Ji-feng Yang,
in hep-ph/9801264 \cite{NLLY}, used a new
Regularization-Renormalization (R-R)
method to calculate the Higgs mass in the Standard Model
to be about 140 GeV.
Guang-jiong Ni has further described the R-R method
in quant-ph/9806009 \cite{NI}.
\vspace{12pt}

When the R-R method of Ni is applied to the D4-D5-E6
physics model,
it is seen that, at tree level,
the mass of the Higgs scalar is about 146 GeV.
\vspace{12pt}

Here is a description of the R-R method of Ni:
\vspace{12pt}

For example , in the calculation of ``self-energy'' of election
in Quantum
ElectroDynamics ( QED) , there are Feynman Diagram Itegrals
(FDI) of the form:

\begin{equation}
I\hspace{0in}=\int \frac{d^4K}{\left( 2\pi \right) ^4}
\frac 1{\left(
K^2-M^2\right) ^2}
\end{equation}

\hspace{0in}\hspace{0in}
\begin{equation}
M^2=p^2x^2+\left( m^2-p^2\right) x
\end{equation}

where $x$ is the Feynman parameter , $K=k-xp.$

Consider the integration range of $K$
to be $\left( -\infty \rightarrow \infty \right) $ .
Since the denominator $%
\left( K^2-M^2\right) ^2\sim K^4$, so the integral diverges as
\begin{equation}
I\sim \int \frac{dK^4}{K^4}\hspace{0in}\sim
\int_0^\Lambda \frac{dK}K
\end{equation}
where a radius $\Lambda $ of large sphere in four dimensional
space is introduced \hspace{0in}and is called as the cutoff
of momentum in
integration . When $\Lambda \rightarrow \infty $ , there is a
logarithmic divergence. In other FDIs, one may encounter
the linear divergence $\left( \sim \Lambda \right) $ and
the quadratic
divergence $\left( \sim \Lambda ^2\right) $\hspace{0in} etc.

Taking the partial derivative of divergent integral I with
respect to the
parameter $M^2$with dimension of mass square,

\begin{equation}
\frac{\partial I}{\partial M^2}=2\int \frac{d^4K}
{\left( 2\pi \right) ^4}%
\frac 1{\left( K^2-M^2\right) ^3}
\end{equation}

As the denominator behaves as $K^6$, the integral
becomes convergent:

\begin{equation}
\frac{\partial I}{\partial M^2}=\frac{-i}
{\left( 4\pi \right) ^2}\frac 1{M^2}
\end{equation}

To get back to $I$, integrate with respect to $M^2$:

\begin{equation}
I=\frac{-i}{\left( 4\pi \right) ^2}
\left( \ln M^2+C_1\right) =\frac{-i}{%
\left( 4\pi \right) ^2}\ln \frac{M^2}{\mu _1^2}
\end{equation}

Here an arbitrary constant of integration $C_1$ appears.
\vspace{12pt}

Rewrite $C_1 = -\ln \mu _1^2$ with $\mu_1$ carrying
a mass dimension so that the argument of logarithmic function is
dimensionless.
\vspace{12pt}

By the chain approximation, derive a renormalized
mass $m_R=m+\delta m$ . When the freely moving particle is on
the mass-shell
, i.e. , $p^2=m^2$ , we have $\left( \alpha \equiv e^2/4\pi
\right) $:

\begin{equation}
\delta m=\frac{\alpha m}{4\pi }
\left( 5-3\ln \frac{m^2}{\mu _1^2}\right)
\end{equation}

The arbitrary constant $\mu _1$ is fixed as follows.
We want the parameter $m$ in the Lagrangian of original theory
to be
understood as the observed mass $m_R$ . Since the mass cannot
be calculated
by perturbative QFT, the mass
can only be fixed by experiment .
Hence , the condition $\delta m=0$ gives

\begin{equation}
\ln \frac{m^2}{\mu _1^2}=5/3,\mu _1=e^{-5/6}m
\end{equation}

The condition $m_R=m$ does not mean that the calculation of
self - energy FDI
is useless. When the motion of particle deviates from
the mass - shell , i.e. ,$p^2\neq m$ , the combination of the
self - energy
formula with other FDI$_s$ in QED gives useful results.
For instance, we are able to calculate quickly the (qualitative)
energy shift of $2S_{1/2}$ state upward with respect to
$2P_{1/2}$ state in
Hydrogen atom being 997 MHz , so called Lamb shift
(with experimental result
1057.8 MHz ). The latter should be viewed as a mass modification
for an electron in a bound state. By using perturbative QFT ,
mass modification can be evaluted, even though mass
generation cannot.

Therefore, we replace the divergence by an arbitrary constant
$\mu _1$ ,
and the constant $\mu _1$ is fixed by the mass $m$ measured in
the
experiment.
\vspace{12pt}

With respect to the trick of taking derivative for reducing
the degree of
divergence, the crucial point that we should act from the
beginning ,
act before the counterterm is introduced, act until the bottom
is reached .
\vspace{12pt}

That is, take the derivative of integral $I$ with respect
to $M^2$

( or to a parameter $\sigma$ added by hand,
say $M^2\rightarrow M^2+\sigma $ )

enough times until it becomes convergent,

then perform the corresponding integrations with respect
to $M^2$

( or $\sigma $ then setting $\sigma \rightarrow 0$ again )

to get back to $I$ .
\vspace{12pt}

Now instead of divergences,there are arbitrary
constants $C_i$ .

Note that each divergence is now resolved into one of the
constants $C_i$ to be fixed, so that each $C_i$ has its own
unique
meaning and role.

\vspace{12pt}

\vspace{12pt}

The difference between the D4-D5-E6 model Higgs mass
of 146 GeV
and the Ni,
Lou, Lu, and Yang calculated value of 138 GeV is due
to these facts:

       the D4-D5-E6 model value of 145.789 GeV (about 146 GeV)
is a tree level value, while Ni,
       Lou, Lu, and Yang calculate the Higgs mass beyond
tree level using a nonperturbative
       Gaussian Effective Potential (GEP) method;
       Ni, Lou, Lu, and Yang use as input parameters:
              the particle masses $m_{W+} = m_{W-}
= 80.359 GeV$
and $m_{Z0} = 91.1884 GeV$,
              the Weinberg angle $sin^{2}\theta_{W}
= s2w = 0.2317$,
and
              the Electromagnetic Fine Structure Constant at
the $Z_{0}$-mass energy level $\alpha_{E(mZ)} = 1 / 128.89$;
       the D4-D5-E6 model uses as input parameter the
vacuum
expectation value of the Higgs scalar
       field $v = 252.514 GeV$, which is based on
the D4-D5-E6 model
identity $v = m_{W+} + m_{W-} + m_{Z0}$,
and which leads to the tree-level
particle masses  $m_{W+} = m_{W-} = 80.326 GeV$
and $m_{Z0} = 91.862 GeV$ as well as the other calculated
particle
masses and force strength constants of the
       D4-D5-E6 model;
Ni, Lou, Lu, and Yang \cite{NLLY} use a T-quark mass
of 175 GeV,
giving them a larger 1-loop level
       fermion contribution than would be the case for
the D4-D5-E6 model which has a T-quark
       mass of 130 GeV.

\subsubsection{Lagrangian Scalar Term-1.}

As shown in chapter 4 of Gockeler and Schucker \cite{GSCH},

  the scalar part of the Lagrangian

$$
\int_{V_{8}} \partial_{8}^{2} \overline{\Phi_{8}} \wedge
\star \partial_{8}^{2} \Phi_{8}
$$

becomes

$$F_{h8} \wedge \star F_{h8}$$

where $F_{8}$ is an 8-dimensional Higgs curvature term.

\vspace{12pt}

 After dimensional reduction to 4-dim SpaceTime,
the scalar $F_{8} \wedge \star F_{8}$ term becomes

$$\int_{V_{4}} \int_{I}
( F_{h44} + F_{h4I} + F_{hII} ) \wedge
\star ( F_{h44} + F_{h4I} + F_{hII} ) = $$

$$= \int_{V_{4}} \int_{I}
F_{h44} \wedge \star F_{h44} +
F_{h4I} \wedge \star F_{h4I} +
F_{hII} \wedge \star F_{hII}$$

where cross-terms are eliminated by antisymmetry
of the wedge $\wedge$ product
and $4$ denotes 4-dim SpaceTime
and $I$ denotes 4-dim Internal Symmetry Space.
\vspace{12pt}

  The Internal Symmetry Space terms $I$ should be integrated
over the 4-dimensional Internal Symmetry Space $I$, to get
3 terms.

\vspace{12pt}

                   The first term is

$$\int_{V_{4}}
F_{h44} \wedge \star F_{h44}$$

Since it, like the weak force curvature term,
is an SU(2) gauge group terms,
this term merges into the SU(2) weak force term

$$\int_{V_{4}}
F_{w} \wedge \star F_{w}$$

 (where w denotes Weak Force).

\subsubsection{Lagrangian Scalar Term-3.}

The third term is

$$\int_{V_{4}} \int_{I}
F_{hII} \wedge \star F_{hII}$$

The third term after integration over the 4-dim Internal
Symmetry Space $I$, produces, by a process similar
to the Mayer Mechanism \cite{MAYER}, terms of the form

\begin{equation}
\int_{V_{4}}
- 2 \mu^2 \Phi ^2 + \lambda \Phi ^4
\end{equation}

where the
wrong-sign $\Lambda \Phi^{4}$ theory potential term
describes the Higgs Mechanism.
\vspace{12pt}

The form and notation above is used by Kane
and Barger and Phillips.
\vspace{12pt}

Ni, and Ni, Lou, Lu, and Yang,
use a different notation

\begin{equation}
V\left( \Phi \right) = - \frac 12\sigma \Phi^2 +
\frac 1{4!}\lambda_{N} \Phi ^4
\end{equation}

Proposition 11.4 of chapter II
of volume 1 of Kobayashi and Nomizu \cite{KN1} states that

$2 F_{hII}(X,Y) = [\Lambda(X), \Lambda(Y)] -
\Lambda([X,Y])$,

where $\Lambda$ takes values in the $SU(2)$ Lie algebra.

\vspace{12pt}

If the action of the Hodge dual $\star$ on $\Lambda$ is
such that

$\star \Lambda = - \Lambda$ and $\star [\Lambda, \Lambda] =
[\Lambda, \Lambda]$,

then
$F_{hII}(X,Y) \wedge
\star F_{hII}(X,Y) =
(1/4)([\Lambda(X), \Lambda(Y)]^{2} - \Lambda([X,Y])^{2} )$.

\vspace{12pt}

If integration of $\Lambda$ over $I$ is
$\int_{I} \Lambda \propto \Phi = (\Phi^{+}, \Phi^{0})$,
where $(\Phi^{+}, \Phi^{0})$ is the complex doublet HIggs
scalar field, then

\vspace{12pt}

$$\int_{I} F_{hII} \wedge
\star F_{hII} =
 (1/4) \int_{I} [\Lambda(X),\Lambda(Y)]^{2} -
\Lambda([X,Y])^{2} = $$

\vspace{12pt}

$$= (1/4) [ \lambda ( \overline{\Phi} \Phi)^{2} - \mu^{2}
\overline{\Phi} \Phi ]$$

\vspace{12pt}

which is the Higgs Mechanism potential term.

\subsubsection{Physical Interpretation and Mass Scale.}

In my notation (and that of Kane \cite{KAN} and
Barger and Phillips \cite{BPH}),
$2 \mu^2$ is the
square $m_{H}^2$ of the tree-level Higgs scalar particle mass,
and
$\Phi$ is the Higgs scalar field,
and its tree-level vacuum expectation value $V$ is given by

                $$ v^2 / 2 = P^2 = M^2 / 2 L$$
or
$$ M^2 = L v^2$$

The tree-level value of the fundamental mass scale vacuum expectation value
$v$
of the Higgs scalar field is set in the D4-D5-E6 model as the
sum
of the tree-level physical masses of the weak bosons
$$v = m_{W+} + m_{W-} + m_{Z0} = 80.326 + 80.326
+ 91.862  = 252.514 GeV$$

so that in the D4-D5-E6 model the
electron mass will be 0.5110 MeV.

The resulting equations, in my notation
(and that of Kane \cite{KAN} and Barger and Phillips
\cite{BPH}), are:

           $$m_{H}^2 = 2 \mu^2$$
$$\mu^2 = \lambda v^2$$
$$m_{H}^2 / v^2 = 2 \lambda$$

In their notation, with $m_{\sigma} = m_{H}$ and
with $\phi_{1}$ as the fundamental Higgs scalar field mass
scale,
Ni, Lou, Lu, and Yang \cite{NLLY} have
$$m_{H}^2 = 2 \sigma$$
$$\phi_{1}^2 = 6 \sigma / \lambda_{N}$$
so that
for the tree-level value of the Higgs scalar particle
mass they have
$$m_{H}^2 / \phi_{1}^2 = \lambda_{N} / 3$$

By combining the non-perturbative Gaussian Effective
Potential (GEP)
approach with their
Regularization-Renormalization (R-R) method,
Ni, Lou, Lu, and Yang \cite{NLLY} find that:
\vspace{12pt}

$m_{H}$ and $\phi_{1}$ are the two fundamental mass scales
of the Higgs mechanism, and
\vspace{12pt}

the fundamental Higgs scalar field mass scale
$\phi_{1}$ of Ni, Lou, Lu, and Yang \cite{NLLY} is
equivalent to
the vacuum expectation value $v$ of the Higgs scalar
field
in my notation (and that of Kane \cite{KAN} and
Barger and Phillips \cite{BPH}),
and
\vspace{12pt}

$\lambda_{N}$ (and the corresponding $\lambda$) can not
only
be interpreted as the Higgs scalar field self-coupling
constant, but also can be interpreted as determining the
invariant ratio between the mass squares of the
Higgs mechanism fundamental mass scales, $m_{H}^2$
and $\phi_{1}^{2} = v^{2}$.
\vspace{12pt}

Since the tree-level value of $\lambda_{N}$
is $\lambda_{N} = 1$,

and since

$$\lambda_{N} / 3 = m_{H}^2 / \phi_{1}^2
= m_{H}^2 / v^2 = 2 \lambda$$

then, at tree-level,

$$\lambda = \lambda_{N} / 6 = 1 / 6$$

so that, at tree-level

$$m_{H}^2 / \phi_{1}^2 = m_{H}^2 / v^2 = 2 / 6 = 1 / 3$$

In the D4-D5-E6 model, the fundamental mass scale vacuum
expectation
value $v$ of the Higgs scalar field is the fundamental
mass parameter that
is to be set to define all other masses by the mass ratio
formulas of the
model, and $v$ is set to be

                       $$v = 252.514 GeV$$

which is the sum
of the tree-level physical masses of the weak bosons

$$v = m_{W+} + m_{W-} + m_{Z0} = 80.326 + 80.326 + 91.862
= 252.514 GeV$$

so that in the D4-D5-E6 model the electron mass will
be 0.5110 MeV.
\vspace{12pt}

Then, the tree-level mass $m_{H}$ of the Higgs scalar
particle is given by

                   $$m_{H} = v / \sqrt{3} = 145.789 GeV$$

Ni, Lou, Lu, and Yang use their Quantum Field Theory
model to
calculate two more important mass scales:
\vspace{12pt}

The Critical Mass (or Energy, or Temperature) $M_{SSB}$
for restoration of the Spontaneous Symmetry Breaking
(SSB) symmetry,
which is $M_{SSB} = m_{H} \sqrt{12 / Ln}$, so that,
for the tree-level value $\lambda_{N} = 1$,

                        $$M_{SSB} = m_{H} \sqrt{12} =
505 GeV$$

The High-Energy Singularity of the Higgs Mechanism
model,
$M_{SING}$, beyond which the Higgs field
vanishes, and
\vspace{12pt}

the Maximum Energy Scale $M_{MAX}$ that can be calculated
in
the Higgs Mechanism model.
\vspace{12pt}

The fact that the Higgs Mechanism model is not calculable
and
the Higgs field vanishes above
Msing and Mmax may justify regarding the Higgs Mechanism
model
as a low energy effective theory,
just as the D4-D5-E6 model is fundamentally a low
(with respect to the Planck energy) energy effective
theory.
\vspace{12pt}

The values calculated by Ni, Lou, Lu, and Yang are

        $$M_{SING} = 0.55 \times 10^{15} GeV$$

$$M_{MAX} = 0.87 \times 10^{15} GeV$$

The Planck energy is

$$M_{PLANCK} = 1.22 \times 10^{19} GeV$$

\subsubsection{Complex Doublet.}

The Higgs scalar field $\Phi$ is a Complex Doublet that
can be expressed in
      terms of a vacuum expectation value $v$ and
a real Higgs field $H$.

The Complex Doublet

$$\Phi = (\Phi^{+}, \Phi^{0}) = (1 / \sqrt{2})
( \Phi_{1} + i \Phi_{2}, \Phi_{3} + i \Phi_{4} ) =
(1 / \sqrt{2}) ( 0, v + H )$$

so that

$$\Phi_{3} = (1 / \sqrt{2}) ( v + H ) $$

where $v$ is the vacuum expectation value and
$H$ is the real surviving Higgs field.

\vspace{12pt}

The value of the fundamental mass scale vacuum
expectation
value $v$ of the Higgs scalar field is in the
D4-D5-E6 physics model set to be 252.514 GeV,
so that the electron mass will turn out to be 0.5110 MeV.

\vspace{12pt}

Now, to interpret the term

$$\int_{I} F_{hII} \wedge
\star F_{hII} =
 (1/4) \int_{I} [\Lambda(X),\Lambda(Y)]^{2} -
\Lambda([X,Y])^{2} = $$

$$= (1/4) [ \lambda ( \overline{\Phi} \Phi)^{2}
- \mu^{2}
\overline{\Phi} \Phi ]$$

in terms of $v$ and $H$,
\vspace{12pt}

note that $\lambda = \mu^2 / v^2$

and that $\Phi = \Phi_{3} = (1 / \sqrt{2}) ( v + H ) $,

so that

$$
F_{hII}(X,Y) \wedge \star F_{hII}(X,Y) =
= (1/4) [ \lambda ( \overline{\Phi} \Phi)^{2}
- \mu^{2}
\overline{\Phi} \Phi ] =
$$

$$
= (1/16) ((\mu^2 / v^2) ( v + H )^4 - (1/8) M^2 ( v + H )^2 =
$$

$$
= (1/4) \mu^2 H^2 - (1/16) \mu^2 v^2 ( 1 - 4 H^3 / v^3
- H^4 / v^4 )
$$

Disregarding some terms in $v$ and $H$,

$$
F_{hII}(X,Y) \wedge \star F_{hII}(X,Y) = (1/4) \mu^2 H^2
- (1/16) \mu^2 v^2
$$

\subsubsection{Lagrangian Scalar Term-2.}

The second term is

$$\int_{V4} \int_{I} F_{h4I} \wedge *F_{h4I}$$

The second term, after integration over the 4-dim Internal
Symmetry Space $I$,
produces, by a process similar to the Mayer Mechanism
\cite{MAYER},
terms of the form

$$\int_{V4} \partial \overline{\Phi} \partial \Phi$$

Proposition 11.4 of chapter II of volume 1 of Kobayashi
and Nomizu \cite{KN1}
states that

$$2 F_{h4I}(X,Y) = [\Lambda(X), \Lambda(Y)] - \Lambda([X,Y])$$

where $\Lambda(X)$ takes values in the SU(2) Lie algebra.
\vspace{12pt}

If the $X$ component of $F_{h4I}(X,Y)$ is in the surviving
4-dim SpaceTime and the $Y$ component of $F_{h4I}(X,Y)$ is
in the 4-dim Internal Symmetry Space $I$,
then the Lie bracket product $[X,Y] = 0$ so
that $\Lambda ([X,Y]) = 0$
and therefore

$$F_{h4I}(X,Y) = (1/2) [\Lambda(X), \Lambda(Y)]
= (1/2) \partial_{X} \Lambda(Y)$$

Integration over Internal Symmetry Space $I$ of

 $$(1/2) \partial_{X} \Lambda(Y)$$

gives $(1/2) \partial_{X} \Phi$.
\vspace{12pt}

Taking into account the Complex Doublet structure of $\Phi$,
the second term is the Integral over 4-dim SpaceTime of

$$
F_{h4I}(X,Y) \wedge \star F_{h4I} =
(1/2) \partial \Phi \wedge *(1/2) \partial \Phi =
(1/4) \partial \Phi \wedge * \partial \Phi =
$$

$$
= (1/4) (1/2) \partial ( v + H ) /\ *\partial ( v + H ) =
(1/8) \partial H \partial H + (some terms in v , H)
$$

Disregarding some terms in $v$ and $H$,

$$F_{h4I}(X,Y) \wedge \star F_{h4I} =
(1/8) \partial H \partial H$$

\subsubsection{Total Higgs Lagrangian.}

Combining the second and third terms,
since the first term is merged into the weak force part
of the Lagrangian:

$$
F_{h4I}(X,Y) \wedge \star F_{h4I} + F_{hII}(X,Y) \wedge
\star F_{hII}(X,Y) =
$$

$$
= (1/8) \partial H \partial H + (1/4) \mu^2 H^2 - (1/16)
\mu^2 v^2 =
$$

$$
= (1/8) ( dH dH + 2 \mu^2 H^2 - (1/2) \mu^2 v^2 )
$$

  This is the form of the Higgs Lagrangian in Barger and
Phillips \cite{BPH}
for a Higgs scalar particle of mass

$$ m_{H} = \mu \sqrt{2} = v / \sqrt{3} = 145.789 GeV$$

\subsubsection{Weak Force.}

What about the Weak Force Strength and Weak Boson Masses?
\vspace{12pt}

In the D4-D5-E6 model, the geometric part of the weak
force strength,
and the geometric weak charge that is its square root, are:

$$\alpha_{W} = 0.2535$$
$$\sqrt{\alpha_{W}} = 0.5035$$

In more customary particle physics notation,
such as that found in Kane \cite{KAN}, there are two weak
charges,
$g_{1}$ and $g_{2}$, such that their squares are weak force strengths.
In the D4-D5-E6 model,
the geometric weak charge is the average of
the customary two weak charges $g_{1}$ and $g_{2}$:

$$\sqrt{\alpha_{W}} = (1/2) ( g_{1} + g_{2} )$$

so that the numerical values are

$$g_{1} = 0.33566$$
$$g_{1}^2 = 0.11267$$
$$g_{2} = 0.66434$$
$$g_{2}^2 = 0.44135$$

Combining some aspects of the D4-D5-E6 model and
some aspects of the customary picture gives
tree-level estimate results that are off by a few percent.
Estimated Weak Boson masses are approximately

$$m_{W} = g_{2} v / 2 = 83.88 GeV$$

$$m_{Z} = \sqrt{ g_{1}^2 + g_{2}^2 } v / 2 = 93.98 GeV$$

Some other relations given by Kane \cite{KAN}, and
the results of using in them some D4-D5-E6 model values,
and the estimates immediately above, are

$$e = \sqrt{4 \pi \alpha_{E}} = 0.30286$$

$$g_{2} = e / \sin{\theta_{W}} = 0.6247$$

$$g_{1} = e / \cos{\theta_{W}} = 0.3463$$

$$G_{F} = \sqrt{2} g_{2}^2 / 8 m_{W}^2 =
1.11 \times 10^{-5} GeV^{-2}$$

For comparison, the D4-D5-E6 model value of the Fermi
constant is

$$G_{F} = \alpha_{W} / M_{W}^2
= 1.188 \times 10^{-5} GeV^{-2}$$

where

$$M_{W} = \sqrt{(m_{W+}^2 + m_{W-}^2 + m_{Z0}^2)} =$$

$$= \sqrt{(80.326^2 + 80.326^2 + 92.862^2)} = 146.09298 GeV$$

Note that $M_{W}$ is very close to
the Higgs scalar particle mass $m_{H} = 145.789 GeV$.

\subsubsection{Higgs, Truth Quark, and Weak Bosons.}

The tree level mass $m_{H}$ of a Higgs scalar,
about 146 GeV,
is somewhat higher than, but roughly similar to,
the tree level Truth quark mass $m_{T}$ of about 130 GeV.
\vspace{12pt}

In the D4-D5-E6 physics model,
the sum of the tree level masses
$m_{W+}^2 + m_{W-}^2 + m_{Z0}^2$
of the 3 weak bosons $W_{+}$, $W_{-}$, and $Z_{0}$,
that is, the physical weak bosons below the Higgs mass
scale,
is the fundamental energy level vacuum expectation
value $v$
of the Higgs scalar field.
\vspace{12pt}

To give the tree-level particle masses  $m_{W+} = m_{W-}
= 80.326 GeV$
and $m_{Z0} = 91.862 GeV$ (as well as the other
calculated
particle masses and force strength constants
of the D4-D5-E6 model),
$v$ is set equal to 252.514 GeV.
\vspace{12pt}

The D4-D5-E6 model assumed value of $v$ is about 80+80+92
= 252 GeV
which is close to the tree level mass
of the truth quark T-T(bar) meson of about 260 GeV.
\vspace{12pt}

$M_{W} = \sqrt{(m_{W+}^2 + m_{W-}^2 + m_{Z0}^2)}$,
the square root
of the sum of the squares of the tree level masses
of the 3 weak bosons $W_{+}$, $W_{-}$, and $Z_{0}$,
that is, the physical weak bosons below the Higgs
mass scale,
is 146.09298 GeV, which is very close to
the mass of the tree level Higgs scalar mass of 145.789 GeV.
\vspace{12pt}

The tree level mass of a pair of Higgs scalars,
about 292 GeV,
is somewhat higher than, but roughly similar to,
the fundamental energy level vacuum expectation value $v$
of the Higgs scalar field, about 252 GeV,
and
the truth quark T-T(bar) meson mass of about 260 GeV.
\vspace{12pt}

\newpage

\subsection{Mathematical Structures.}

The D4-D5-E6 physics model emerges from bits
just as it does from the points of
Simplex Physics above the Planck Energy
which is similar to its emergence
from the arrows of quantum set theory
and from the structure of Metaclifford algebras.
\vspace{12pt}

The mathematical structures used in the D4-D5-E6 model
are
much like the structures of:
\vspace{12pt}

IFA, whose $256 = 2^8 = 2^4 \times 2^4 = 16 \times 16$
elements correspond to the Cl(8) Clifford Algebra;
\vspace{12pt}

Wei Qi, whose board grid and stones correspond to
the Spin(8) vector HyperDiamond lattice;
\vspace{12pt}

I Ching, whose 28 antisymmetrized off-diagonal hexagrams,
out of the total $64 = 2^6 = 2^3 \times 2^3 = 8 \times 8$ hexagrams,
correspond to the 28Spin(8) bivector gauge bosons;
\vspace{12pt}

Tai Hsuan Ching, whose $81 = 3^{4}$ ternary power-of-3
structure
corresponds to 16 first-generation Spin(8) spinor
fermions,
and whose ternary structure corresponds to the
3-generation structure
of fermions in the D4-D5-E6 model; and
\vspace{12pt}

Tarot, whose 78 cards correspond to 78-dimensional E6.
\vspace{12pt}

\newpage

\subsection{Outline of Chapters.}

The purpose of this paper is to describe the model
in some detail.
\vspace{12pt}

After this introductory overview,
\vspace{12pt}

{\bf Chapter 2} describes the construction of discrete
Clifford algebras
from set theory and some simple natural operations.
\vspace{12pt}

Begin with set theory to get the Natural Numbers $N$.
\vspace{12pt}

Then use reflection to get the integers $Z$.
\vspace{12pt}

Then use the set of subsets and the XOR operator to get
the Discrete Clifford Group ($DClG(n)$).
\vspace{12pt}

$DClG(n)$ is extended to its $Z$-Group Algebra,
thus producing a discrete real Clifford Algebra ($DCl$(0,n))
over the $Z_{n}$ hypercubic lattice.
\vspace{12pt}

For some $n$, $DCl$(0,n) is naturally extended from
the $Z_{n}$ hypercubic lattice to larger lattices,
such as the $D_{4}$ quaternionic integer lattices for $n=4$
and an $E_{8}$ octonionic integer lattices for $n=8$.
\vspace{12pt}

The real Clifford Algebras have periodicity 8,
so the fundamental real Clifford Algebra produced
by this process is $DCl(0,8)$.
\vspace{12pt}

The vector, +half-spinor and -half-spinor
representations $DCl(0,8)$ are all isomorphic by triality to
the discrete integral octonions.
\vspace{12pt}

{\bf Chapter 3} describes the octonionic structure
of $E_{8}$ lattices.
\vspace{12pt}

{\bf Chapter 4} describes the scalar representation
of $DCl(0,8)$
as physically representing the Higgs scalar particle;
\vspace{12pt}

the vector representation of $DCl(0,8)$
as an $E_{8}$ lattice physically representing
an 8-dimensional spacetime;
\vspace{12pt}

the bivector representation of $DCl(0,8)$
as having 28 basis bivectors that represent
the 28 gauge boson infinitesimal generators
of a $Spin(0,8)$ gauge group;
\vspace{12pt}

and the two half-spinor representations of $DCl(0,8)$
as two $E_{8}$ lattices, in which the 8 octonion basis
vectors
of each physically represent 8 fundamental
first-generation
fermion particles (neutrino; red, blue, green up quarks;
red, blue green down quarks; electron) and
8 fermion antiparticles.
\vspace{12pt}

{\bf Chapter 5} describes a 4-dimensional HyperDiamond
lattice spacetime
that comes from a 4-dimensional sub-lattice of
the $E_{8}$
lattice spacetime.
\vspace{12pt}

{\bf Chapter 6} describes a 4-dimensional internal
symmetry space
that comes from the rest of the original $E_{8}$
lattice spacetime.
A separate copy of the internal symmetry space lives
on each vertex
of the spacetime 4-dim HyperDiamond lattice.
Each copy of the internal symmetry space looks
like a 4-dim HyperDiamond lattice.
\vspace{12pt}

{\bf Chapter 7} describes how a Feynman Checkerboard
construction
on the HyperDiamond structures gives
the physics of the Standard Model plus Gravity.
\vspace{12pt}

{\bf Chapter 8} describes the numerical calculation of
charge as the amplitude for a particle to emit a gauge
boson,
with the force strength constant being the square of the charge.
\vspace{12pt}

{\bf Chapter 9} describes the numerical calculation of
particle mass as the amplitude for a particle to
change
direction.
\vspace{12pt}

{\bf Chapter 10} describes HyperDiamond Feynman
Checkerboard configurations
that represent protons as triples of confined quarks,
pions as confined quark-antiquark pairs,
and physical gravitons as quadruples of confined
Spin(0,5) gravitons.
\vspace{12pt}

{\bf Appendix A} describes some earlier papers,
including some
errata for them.
\vspace{12pt}

The $D_{4}-D_{5}-E_{6}$ model that is described in
some detail
on the World Wide Web at URL
\vspace{12pt}

http://xxx.lanl.gov/abs/hep-ph/9501252
\vspace{12pt}

is a continuum version of the
HyperDiamond Feynman Checkerboard model at URL
\vspace{12pt}

http://xxx.lanl.gov/abs/hep-ph/9501252
\vspace{12pt}

Both of those papers, and all my papers written prior
to June 1998, contain an incorrect value (about 260 GeV)
of the Higgs scalar mass, which should be about 146 GeV.
\vspace{12pt}

The $D_{4}-D_{5}-E_{6}$ model is also described on the
World Wide Web at URLs
\vspace{12pt}

http://galaxy.cau.edu/tsmith/d4d5e6hist.html
\vspace{12pt}

http://www.innerx.net/personal/tsmith/d4d5e6hist.html
\vspace{12pt}

Briefly, roughly, and non-rigorously,
the $D_{4}-D_{5}-E_{6}$ model is
constructed from $E_{6}$, $D_{5} = Spin(10$, and $D_{4}
= Spin(8)$:
\vspace{12pt}

The first generation of fermions are constructed
from $E_{6} / Spin(10) \times U(1)$, whose dimension
is 78-45-1=32,
the real dimensionality of a bounded complex domain
whose Shilov boundary has real dimension 16=8+8 for
8 fermion particles and 8 for antiparticles;
\vspace{12pt}

An 8-dimensional spacetime is constructed
from $Spin(10) / Spin(8) \times U(1)$, whose
dimension is 45-28-1=16, the real dimensionality
of a bounded complex domain whose Shilov boundary
has real dimension 8 for 8-diemnsional spacetime;
\vspace{12pt}

28 gauge bosons are constructed directly from
28-dimensional $Spin(8)$.
\vspace{12pt}

Then reduce spacetime from 8 dimensions to
4 dimensions,
by choosing a quaternionic subspace
of octonionic 8-dimensional spacetime.
\vspace{12pt}

The result of that spacetime symmetry breaking is:
\vspace{12pt}

The fermions get 3 generations, corresponding
to $E_{6}$, $E_{7}$, and $E_{8}$;
\vspace{12pt}

The 28 = 16+12 gauge bosons split into two parts:
\vspace{12pt}

16 of them form $U(4)$, which is $U(1) \times SU(4)$,
the $U(1)$ to give a complex phase to propagator amplitudes,
the $SU(4) = Spin(6)$ to give the conformal group,
the $Spin(6)$ (compact version of Spin(4,2) conformal group
has 5 dimensions of conformal and scale
transformations that give a mass scale and a Higgs scalar
\vspace{12pt}

and
has 10 dimensions that give the $Spin(5)$ deSitter group,
which is gauged to give Gravity;
\vspace{12pt}

12 = 8+3+1 of them form $SU(3) \times SU(2) \times U(1)$
of the Standard Model.
\vspace{12pt}

The D4-D5-E6 model coset spaces $E6 / (D5 \times U(1))$ and
$D5 / (D4 \times U(1))$ are Conformal Spaces.
\vspace{12pt}

You can continue the chain to $D4 / (D3 \times U(1))$
where D3 is the 15-dimensional Conformal Group whose
compact version is Spin(6),
\vspace{12pt}

and to $D3 / (D2 \times U(1))$ where D2 is the
6-dimensional Lorentz Group whose compact version is Spin(4).
\vspace{12pt}

Electromagnetism, Gravity, and the Zero Point Fluctuation
of the vacuum
all have in common the symmetry of the 15-dimensional D3
Conformal Group
whose compact version is Spin(6),
as can be seen by the following structures with D3 Conformal
Group symmetry:
\vspace{12pt}

Maxwell's equations of Electromagnetism;
\newline

Gravity derived from the Conformal Group using
the MacDowell-Mansouri mechanism;
\newline

the Quantum Theoretical Hydrogen atom; and
\newline

the Lie Sphere geometry of SpaceTime Correlations in the Many-Worlds picture.
\vspace{12pt}

Further, the 12-dimensional Standard Model Lie Algebra
$U(1) \times SU(2) \times SU(3)$ may be related to the D3
Conformal Group Lie Algebra in the same way
that the 12-dimensional Schrodinger Lie Algebra is
related to the D3 Conformal Group Lie Algebra.
\vspace{12pt}

\newpage

\section{From Sets to Clifford Algebras.}

\subsection{Sets, Reflections, Subsets, and XOR.}

Start with von Neumann's Set Theoretical definition of
the Natural Numbers $N$:
\vspace{12pt}

$0 = \emptyset , 1= \{ \emptyset \} , 2 = \{ \emptyset ,
\{ \emptyset \} \} , ... , n + 1 = n \cup  \{ n \}  $ . \\
\vspace{12pt}

in which each Natural Number $n$ is a set of $n$ elements.
\vspace{12pt}

By reflection through zero, extend the Natural Numbers $N$
to include the negative numbers,
thus getting an Integral Domain,
the Ring $Z$.
\vspace{12pt}

Now, following the approach of Barry Simon \cite{Simon},
consider a set $S_{n} = \{ e_{1}, e_{2}, ... , e_{n} \} $
of $n$ elements.
\vspace{12pt}

Consider the set $2^{S_{n}}$ of all of its $2^{n}$ subsets,
with a product on $2^{S_{n}}$ defined
as the symmetric set difference $XOR$.
\vspace{12pt}

Denote the elements of $2^{S_{n}}$ by $m_{A}$ where $A$!is
in $;^{S_{n}}$.
\vspace{12pt}

\subsection{Discrete Clifford Algebras.}

To go beyond set theory to Discrete Clifford Algebras,
enlarge $2^{S_{n}}$ to $DClG(n)$ by:
\vspace{12pt}

order the basis elements of $S_{n}$,
\vspace{12pt}

and then give each element of $2^{S_{n}}$ a sign,
either +1 or -1, so that $DClG(n)$ has $2^{n + 1}$ elements.
\vspace{12pt}

This amounts to orientation of the signed unit basis of $S_{n}$.
\vspace{12pt}

Then define a product on $DClG(n)$ by
\vspace{12pt}

$(x_{1} e_{A}) (x_{2} e_{B}) = x_{3} e_{A \: XOR \: B})$
\vspace{12pt}

where $A$ and $B$ are in $2^{S_{n}}$ with ordered elements,
and $x_{1}$, $x_{2}$, and $x_{3}$ determine the signs.
\vspace{12pt}

For given $x_{1}$ and $x_{2}$, $x_{3} = x_{1} x_{2} x(A,B)$
where $x(A,B)$ is a function that determines sign by
using the rules
\vspace{12pt}

$e_{i} e_{i} = +1$ for $i$ in $S_{n}$
\vspace{12pt}

and $e_{i} e_{j} = - e_{j} e_{i}$ for $i \neq j$ in $S_{n}$ ,
\vspace{12pt}

then writing $(A,B)$ as an ordered set of elements of $S_{n}$,
\vspace{12pt}

then using  $e_{i} e_{j} = - e_{j} e_{i}$ to move
each of the $B$-elements to the left until it:
\vspace{12pt}

either meets a similar element and then cancelling it
with the similar element by using $e_{i} e_{i} = +1$
\vspace{12pt}

or it is in between two $A$-elements in the proper order.
\vspace{12pt}

$DClG(n)$ is a finite group of order $2^{n + 1}$.
\vspace{12pt}

It is the Discrete Clifford Group of $n$ signed
ordered basis elements of $S_{n}$.
\vspace{12pt}

Now we can construct a discrete Group Algebra of $DClG(n)$
by extending $DClG(n)$ by the integral domain ring $Z$
\vspace{12pt}

and using the relations
\vspace{12pt}

$e_{i} e_{j} + e_{j} e_{i} =  2  \delta(i,j) 1$
\vspace{12pt}

where $\delta(i,j)$ is the Kronecker delta.
\vspace{12pt}

Since $DClG(n)$ is of order $2^{n + 1}$,
and since two of its elements are $-1$ and $+1$,
which act as scalars, the discrete Group Algebra of $DClG(n)$
is $2^{n}$-dimensional.  The vector space on which it
acts is the $n$-dimensional hypercubic lattice $Z^{n}$.
\vspace{12pt}

The discrete Group Algebra of discrete Clifford Group $DClG(n)$
is the discrete Clifford Algebra $DCl(n)$.
\vspace{12pt}

Here is an explicit example showing how to assign
the elements of the Clifford Group
to the basis elements of the Clifford Algebra Cl(3):
\vspace{12pt}

First, order the $2^{3+1} = 16$ group strings into rows lexicographically:

\begin{equation}
\begin{array}{|c|}
\hline
0000\\
0001\\
0010\\
0011\\
0100\\
0101\\
0110\\
0111\\
1000\\
1001\\
1010\\
1011\\
1100\\
1101\\
1110\\
1111\\
\hline
\end{array}
\end{equation}

Then discard the first bit of each string,
because it corresponds to sign which is redundant
in defining the Algebra basis.
This reduces the number of different strings to $2^3 = 8$:

\begin{equation}
\begin{array}{|c|}
\hline
000\\
001\\
010\\
011\\
100\\
101\\
110\\
111\\
\hline
\end{array}
\end{equation}

Then separate them into columns by how many 1's they have:

\begin{equation}
\begin{array}{|c|c|c|c|}
\hline
000 &     &     &     \\
    & 001 &     &     \\
    & 010 &     &     \\
    &     & 011 &     \\
    & 100 &     &     \\
    &     & 101 &     \\
    &     & 110 &     \\
    &     &     & 111 \\
\hline
\end{array}
\end{equation}

Now they are broken down into the 1 3 3 1 graded pattern
of the Clifford Algebra Cl(3).
\vspace{12pt}

The associative Cl(3) product can be deformed
into the 1 x 1 Octonion nonassociative product
by changing EE from 1 to -1,
and IJ from K to -k, JK from I to -i, KI from J to -j,
and changing the cross-terms accordingly.

If you want to make an Octonion basis without the graded structure,
and with the 7 imaginary octonions all on equal footing,
all you have to do is to assign them, one-to-one in the order
starting from the left column and from the top of each row,
to the 8 Octonion Algebra strings:

\begin{equation}
\begin{array}{|c|c|c|}
\hline
 1  & 000 & 0000000 \\
 I  & 001 & 1000000 \\
 J  & 010 & 0100000 \\
 K  & 100 & 0010000 \\
 i  & 110 & 0001000 \\
 j  & 101 & 0000100 \\
 k  & 011 & 0000010 \\
 E  & 111 & 0000001 \\
\hline
\end{array}
\end{equation}

To give an example of how to write an octonion product
in terms of XOR operations,
look at the 7 associative triangles:

\begin{equation}
\begin{array}{|c|}
\hline
ijk\\
JiK\\
jIK\\
JIk\\
IEi\\
JEj\\
KEk\\
\hline
\end{array}
\end{equation}

which, in string terms,
are each represented by 3 element strings
and 1 Asssociative Triangle string:

\begin{equation}
\begin{array}{|c|c|c|c|}
\hline
Octonion & 3 Elements & Associative & Coassociative \\
         &            & Triangle    & Square        \\
\hline
         & i-0001000  &             &               \\
   I     & J-0100000  &  0111000    &  1000111      \\
         & K-0010000  &             &               \\
\hline
         & I-1000000  &             &               \\
   J     & j-0000100  &  1010100    &  0101011      \\
         & K-0010000  &             &               \\
\hline
         & I-1000000  &             &               \\
   K     & J-0100000  &  1100010    &  0011101      \\
         & k-0000010  &             &               \\
\hline
         & E-0000001  &             &               \\
   i     & I-1000000  &  1001001    &  0110110      \\
         & i-0001000  &             &               \\
\hline
         & E-0000001  &             &               \\
   j     & J-0100000  &  0100101    &  1011010      \\
         & j-0000100  &             &               \\
\hline
         & E-0000001  &             &               \\
   k     & K-0010000  &  0010011    &  1101100      \\
         & k-0000010  &             &               \\
\hline
         & i-0001000  &             &               \\
   E     & j-0000100  &  0001110    &  1110001      \\
         & k-0000010  &             &               \\
\hline
\end{array}
\end{equation}

Here is Onar Aam's method for calculating the octonion
product $a$b
of octonion basis elements $a$ and $b$ in terms
of these strings:
\vspace{12pt}

To multiply using triangles,
note that there are 7 octonion imaginary elements
and
that $XOR$ of two triangles give a square
and
that the Hodge dual (within imaginary octonions) of
that square is the triangle that
represents the product of the two triangles,
so that:

$$ab = \star (a XOR b)$$

For instance,
here is an example of multiplying by triangles
(up to +/- sign determined by ordering):

$$EI = \star ( 0001110 XOR 0111000 ) = \star (0110110)
= 1001001 = i$$

On the other hand,
the $XOR$ of two squares is a square,
so that
multiplying by squares simply becomes an $XOR$.
and we have (up to +/- sign determined by ordering):

$$ij = 0110110 XOR 1011010 = 1101100 = k$$

The Clifford Algebra product $\cdot$ combines
the vector space exterior $\wedge$ product
and the vector space interior product $\vdash$
so that, if $a$ is a 1-vector and $B$ is a k-vector,

$$a \cdot B  =  a \wedge B  -  a \vdash B$$

The Clifford Algebra has the same graded structure
as the exterior $\wedge$ algebra of the vector space,
and
the underlying exterior product antisymmetry rule that,
for p-vector $A$ and q-vector $B$

$$A \wedge B  =  (-1)^{pq}  B \wedge A$$

The associative Cl(3) product can be deformed
into the 1 x 1 Octonion nonassociative product
by changing $EE$ from 1 to -1,
and $IJ$ from $K$ to $-k$, $JK$ from $I$ to $-i$,
$KI$ from $J$ to $-j$,
and changing the cross-terms accordingly.
\vspace{12pt}

A fundamental reason for the deformation is that
the graded structure of the Clifford algebra gives it
the underlying exterior product antisymmetry rule that
while, for Octonions,
you want for all unequal imaginary octonions to have
the antisymmetry rule  $AB  =  - BA$
and for equal ones to have  $AA = -1$.
\vspace{12pt}

\vspace{12pt}

The discrete Clifford Algebra $DCl(n)$ acts on
the $n$-dimensional hypercubic lattice $Z^{n}$.
\vspace{12pt}

\subsection{Real Clifford and Division Algebras.}

Compare the Discrete Clifford Algebra, with
the Real Group Algebra of $DClG(n)$ made by
extending it by the real numbers $R$, which is the usual real
Euclidean Clifford Algebra $Cl(n)$ of dimension $2^{n}$,
acting on the real $n$-dimensional vector space $R^{n}$.)
\vspace{12pt}

The empty set $\emptyset$ corresponds to a vector
space of dimension -1,
so that $Cl(-1)$ corresponds to the VOID.
\vspace{12pt}

$Cl(0)$ has dimension $2^{0} = 1$  and
corresponds to the real numbers and to time.
Its even subalgebra is EMPTY.
$Cl(0)$ is the $1 \times 1$ real matrix algebra.
\vspace{12pt}

$Cl(1)$ has dimension $2^{1} = 2 = 1 + 1 = 1 + i $  and
corresponds to the complex numbers and to 2-dim space-time.
Its even subalgebra is $Cl(0)$.
$Cl(1)$ is the $1 \times 1$ complex matrix algebra.
\vspace{12pt}

$Cl(2)$ has dimension
$2^{2} = 4 = 1 + 2 + 1 = 1 + \{ j,k \} + i $  and
corresponds to the quaternions and to 4-dim space-time.
Its even subalgebra is $Cl(1)$.
$Cl(2)$ is the $1 \times 1$ quaternion matrix algebra.
\vspace{12pt}

$Cl(3)$ has dimension $2^{3} = 8 = 1 + 3 + 3 + 1 =
1 + \{ I,J,K \} + \{ i,j,k \} + E $  and
corresponds to the octonions and to 8-dim space-time.
Its even subalgebra is $Cl(2)$.
$Cl(3)$ is the direct sum of two $1 \times 1$
quaternion matrix algebras.
\vspace{12pt}

The associative Cl(3) product can be deformed
into the 1 x 1 Octonion nonassociative product
by changing $EE$ from 1 to -1,
and $IJ$ from $K$ to $-k$, $JK$ from $I$ to $-i$,
$KI$ from $J$ to $-j$,
and changing the cross-terms accordingly.
\vspace{12pt}

A fundamental reason for the deformation is that
the graded structure of the Clifford algebra gives it
the underlying exterior product antisymmetry rule that
while, for Octonions,
you want for all unequal imaginary octonions to have
the antisymmetry rule  $AB  =  - BA$
and for equal ones to have  $AA = -1$.
\vspace{12pt}

There are a number of choices you can make
in writing the octonion multiplication table:
\vspace{12pt}

Given $1$, $i$, and $j$,
there are 2 inequivalent quaternion multiplication tables,
one with $ij = k$ and the reverse with $ji = k$, or $ij = -k$.
\vspace{12pt}

To get an octonion multiplication table,
start with an orthonormal basis of 8 octonions
$\{ 1,i,j,k,E,I,J,K \}$, and
\vspace{12pt}

pick a scalar real axis $1$
\vspace{12pt}

and pick (2 sign choices) a pseudoscalar axis $E$ or $-E$.
\vspace{12pt}

Then you have 6 basis elements to designate as $i$ or $-i$,
which is 6 element choices and 2 sign choices.
\vspace{12pt}

Then you have 5 basis elements to designate as $j$ or $-j$,
which is 5 element choices and 2 sign choices.
\vspace{12pt}

Then the underlying quaternionic product fixes $ij$
as $k$ or $-k$,
which is 2 sign choices.
\vspace{12pt}

How many inequivalent octonion multiplication tables are there?
\vspace{12pt}

You had 6 i-element choices, 5 j-element choices,
and 4 sign choices,

for a total of $6 \times 5 \times 2^4 = 30 \times
16 = 480$ octonion products.
\vspace{12pt}

$Cl(4)$ has dimension $2^{4} = 16 = 1 + 4 + 6 + 4 + 1 =$
\vspace{12pt}

$= 1 + \{ S,T,U,V \} + \{ i,j,k,I,J,K \} + \{ W,X,Y,Z \} + E $
\vspace{12pt}

$Cl(4)$ corresponds to the sedenions.
Its even subalgebra is $Cl(3)$.
$Cl(4)$ is the $2 \times 2$ quaternion matrix algebra.
\vspace{12pt}

$Cl(4)$ is the first Clifford algebra that is NOT
made of $1 \times 1$ matrices or the direct sum of
$1 \times 1$ matrices,
and the $Cl(4)$ sedenions do NOT form a division algebra.
\vspace{12pt}

For more details see these WWW URLs,
web pages, and references therein:

Clifford Algebras:
\vspace{12pt}

http://galaxy.cau.edu/tsmith/clfpq.html
\vspace{12pt}

http://www.innerx.net/personal/tsmith/clfpq.html
\vspace{12pt}

McKay Correspondence:
\vspace{12pt}

http://galaxy.cau.edu/tsmith/DCLG-McKay.html
\vspace{12pt}

http://www.innerx.net/personal/tsmith/DCLG-McKay.html
\vspace{12pt}

Octonions:
\vspace{12pt}

http://galaxy.cau.edu/tsmith/3x3OctCnf.html
\vspace{12pt}

http://www.innerx.net/personal/tsmith/3x3OctCnf.html
\vspace{12pt}

Sedenions:
\vspace{12pt}

http://galaxy.cau.edu/tsmith/sedenion.html
\vspace{12pt}

http://www.innerx.net/personal/tsmith/sedenion.html
\vspace{12pt}

Cross-Products:
\vspace{12pt}

http://galaxy.cau.edu/tsmith/clcroct.html
\vspace{12pt}

http://www.innerx.net/personal/tsmith/clcroct.html
\vspace{12pt}

NonDistributive Algebras:
\vspace{12pt}

http://galaxy.cau.edu/tsmith/NDalg.html
\vspace{12pt}

http://www.innerx.net/personal/tsmith/NDalg.html
 \vspace{12pt}

\newpage

\subsection{Discrete Division Algebra Lattices.}

The discrete Clifford Algebras $DCl(n)$ can be
extended from hypercubic lattices to lattices
based on the Discrete Division Algebras.
\vspace{12pt}

For $n = 2$,
the $2$-dimensional hypercubic lattice $Z^{2}$ can be
thought of as the Gaussian lattice of the complex numbers.
\vspace{12pt}

For $n$ greater than 2,
the action of the discrete Clifford Algebra $DCl(n)$ on
the $n$-dimensional hypercubic lattice $Z^{n}$ can be
extended to action on the $D_{n}$ lattice
with $2n(n - 1)$ nearest neighbors to the origin,
corresponding to the second-layer, or norm-square 2,
vertices of the hypercubic lautice $Z_{n}$.
The $D_{n}$ lattice is called the Checkerboard lattice,
because it can be represented as one half of the vertices
of $Z^{n}$.
\vspace{12pt}

For $n = 4$,
the extension of the $4$-dimensional hypercubic lattice $Z^{4}$
to the $D_{4}$ lattice produces the lattice of
integral quaternions,
with 24 vertices nearest the origin, forming a 24-cell.
\vspace{12pt}

For $n = 8$,
the $8$-dimensional lattice $D_{8}$ can be
extended to the $E_{8}$ lattice of integral octonions,
with 240 vertices nearest the origin, forming
a Witting polytope,
by fitting together two copies of the $D_{8}$ lattice,
each of whose vertices are at the center of the holes
of the other.
\vspace{12pt}

For $n = 16$,
the $16$-dimensional lattice $D_{16}$ can be
extended to the $\Lambda_{16}$ Barnes-Wall lattice.
with 4,320 vertices nearest the origin.
\vspace{12pt}

For $n = 24$,
the $24$-dimensional lattice $D_{24}$ can be
extended to the $\Lambda_{24}$ Leech lattice.
with 196,560 vertices nearest the origin.
\vspace{72pt}

\subsection{Spinors.}

The discrete Clifford Group $DClG(n)$ is a subgroup
of the discrete Clifford Algebra $DCl(n)$.
\vspace{12pt}

There is a $1-1$ correspondence between
the representations of $DCl(n)$ and
those representations of $DClG(n)$ such that $U(-1) = -1$.
\vspace{12pt}

$DClG(n)$ has $2^{n}$ 1-dimensional representations,
each with $U(-1) = +1$.
\vspace{12pt}

The irreducible representations of $DClG(n)$
with dimension greater than 1 have $U(-1) = -1$,
and are representations of $DCl(n)$.
\vspace{12pt}

If $n$ is even,
there is one such representation, of degree $2^{n/2}$,
the full spinor representation of $DCl(n)$.
It is reducible into two half-spinor representations,
each of degree $2^{(n - 1)/2}$.
\vspace{12pt}

If $n$ is odd,
there are two such representations, each of degree
$2^{(n - 1)/2}$,
One of them is the spinor representation of $DCl(n)$.
\vspace{12pt}

\subsection{Signature.}

So far, we have been discussing only Euclidean space with
positive definite signature, $DCl(n) = DCl(0,n)$.  Symmetries
for the general signature cases include:
\vspace{12pt}

$DCl(p-1,q) = DCl(q-1,p)$;
\vspace{12pt}

The even subalgebras of $DCl(p,q)$ and $DCl(q,p)$ are
isomorphic;
\vspace{12pt}

$DCl(p,q)$ is isomoprphic to both the even subalgebra of $DCl(p+1,q)$
and the even subalgebra of $DCl(p,q+1)$.
\vspace{12pt}

Signature is not meaningful for complex vector spaces.
The complex Clifford algebra $DCl(2p)_{\bf{C}}$ is the
complexification $DCl(p,p) \otimes_{\bf{R}} {\bf{C}}$
\vspace{12pt}

\subsection{Periodicity 8.}

$DCl(p,q)$ has the periodicity properties:
\vspace{12pt}

$DCl(n,n+8) = DCl(n,n) \otimes M({\bf{R}},16)
= DCl(n,n) \otimes DCl(0,8) = DCl(n+8,n)$
\vspace{12pt}

$DCl(n-4,n+4) = DCl(n,n)$
$DCl(n,n+8) = DCl(n,n) \otimes M({\bf{R}},16) = DCl(n,n)
\otimes DCl(0,8) = DCl(n+8,n)$
\vspace{12pt}

Therefore any discrete Clifford algebra $DCl(p,q)$ of any size
can first be embedded in a larger one with $p$ and $q$
multiplex of 8,
and then the larger one can be "factored" into
\vspace{12pt}

$DCl(0,p) \otimes DCl(0,8) \otimes ... \otimes DCl(0,8)$
\vspace{12pt}

so that the fundamental building block of the real discrete
Clifford algebras is $DCl(0,8)$.
\vspace{12pt}

The vector, +half-spinor, and -half-spinor representations
\newline

of $DCl(0,8)$ are each 8-dimensional
\newline

and can be represented by an octonionic E8 lattice.
\vspace{12pt}

\newpage

\subsection{Many-Worlds Quantum Theory.}

To see how Many-Worlds Quantum Theory arises naturally
in the D4-D5-E6 HyperDiamond Feynman Checkerboard physics model,
note that the model is basically built by using
the discrete Clifford Algebra $DCL(0,8)$ as its basic
building block,
due to the Periodicity 8 property, so that the model looks
like a tensor product of Many Copies of $DCl(0,8)$:

$$DCl(0,p) \otimes DCl(0,8) \otimes ... \otimes DCl(0,8)$$

How do the Many Copies of $DCl(0,8)$ fit together?
\vspace{12pt}

Consider that each Copy of $DCl(0,8)$ has graded structure:

\begin{equation}
\begin{array}{ccccccccc}
    1 &  8 & 28 & 56 & 70 & 56 & 28 &  8 &  1  \\
\end{array}
\end{equation}

The vector 8 space corresponds to an 8-dimensional spacetime
that is a discrete E8 lattice.
\vspace{12pt}

Take any two Copies of $DCl(0,8)$ and consider the origin
of the E8 lattice of each Copy.
\vspace{12pt}

From each origin,
there are 240 links to nearest-neighbor vertices.
\vspace{12pt}

The two Copies naturally fit together
if the
origin of the E8 lattice of the vector 8 space of one Copy
and the
origin of the E8 lattice of the vector 8 space of the other Copy
are
nearest neighbors,
one at each end of a single link in the E8 lattice.
\vspace{12pt}

If you start with a Seed Copy of $DCl(0,8)$,
and repeat the fitting-together process with other copies,
the result is one large E8 lattice spacetime,
with one Copy of $DCl(0,8)$ at each vertex.
\vspace{12pt}

Since there are 7 TYPES OF E8 LATTICE,
7 different types of E8 lattice spacetime neighborhoods
can be constructed.
\vspace{12pt}

WHAT HAPPENS at boundaries of different E8 neighborhoods?
\vspace{12pt}

ALL the E8 lattices have in common
links of the form
\vspace{12pt}

   $$ \pm V $$
\vspace{12pt}

(where $V = 1,i,j,k,E,I,J,K$)
\vspace{12pt}

but they DO NOT AGREE for all links of the form
\vspace{12pt}

$$( +\pm W  \pm X  \pm Y  \pm Z ) / 2 $$
\vspace{12pt}

(where $W = 1,E;  X = i,I; Y = j,J; Z = k,K$)
\vspace{12pt}

ALL the E8 lattices BECOME CONSISTENT
if they are
DECOMPOSED into two 4-dimensional HyperDiamond lattices
so that
\vspace{12pt}

$$E8 = 8HD = 4HDa + 4HDca$$

where $4HDa$ is
the 4-dimensional associative Physical Spacetime
and
$4HDca$ is
the 4-dimensional coassociative Internal Symmetry space.
\vspace{12pt}

Therefore,
the D4-D5-E6 HyperDiamond Feynman Checkerboard model
is physically represented on a $4HD$ lattice Physical Spacetime,
with
an Internal Symmetry space that is also another $4HD$.
\vspace{12pt}

BIVECTOR GAUGE BOSON STATES ON LINKS:
\vspace{12pt}

The bivector 28 space corresponds to the 28-dimensional
D4 Lie algebra Spin(0,8), which,
after Dimensional Reduction of Physical Spacetime,
corresponds to 28 gauge bosons:
\vspace{12pt}

    12 for the Standard Model,
\vspace{12pt}

    15 for Conformal Gravity and the Higgs Mechanism, and
\vspace{12pt}

     1 for propagator phase.
\vspace{12pt}

Define a Bivector State of a given Copy of $DCl(0,8)$
to be a configuration of the 28 gauge bosons at its vertex.
\vspace{12pt}

Now look at any link in the E8 lattice,
and at the two Copies of $DCl(0,8)$ at each end.
\vspace{12pt}

The gauge boson state on that link is given by
the Lie algebra bracket product
of
the Bivector States of the two Copies
of $DCL(0,8)$ at each end.
\vspace{12pt}

Now define
the Total Superposition Bivector State of a given Copy
of $DCl(0,8)$
to be the superposition of all configurations
of the 28 gauge bosons at its vertex
\vspace{12pt}

and
the Total Superposition Gauge Boson State on a link
to be the superposition of all gauge boson states on that link.
\vspace{12pt}

SPINOR FERMION STATES AT VERTICES:
\vspace{12pt}

The 8+8 = 16 fermions corresponding to spinors
do not correspond to any single grade of $DCl(0,8)$
but correspond to the entire Clifford algebra as a whole.
\vspace{12pt}

Its total dimension is  $2^8 = 256 = 16 \times 16$
\vspace{12pt}

and
there are, in the first generation,
\vspace{12pt}

8 half-spinor fermion particles and
\vspace{12pt}

8 half-spinor fermion antiparticles,
\vspace{12pt}

for a total of 16 fermions.
\vspace{12pt}

Dimensional Reduction of Physical Spacetime produces
3 Generations of spinor fermion particles and antiparticles.

Define a Spinor Fermion State at a vertex
occupied by a given Copy of $DCl(0,8)$
to be a configuration of the spinor fermion
particles and antiparticles of all 3 generations
at its vertex.
\vspace{12pt}

Now define
the Total Superposition Spinor Fermion State at a vertex
to be the superposition of all Spinor fermion states at
that vertex.
\vspace{12pt}

Now we have:
\vspace{12pt}

$4HD$ HyperDiamond lattice Physical Spacetime
\vspace{12pt}

and
for each link, a Total Superposition Gauge Boson State
\vspace{12pt}

and
for each vertex, a Total Superposition Spinor Fermion State.
\vspace{12pt}

Now:
\vspace{12pt}

Define Many-Worlds Quantum Theory by specifying
each of its many Worlds, as follows:
\vspace{12pt}

Each World of the Many-Worlds is determined by:
\vspace{12pt}

for each link, picking one Gauge Boson State from the
Total Superposition
\vspace{12pt}

and
\vspace{12pt}

for each vertex, picking one Spinor Fermion State from
the Total Superposition.
\vspace{12pt}

To get an idea of how to think about
the D4-D5-E6 HyperDiamond Feynman Checkerboard lattice model,
here is a rough outline of how the Uncertainty Principle works:
\vspace{12pt}

Do NOT (as is conventional) say that a particle is
sort of "spread out" around a given location
in a given space-time
due to "quantum uncertainty".
\vspace{12pt}

Instead, say that the particle is really at
a point in space-time
\vspace{12pt}

BUT that the "uncertainty spread" is not a property of the
particle, but is due to dynamics of the space-time,
in which particle-antiparticle pairs x-o are being created
sort of at random.  For example, in one of the Many-Worlds,
the spacetime might not be just an empty vertex
\vspace{12pt}

but would have created a particle-antiparticle pair
\vspace{12pt}

If the original particle is where we put it to start with,
then in this World we would have the original particle
plus a new particle plus a new antiparticle.
\vspace{12pt}

Now, if the new antiparticle annihilates the original particle,
we would see a particle at the position of the new particle,
\vspace{12pt}

and, since the particles are indistinguishable from each other,
it would APPEAR that the original particle was at the different
location of the new particle, and the probabilities of
such appearances would
look like the conventional uncertainty in position.
\vspace{12pt}

In the D4-D5-E6 model, correlated states, such as
a particle-antiparticle pair coming from the non-trivial vacuum,
or an amplitude for two entangled particles,
extend over a part of the lattice that includes both particles.
The stay in the same World of the Many-Worlds
until they become uncorrelated.
\vspace{12pt}

\newpage

\section{Octonions and $E_{8}$ lattices.}

The $E_{8}$ lattice is made up of one
hypercubic Checkerboard $D_{8}$ lattice plus
another $D_{8}$ shifted by a glue vector
(1/2, 1/2, 1/2, 1/2, 1/2, 1/2, 1/2, 1/2).
\vspace{12pt}

If octonionic coordinate are chosen so that
a given minimal vector in $E_{8}$ is +1,
the vectors in $E_{8}$ that are perpendicular to +1 make up
a spacelike $E_{7}$ lattice.
\vspace{12pt}

The $E_{8}$ lattice nearest neighbor vertices have
only 4 non-zero coordinates,
like 4-dimensional spacetime with speed of light
$c$ = $\sqrt{3}$,
\vspace{12pt}

rather than 8 non-zero coordinates,
like 8-dimensional spacetime with speed of
light $c$ = $\sqrt{7}$,
so the $E_{8}$ lattice light-cone structure appears to be
4-dimensional rather than 8-dimensional.
\newline
\vspace{12pt}

To build the $E_{8}$ Lattice:
\vspace{12pt}

Begin with an 8-dimensional octonionic spacetime $R^{8}$,
where a basis for the octonions is $\{ 1,i,j,k,E,I,J,K \}$ .
\vspace{12pt}

The vertices of the $E_{8}$ lattice are of the form
\vspace{12pt}

$(a_{0}1 + a_{1}E + a_{2}i + a_{3}j + a_{4}I + a_{5}K
+ a_{6}k + a_{7}J)/2$ ,
\vspace{12pt}

where the $a_{i}$ may be either all even integers,
all odd integers,
\vspace{12pt}

or four of each (even and odd),
\vspace{12pt}

with residues mod 2 in the four-integer cases being
\vspace{12pt}

$(1;0,0,0,1,1,0,1)$
\vspace{12pt}

or $(0;1,1,1,0,0,1,0)$
\vspace{12pt}

or the same with the last seven cyclically permuted.
$E_{8}$ forms an integral domain of integral octonions.
\vspace{12pt}

The $E_{8}$ lattice integral domain has 240 units:
\vspace{12pt}

$ \pm 1, \pm i, \pm j, \pm k \pm E, \pm I, \pm J \pm K,$
\vspace{12pt}

$( \pm 1 \pm I \pm J \pm K)/2, ( \pm E \pm i \pm j \pm k)/2,$
\vspace{12pt}

and the last two with cyclical permutations of
\vspace{12pt}

$\{ i,j,k,E,I,J,K \}$ in the order $(E, i, j, I, K, k, J)$.
\vspace{12pt}

The cyclical permutation $(E, i, j, I, K, k, J)$
preserves the integral domain $E_{8}$,
but is not an automorphism of the octonions
since it takes the associative triad $\{ i,j,k \}$
into the anti-associative triad $\{ j,ie,je \}$.
\vspace{12pt}

The cyclical permutation $(E, I, J, i, k, K, j)$
is an automorphism of the octonions but
takes the $E_{8}$ integral domain defined above
into another of seven integral domains.
\vspace{12pt}

Denote the integral domain described above as $7E_{8}$,
\vspace{12pt}

and the other six by $iE_{8}$ , $i = 1, ... , 6$.
\vspace{12pt}

\newpage

The 240 units of the $7E_{8}$ lattice corresponding to
the integral domain $7E_{8}$ represent the 240 lattice points
in the shell at unit distance (also commonly normalized as 2):
\vspace{12pt}

	      $$\pm 1,  \pm i,  \pm j,  \pm k,  \pm E,
\pm I,  \pm J,  \pm K,$$
$$(\pm 1 \pm I \pm J \pm K)/2$$
$$(\pm e \pm i \pm j \pm k)/2$$
$$(\pm 1 \pm K \pm E \pm k)/2$$
$$(\pm i \pm j \pm I \pm J)/2$$
$$(\pm 1 \pm k \pm i \pm J)/2$$
$$(\pm j \pm I \pm K \pm E)/2$$
$$(\pm 1 \pm J \pm j \pm E)/2$$
$$(\pm I \pm K \pm k \pm i)/2$$
$$(\pm 1 \pm E \pm I \pm i)/2$$
$$(\pm K \pm k \pm J \pm j)/2$$
$$(\pm 1 \pm i \pm K \pm j)/2$$
$$(\pm k \pm J \pm E \pm I)/2$$
$$(\pm 1 \pm j \pm k \pm I)/2$$
$$(\pm J \pm E \pm i \pm K)/2$$

\vspace{12pt}

\newpage

	The other six integral domains  $iE_{8}$ are:
\vspace{12pt}

$1E_{8}$:
\vspace{12pt}

$$\pm 1,  \pm i,  \pm j,  \pm k,  \pm E,  \pm I,  \pm J,
\pm K, $$
$$(\pm 1 \pm J \pm i \pm j)/2$$
$$(\pm k \pm E \pm I \pm K)/2$$
$$(\pm 1 \pm j \pm I \pm K)/2$$
$$(\pm i \pm k \pm E \pm J)/2$$
$$(\pm 1 \pm K \pm k \pm i)/2$$
$$(\pm j \pm E \pm I \pm J)/2$$
$$(\pm 1 \pm i \pm E \pm I)/2$$
$$(\pm j \pm k \pm J \pm K)/2$$
$$(\pm 1 \pm I \pm J \pm k)/2$$
$$(\pm i \pm j \pm E \pm K)/2$$
$$(\pm 1 \pm k \pm j \pm E)/2$$
$$(\pm i \pm I \pm J \pm K)/2$$
$$(\pm 1 \pm E \pm K \pm J)/2$$
$$(\pm i \pm j \pm k \pm I)/2$$

\vspace{12pt}

\newpage

$2E_{8}$:
\vspace{12pt}

$$\pm 1,  \pm i,  \pm j,  \pm k,  \pm E,  \pm I,  \pm J,
\pm K,$$
$$(\pm 1 \pm i \pm k \pm E)/2$$
$$(\pm j \pm I \pm J \pm K)/2$$
$$(\pm 1 \pm E \pm J \pm j)/2$$
$$(\pm i \pm k \pm I \pm K)/2$$
$$(\pm 1 \pm j \pm K \pm k)/2$$
$$(\pm i \pm E \pm I \pm J)/2$$
$$(\pm 1 \pm k \pm I \pm J)/2$$
$$(\pm i \pm j \pm E \pm I)/2$$
$$(\pm 1 \pm J \pm i \pm K)/2$$
$$(\pm j \pm k \pm E \pm I)/2$$
$$(\pm 1 \pm K \pm E \pm I)/2$$
$$(\pm i \pm j \pm k \pm J)/2$$
$$(\pm 1 \pm I \pm j \pm i)/2$$
$$(\pm k \pm E \pm J \pm K)/2$$

\vspace{12pt}

\newpage

$3E_{8}$:
\vspace{12pt}

$$\pm 1,  \pm i,  \pm j,  \pm k,  \pm E,  \pm I,  \pm J,
\pm K,$$
$$(\pm 1 \pm k \pm K \pm I)/2$$
$$(\pm i \pm j \pm E \pm J)/2$$
$$(\pm 1 \pm I \pm i \pm E)/2$$
$$(\pm j \pm k \pm J \pm K)/2$$
$$(\pm 1 \pm E \pm j \pm K)/2$$
$$(\pm i \pm k \pm I \pm J)/2$$
$$(\pm 1 \pm K \pm J \pm i)/2$$
$$(\pm j \pm k \pm E \pm I)/2$$
$$(\pm 1 \pm i \pm k \pm j)/2$$
$$(\pm e \pm I \pm J \pm K)/2$$
$$(\pm 1 \pm j \pm I \pm J)/2$$
$$(\pm i \pm k \pm E \pm K)/2$$
$$(\pm 1 \pm J \pm E \pm k)/2$$
$$(\pm i \pm j \pm I \pm K)/2$$

\vspace{12pt}

\newpage

 $4E_{8}$:
\vspace{12pt}

$$\pm 1,  \pm i,  \pm j,  \pm k,  \pm E,  \pm I,  \pm J,
\pm K,$$
$$(\pm 1 \pm K \pm j \pm J)/2$$
$$(\pm i \pm k \pm E \pm I)/2$$
$$(\pm 1 \pm J \pm k \pm I)/2$$
$$(\pm i \pm j \pm E \pm K)/2$$
$$(\pm 1 \pm I \pm E \pm j)/2$$
$$(\pm i \pm k \pm J \pm K)/2$$
$$(\pm 1 \pm j \pm i \pm k)/2$$
$$(\pm e \pm I \pm J \pm K)/2$$
$$(\pm 1 \pm k \pm K \pm E)/2$$
$$(\pm i \pm j \pm I \pm J)/2$$
$$(\pm 1 \pm E \pm J \pm i)/2$$
$$(\pm j \pm k \pm I \pm K)/2$$
$$(\pm 1 \pm i \pm I \pm K)/2$$
$$(\pm j \pm k \pm E \pm J)/2$$

\vspace{12pt}

\newpage

$5E_{8}$:
\vspace{12pt}

$$\pm 1,  \pm i,  \pm j,  \pm k,  \pm E,  \pm I,  \pm J,
\pm K,$$
$$(\pm 1 \pm j \pm E \pm i)/2$$
$$(\pm k \pm I \pm J \pm K)/2$$
$$(\pm 1 \pm i \pm K \pm J)/2$$
$$(\pm j \pm k \pm E \pm I)/2$$
$$(\pm 1 \pm J \pm I \pm E)/2$$
$$(\pm i \pm j \pm k \pm K)/2$$
$$(\pm 1 \pm E \pm k \pm K)/2$$
$$(\pm i \pm j \pm I \pm J)/2$$
$$(\pm 1 \pm K \pm j \pm I)/2$$
$$(\pm i \pm k \pm E \pm J)/2$$
$$(\pm 1 \pm I \pm i \pm k)/2$$
$$(\pm j \pm E \pm J \pm K)/2$$
$$(\pm 1 \pm k \pm J \pm j)/2$$
$$(\pm i \pm E \pm I \pm K)/2$$

\vspace{12pt}

\newpage

 $6E_{8}$:
\vspace{12pt}

$$\pm 1,  \pm i,  \pm j,  \pm k,  \pm E,  \pm I,  \pm J,
\pm K,$$
$$(\pm 1 \pm E \pm I \pm k)/2$$
$$(\pm i \pm j \pm J \pm K)/2$$
$$(\pm 1 \pm k \pm j \pm i)/2$$
$$(\pm e \pm I \pm J \pm K)/2$$
$$(\pm 1 \pm i \pm J \pm I)/2$$
$$(\pm j \pm k \pm E \pm K)/2$$
$$(\pm 1 \pm I \pm K \pm j)/2$$
$$(\pm i \pm k \pm E \pm J)/2$$
$$(\pm 1 \pm j \pm E \pm J)/2$$
$$(\pm i \pm k \pm I \pm K)/2$$
$$(\pm 1 \pm J \pm k \pm K)/2$$
$$(\pm i \pm j \pm E \pm I)/2$$
$$(\pm 1 \pm K \pm i \pm E)/2$$
$$(\pm j \pm k \pm I \pm J)/2$$

\vspace{12pt}

\newpage

	The vertices that appear in more than one lattice are:
\vspace{12pt}

$\pm 1,  \pm i,  \pm j,  \pm k,  \pm E,  \pm I,  \pm J,
\pm K$ in all of them;
\vspace{12pt}

$(\pm 1 \pm i \pm j \pm k)/2$ and $(\pm e \pm I \pm J
\pm K)/2$ in $3E_{8}$, $4E_{8}$, and $6E_{8}$;
\vspace{12pt}

$(\pm 1 \pm i \pm E \pm I)/2$ and $(\pm j \pm k \pm J
\pm K)/2$ in $7E_{8}$, $1E_{8}$, and $3E_{8}$;
\vspace{12pt}

$(\pm 1 \pm j \pm E \pm J)/2$ and $(\pm i \pm k\pm I \pm K)/2$
in $7E_{8}$, $2E_{8}$, and $6E_{8}$;
\vspace{12pt}

$(\pm 1 \pm k \pm E \pm K)/2$ and $(\pm i \pm j \pm I \pm J)/2$
in $7E_{8}$, $4E_{8}$, and $5E_{8}$;
\vspace{12pt}

$(\pm 1 \pm i \pm J \pm K)/2$ and $(\pm j \pm k \pm E \pm I)/2$
in $2E_{8}$, $3E_{8}$, and $5E_{8}$;
\vspace{12pt}

$(\pm 1 \pm j \pm I \pm K)/2$ and $(\pm i \pm k\pm e \pm J)/2$
in $1E_{8}$, $5E_{8}$, and $6E_{8}$;
\vspace{12pt}

$(\pm 1 \pm k \pm I \pm J)/2$ and $(\pm i \pm j \pm E \pm K)/2$
in $1E_{8}$, $2E_{8}$, and $4E_{8}$;
\vspace{12pt}

The 240 unit vertices in the $E_{8}$ lattices do not include
any of the 256 $E_{8}$ light cone vertices,
of the form $(\pm 1 \pm i \pm j \pm k \pm E \pm I \pm J
\pm K)/2$.
\vspace{12pt}

They appear in the next layer out from the origin,
at radius sqrt 2, which layer contains in all 2160 vertices.
\vspace{12pt}

The E8 lattice is, in a sense,
fundamentally 4-dimensional.
\vspace{12pt}

For instance:
\vspace{12pt}

The E8 lattice nearest neighbor vertices have
only 4 non-zero coordinates,
like
4-dimensional spacetime with speed of light  $c = \sqrt{3}$,
rather than 8 non-zero coordinates,
like
8-dimensional spacetime with speed of light $c = \sqrt{7}$,
so
the E8 lattice light-cone structure appears to be
4-dimensional rather than 8-dimensional.
\vspace{12pt}

The E8 lattice can be represented by quaternionic
icosians, as described by Conway and Sloane \cite{CON}.
\vspace{12pt}

The E8 lattice can be constructed, using the Golden ratio,
from the D4 lattice, which has a 24-cell nearest neighbor polytope.
The construction starts with the 24 vertices of a 24-cell,
then adds Golden ratio points on each of the 96 edges of
the 24-cell, then extends the space to 8 dimensions
by considering the algebraicaly independent $\sqrt{5}$
part of the coordinates to be geometrically independent,
and
finally doubling the resulting 120 vertices in 8-dimensional
space by considering both the D4 lattice and
its dual D4*
to get the 240 vertices of the E8 lattice nearest neighbor
polytope (the Witting polytope).
\vspace{12pt}

The 240-vertex Witting polytope,
the E8 lattice nearest neighbor polytope,
most naturally lives in 4 complex dimensions,
where it is self-dual, rather than in 8 real dimensions.
\vspace{12pt}

\newpage

\section{$E_{8}$ Spacetime and Particles.}

The 256-dimensional Clifford algebra $DCl(0,8)$
has graded structure
\vspace{12pt}

\[
\begin{array}{ccccccccccccccccccccc}
&&{\bf1}&&{\bf8}&&{\bf28}&&56&&70&&56&&28&&8&&1&&\\
\end{array}
\]

The grade-0 scalar of $DCl(0,8)$ is 1-dimensional,
representing the Higgs scalar field.
\vspace{12pt}

The grade-1 vectors of $DCl(0,8)$ are 8-dimensional,
representing spacetime prior to dimensional reduction.
\vspace{12pt}

The grade-2 bivectors of $DCl(0,8)$ are 28-dimensional,
representing a $Spin(0,8)$ gauge group prior to
dimensional reduction.
\vspace{12pt}

The entire 256-dimensional $DCl(0,8)$ can be
represented by $16x16$ matrices.
\vspace{12pt}

Each row or column of $DCl(0,8)$ is a 16-dimensional
minimal left or right ideal of $DCl(0,8)$,
and can be represented by two integral octonions.
\vspace{12pt}

Each of the two integral octonions in a row minimal
left ideal is a half-spinor, one +half-spinor and
the other a mirror image -half-spinor.
\vspace{12pt}

The basis elements for the +half-spinor integral octonions
correspond to first-generation fermion particles.
\vspace{12pt}

\begin{equation}
\begin{array}{|c|c|} \hline
Octonion  & Fermion \: Particle \\
basis \: element & \\ \hline
1 & e-neutrino   \\ \hline
i & red \: up \: quark \\ \hline
j & green \: up \: quark \\ \hline
k & blue \: up \: quark  \\ \hline
E & electron \\ \hline
I & red \: down \: quark  \\ \hline
J & green \: down \: quark  \\ \hline
K & blue \: down \: quark  \\ \hline
\end{array}
\end{equation}

\vspace{12pt}

The basis elements for the -half-spinor integral octonions
correspond to the first-generation fermion antiparticles.
\vspace{12pt}

The column minimal right ideal half-spinors correspond
to the Clifford algebra gammas of spacetime transformations.
\vspace{12pt}

\vspace{12pt}

\vspace{12pt}

In calculations, it is sometimes convenient to use
the volumes of compact manifolds that represent spacetime,
internal symmetry space, and fermion representation space.
\vspace{12pt}

The compact manifold that represents 8-dim spacetime
is ${\bf R}P^1 \times S^7$, the Shilov boundary of
the bounded complex
homogeneous domain that corresponds to
$Spin(10) / (Spin(8) \times U(1))$.
\vspace{12pt}

The compact manifold that represents 4-dim spacetime
is ${\bf R}P^1 \times S^3$, the Shilov boundary
of the bounded complex
homogeneous domain that corresponds to
$Spin(6) / (Spin(4) \times U(1))$.
\vspace{12pt}

The compact manifold that represents 4-dim internal
symmetry space
is ${\bf R}P^1 \times S^3$, the Shilov boundary
of the bounded complex
homogeneous domain that corresponds to
$Spin(6) / (Spin(4) \times U(1))$.
\vspace{12pt}

The compact manifold that represents
the 8-dim fermion representation space
is ${\bf R}P^1 \times S^7$, the Shilov boundary
of the bounded complex
homogeneous domain that corresponds to
$Spin(10) / (Spin(8) \times U(1))$.
\vspace{12pt}

The manifolds  ${\bf R}P^1 \times S^3$  and
${\bf R}P^1 \times S^7$ are
homeomorphic to  $S^1 \times S^3$  and   $S^1 \times S^7$,
which are untwisted trivial sphere bundles over $S^1$.
The corresponding twisted sphere bundles are
the generalized Klein bottles
$Klein(1,3) Bottle$  and  $Klein(1,7) Bottle$.
\vspace{12pt}

\subsection{Row and Column Spinors.}

The entire 256-dimensional DCl(0,8) can be
represented by 16x16 real matrices.
\vspace{12pt}

Each column or row of DCl(0,8) is a 16-dimensional
minimal left or right ideal of DCl(0,8),
and can be represented by two integral octonions.
\vspace{12pt}

The column minimal ideal spinors, as described above,
correspond to fermion particles and antiparticles,
\vspace{12pt}

while the row minimal ideals correspond to SpaceTime
Dirac gammas.
\vspace{12pt}

Each of the two integral octonions in a row minimal
right ideal is a half-spinor, one +half-spinor and
the other a mirror image -half-spinor.
\vspace{12pt}

The row minimal right ideal half-spinors correspond
to the Clifford algebra gammas of spacetime transformations.
Each of the two integral octonions in a row minimal
right ideal is a half-spinor, one +half-spinor and
the other a mirror image -half-spinor.
\vspace{12pt}

The 8 basis elements for the +half-spinor integral octonions
and for the -half-spinor integral octonions each correspond,
by triality, to the 8 basis elements of the Cl(8) vector space,
and therefore to the 8-dimensional spacetime gamma matrices:
\vspace{12pt}

After Dimensional Reduction to 4-dimensional spacetime,
the 8 Gammas of the 8-dimensional Octonionic vector
space of Cl(8)
are replaced by 4 Gammas of the 4-dimensional SpaceTime,
which can be seen to have Quaternionic structure.
\vspace{12pt}

Physical 4-dimensional spinors live in the even subalgebra
of the Cl(4) Clifford algebra, and so are either bivectors,
or bivectors plus the scalar or the pseudoscalar.
\vspace{12pt}

Since bivectors generate rotations (or boosts),
physical spinors transform under transformations of spacetime
like spinning (rotating) spheres.
\vspace{12pt}

Since gravity can be represented in 4-dimensional spacetime
by geometric transformations of spacetime,
the interaction between spinors and gravity can
be represented by curvature in a Clifford Manifold,
as done by William Pezzaglia in gr-qc/9710027 in his
derivation of the Papapetrou Equations.
Pezzaglia's picture may be consistent with
the picture of G. Sardanashvily and of the D4-D5-E6 model
with respect to Gravity and the Higgs Mechanism.
\vspace{12pt}

For a more conventional derivation of the Papapetrou Equations,
see, for example, Geometric Quantization, by N. Woodhouse,
second edition Clarendon Press 1992, pages 127-129.
\vspace{12pt}

To me, the Clifford method of William Pezzaglia semms
clearer than the more conventional derivation,
particularly with respect to fundamental fermions
and massless (at tree level) neutrinos.
\vspace{12pt}

The Papapetrou Equations are important because
they show that gravity acts on spinor particles,
even if they are massless.
\vspace{12pt}

Torsion comes from spin, and both are related to
physics models using Conformal Weyl curvature.
\vspace{12pt}

\subsection{Gauge Bosons in 4-dim SpaceTime.}

After Dimensional Reduction to 4-dimensional spacetime,
Bivector Gauge Bosons and the Higgs Scalar take the
form of the Standard Model plus Gravity.
\vspace{12pt}

You can use Clifford Algebras to see how Dimensional Reduction
affects the bivector Gauge Bosons and the Higgs scalar,
\vspace{12pt}

First go from DCl(0,8) to its even subalgebra DCle(0,8)
which is the Clifford algebra DCl(0,7).
\vspace{12pt}

Then go to its even subalgebra DCle(0,7)
which is the Clifford algebra DCl(0,6).
\vspace{12pt}

Then go to its even subalgebra DCle(0,6),
which is the Clifford algebra DCl(0,5):
\vspace{12pt}

The two components of DCl(0,5) can be regarded
as Real and Imaginary parts.
\vspace{12pt}

The 4x4 = 16 components of the Real part of DCl(0,5) are:
\vspace{12pt}

 1-dimensional U(1) propagator phase (light blue); and
\vspace{12pt}

15-dimensional Conformal group that gauges to produce Gravity
   from
\vspace{12pt}

the MacDowell-Mansouri mechanism and 10 Poincare group
   generators (red), with gravitons that see their Symmetry Space
   of Spacetime according to their group symmetry,
   and
\vspace{12pt}

5 Conformal degrees of freedom (brown) that,
           when gauge-fixed, combine with the scalar to
           produce the Higgs mechanism;
\vspace{12pt}

The 1 + 12 + 3 components of the Imaginary part of DCl(0,5) are:
\vspace{12pt}

 1-dimensional Higgs scalar, that combines with the gauge-fixed
           Conformal degrees of freedom to
           produce the Higgs mechanism;
\vspace{12pt}

12 Standard Model gauge bosons, 3 for weak SU(2) (red),
           9 for 8 SU(3) gluons, and 1 U(1) photon (yellow),
           that see Internal Symmetry Space according
           to their group symmetry;
           and
\vspace{12pt}

 3 components that are 4-vectors (gray), not bivectors,
           and do not contribute to
           the physics of gauge bosons and the scalar.
\vspace{12pt}

The Gauge Boson dimensional reduction process
can also be described by Division
Algebras:
\vspace{12pt}

       Start with the 28 generators of the Spin(8)
Lie Algebra
acting on its 8-dimensional Vector space.
\vspace{12pt}

       Reduce the 8-dimensional Vector space
to 4-dimensional Spacetime.
\vspace{12pt}

       Reduce Spin(8) to U(4) = U(1)xSU(4) = U(1)xSpin(6),
where the U(1) represents the phase of propagators
and the Spin(6) = Spin(4,2)
acts as the Conformal Group (or its Euclidean version)
on 4-dimensional Spacetime that produce Gravity and
the Higgs Mechanism.
\vspace{12pt}

       Since U(4) has 16 generators that act
on Spacetime as phases
or as Conformal Group elements,
that leaves 28 - 16 = 12 Spin(8) generators
to be accounted for.
\vspace{12pt}

       The 12 remaining generators should not act
on Spacetime,
but on the 4-dimensional Internal Symmetry Space.
\vspace{12pt}

The 4-dimensional Spacetime is a subspace
of the 6-dimensional Spin(6) Vector space.
\vspace{12pt}

The Internal Symmetry Space on which the 12 remaining generators
act should correspond to the Spinor representation space
of Spin(6).
\vspace{12pt}

       Since the Spin(6) Clifford Algebra Cl(6) is
the 8x8 real matrix algebra M(8,R), the full
       Spinor space of Spin(6) is 8-dimensional.
\vspace{12pt}

       If the full 8-dimensional Spinor space of Cl(6) is
given Octonionic structure, then it should transform under
both G2,
the automorphism group of the Octonions,
and Spin(6)=SU(4).
\vspace{12pt}

Octonionic Cl(6) Spinor space is 1 copy of
the Quaternions,
and corresponds to a full 8-dimensional
real generalized Spinor space of Cl(6).
\vspace{12pt}

14-dimensional G2 and 15-dimensional
Spin(6)=SU(4)
both have 8-dimensional SU(3) as a subgroup,
and SU(3) is the intersection of G2 and Spin(6)=SU(4).
\vspace{12pt}

SU(3) acts globally on the full CP2 Internal Symmetry Space
through the fibration SU(3)/S(U(1)xU(2)) = CP2.
\vspace{12pt}

SU(3) is made up of 8 of the remaining 12 Spin(8) generators,
and physically is the local gauge symmetry of the Color Force.
\vspace{12pt}

       If the full 8-dimensional Spinor space of Cl(6) is
given Quaternionic structure, then it should transform under
both Sp(1)=SU(2)=Spin(3)=S3, the automorphism group
of the Quaternions, and Spin(6)=SU(4).
\vspace{12pt}

Quaternionic Cl(6) Spinor space is 2 copies of the Quaternions,
and corresponds to 2 4-dimenisional real generalized
half-Spinor spaces of Cl(6).
\vspace{12pt}

3-dimensional SU(2) is a subgroup of Spin(6)=SU(4).
\vspace{12pt}

SU(2) acts globally on a CP1 = S2 subspace of CP2 Internal
Symmetry Space through the fibration SU(2)/U(1) = S2.
2 copies of SU(2)
correspond to the 2 copies of the Quaternions in Cl(6)
Spinor space and
to the 2 copies of S2 needed to make a
4-dimensional Internal Symmetry Space.
\vspace{12pt}

              If the 2 copies of SU(2) are synchronized so as
to be consistent throughout the full Cl(6) Spinor Space and
the full 4-dimensional Internal Symmetry Space,
then SU(2) is made up
of 3 of the remaining 12 - 8 = 4 Spin(8) generators,
and physically is the local gauge symmetry of the Weak force.
\vspace{12pt}

       If the full 8-dimensional Spinor space of Cl(6) is
given Complex structure, then it should transform
under both U(1)=S1,
the automorphism group of the Complex numbers,
and Spin(6)=SU(4).
\vspace{12pt}

              Complex Cl(6) Spinor space is 4 copies
of the Complex numbers,
and corresponds to the 4 Complex dimensions of the full
(spinor pus conjugate spinor) Cl(6) Spinor space derived from
the even subalgebra Cl(5) = M(4,C) of the Cl(6)
Clifford algebra.
\vspace{12pt}

1-dimensional U(1) is a subgroup of Spin(6)=SU(4).
U(1)=S1 acts globally on a S1 subspace of CP2 Internal
Symmetry Space. 4 copies of U(1) correspond to the 4 copies of
the Complex numbers in Cl(6) Spinor space and to the
4 copies of S1
needed to make a 4-dimensional Internal Symmetry Space.
\vspace{12pt}

              If the 4 copies of U(1) are synchronized so as
to be consistent throughout the full Cl(6) Spinor Space and
the full 4-dimensional Internal Symmetry Space, then:
\vspace{12pt}

U(1) is made up of the 1 remaining 12 - 8 - 3 = 1 Spin(8)
generator, and physically is the local gauge symmetry of Electromagnetism.
\vspace{12pt}

The Standard Model U(1)xSU(2)xSU(3) Lie algebra structure
acts on the 8-dimensional generalized
half-Spinor space of Cl(6) through:
\vspace{12pt}

       1 copy of Octonionic SU(3), acting on the Octonions O.
\vspace{12pt}

       2 synchronized copies of Quaternionic SU(2),
acting as SU(2) on the Quaternions Q.
\vspace{12pt}

       4 synchronized copies of Complex U(1),
acting as U(1) on the Complex numbers C.
\vspace{12pt}

If each of U(1), SU(2), and SU(3) are considered to act
respectively on C, Q, and O, then the Standard Model Lie algebra
U(1)xSU(2)xSU(3) can be considered as acting on
the 2x4x8=64-dimensional
tensor product space T = C x Q x O
(which is a non-alternative algebra).
\vspace{12pt}

Using T = C x Q x O to describe the Standard Model is
the idea of Geoffrey Dixon.
\vspace{12pt}

To see how his approach works, let the basis for the
Complex numbers be {1,i},
the basis for the Quaternions be {1,i,j,k},
and the basis for the Octonions be {1,i,j,k,E,I,J,K},
and denote by q an imaginary unit quaternion, and
then decompose the Identity of T by using
\vspace{12pt}

       L0 = ( 1 + i i ) / 2
\vspace{12pt}

       L1 = ( 1 - i i ) / 2
\vspace{12pt}

       L2 = ( 1 + i q ) / 2
\vspace{12pt}

       L3 = ( 1 - i q ) / 2
\vspace{12pt}

       R+ = ( 1+ i E ) / 2
\vspace{12pt}

       R- = ( 1 - i E ) / 2
\vspace{12pt}

to form 4 orthogonal associative primitive idempotent
projection operators
\vspace{12pt}

       D0 = L0 R+ = ( 1 + i i ) ( 1+ i E ) / 4 =
( 1 + i i + i E - i E ) / 4
\vspace{12pt}

       D1 = L1 R+ = ( 1 - i i ) ( 1+ i E ) / 4 =
( 1 - i i + i E + i E ) / 4
\vspace{12pt}

       D2 = L2 R- = ( 1 + i q ) ( 1- i E ) / 4 =
( 1 + i q - i E + q E ) / 4
\vspace{12pt}

       D3 = L3 R- = ( 1 - i q ) ( 1- i E ) / 4 =
( 1 - i q - i E - q E ) / 4
\vspace{12pt}

whose symmetries are
\vspace{12pt}

       D0 - U(1)xSU(2)xSU(3)
\vspace{12pt}

       D1 - U(1)xSU(2)xSU(3) x Ui(1)
\vspace{12pt}

       D2 - U(1)xSU(2)xSU(3) x Uq(1) x SUq(2)
\vspace{12pt}

       D3 - U(1)xSU(2)xSU(3) x UE(1) x SUq(2)
\vspace{12pt}

SUq(2) is an SU(2) symmetry due to the variability of
the imaginary unit Quaternion q over the
entire S3 = SU(2) of imaginary unit Quaternions.
If q were fixed (with respect to i) then the 2 SU(2)
symmetries would be synchronized, and
the resulting symmetries would be
\vspace{12pt}

       D0 - U(1)xSU(2)xSU(3)
\vspace{12pt}

       D1 - U(1)xSU(2)xSU(3) x Ui(1)
\vspace{12pt}

       D2 - U(1)xSU(2)xSU(3) x Uq(1)
\vspace{12pt}

       D3 - U(1)xSU(2)xSU(3) x UE(1)
\vspace{12pt}

Ui(1), Uq(1), and UE(1) are U(1) symmetries due to
the variability
of each imaginary unit
Quaternions over the S1 spanned by it
in the parallelizable S3,
and to the variability of the imaginary
unit Octonion over the S1 spanned by it
in the parallelizable S7.
If the imaginary unit Quaternions
and Octonion were fixed, then the 4 U(1)
symmetries would be synchronized,
and the resulting
symmetries would be
\vspace{12pt}

       D0 - U(1)xSU(2)xSU(3)
\vspace{12pt}

       D1 - U(1)xSU(2)xSU(3)
\vspace{12pt}

       D2 - U(1)xSU(2)xSU(3)
 \vspace{12pt}

      D3 - U(1)xSU(2)xSU(3)
\vspace{12pt}

which is the Standard Model Lie algebra U(1)xSU(2)xSU(3).
\vspace{12pt}

In his book, Geoffrey Dixon noted that his approach,
like mine, produced the Lie algebra
\vspace{12pt}

U(1)xU(1)xU(1)xU(1) x SU(2)xSU(2) x SU(3)
\vspace{12pt}

prior to synchronization of the U(1)'s and the SU(2)'s.
\vspace{12pt}

Note - My terms, generalized Spinor and generalized
half-Spinor space of Cl(6),
are not standard mathematical terms but are used here
because
they seem to me to be useful.
\vspace{12pt}

From a more geometric Weyl Group - Root Vector Space point
to view, the Gauge
Boson dimensional reduction process can also be described
this way:
\vspace{12pt}

Start with the 24-cell whose vertices are
the 24-vertex root vectors of the Spin(8) Lie
 algebra in 4-dimensional root vector space
with 4 origin root vectors.
\vspace{12pt}

The 12 vertices of the central cuboctahedron of the 24-cell
plus the 4 origin root vectors correspond to the
U(4) = U(1)xSU(4) = U(1)xSpin(6) Lie subalgebra,
where the U(1) represents the phase of
propagators and the Spin(6) = Spin(4,2) acts as the Conformal Group
(or its Euclidean version) on 4-dimensional Spacetime
that produce Gravity and the Higgs Mechanism. Since
the 4 origin root vectors of Spin(8) are included here,
the U(1)xSpin(6) Lie subalgebra acts
on 4-dimensional Spacetime.
\vspace{12pt}

The remaining 12 vertices are in the two octahedra of
the 24-cell.
Since they do not include any
of the 4 origin root vectors of Spin(8),
they do not represent Lie algebras that act on
4-dimensional Spacetime,
but rather represent Lie algebras that act on CP2 Internal
Symmetry Space.
\vspace{12pt}

The 6 vertices of one octahedron can be projected
into interpenetrating triangles in a
plane corresponding to the 6 outer vertices
of the SU(3) Lie algebra root vectors in
2-dimensional space:
\vspace{12pt}

2 opposite vertices of the other octahedron correspond
to the 2 center vertices of the SU(3)
Lie algebra root vector space in 2-dimensional space,
giving us the entire root vector
diagram of 8-dimensional SU(3).
\vspace{12pt}

       The remaining 4 vectors of the (blue) octahedron
correspond to the 2 outer vertices and 2 center vertices
of the U(2) Lie Algebra.
\vspace{12pt}

       Now we have decomposed the 28 generators of Spin(8) into:
\vspace{12pt}

              16 generators of U(4) = U(1)xSU(4) = U(1)xSpin(6)
for propagator phase, Gravity, and the Higgs Mechanism,
acting on 4-dimensional Spacetime.
\vspace{12pt}

              8 generators of SU(3) for the Color Force,
acting on 4-dimensional Internal Symmetry Space.
\vspace{12pt}

              4 generators of U(2) = U(1)xSU(2) for Electromagnetism
and the Weak Force, acting on 4-dimensional Internal
Symmetry Space.
\vspace{12pt}

\subsection{Cl(8) Multivectors of grades 3, 4, 5, 6, 7, 8.}

As to the remaining Multivectors of DCl(0,8):
\vspace{12pt}

35 of the 70 4-vectors in the 4-grade subspace are the
symmetric parts of
the +1 8x8 component of even subalgebraof DCl(8) that
remain after taking
out the 1-dimensional scalar 0-vector Higgs scalar.
\vspace{12pt}

The 1-dimensional scalar 0-vector representing the Higgs scalar
can be thought of as the trace of the full symmetric
1+35=36-dimensional space of symmetric 8x8 real matrices.
\vspace{12pt}

The 35  4-vectors are the traceless symmetric 8x8 matrices.
They are related to the coassociative 4-form that is fixed
in the dimensional reduction process to determine the
internal symmetry space.  At our low energy levels,
below the Planck-scale at which dimensional reduction occurs
in the D4-D5-E6 model, the 35 4-vectors do not play a
dynamical role that we can test experimentally here and now.
\vspace{12pt}

The 56-dimensional trivector 3-grade subspace is related
to the structure of
3+1 = 4-dimensional subspaces of 1+7 = 8-dimensional spacetime
that are connected with the E8 HyperDiamond lattice
links that are (normalized) sums of 4 of the basis octonions.
\vspace{12pt}

To reduce the dimension of spacetime to 1+3=4 dimensions,
an associative 3-form is used.  This effectively fixes a
particular trivector, so the 56 trivectors do not play
a dynamical role in the 4-dimensional phase
of the D4-D5-E6 model.
\vspace{12pt}

The other 35  56  28   8   1  are (by Hodge * duality)
located symmetrically on the opposite side of the
16x16 DCl(8)matrix.
\vspace{12pt}

By Hodge duality, they can be viewed as the structures of
the D4-D5-E6 model that act on the antiparticle
half-spinor fermions,
while the  1   8  28  56  35  can be viewed as acting
on the particle half-spinor fermions.
\vspace{12pt}

\subsection{Subtle Triality Supersymmetry.}

The Dynkin diagram for Spin(8) has 1 central vertex
with 3 lines extending from it to 3 external vertices.
\vspace{12pt}

Each vertex represents a representation of Spin(8),
with the center vertex (Spin(8)) corresponding
to the 28-dimensional adjoint representation
that I identified with gauge bosons.
\vspace{12pt}

The three representations for spacetime, fermion particles,
and fermion antiparticles are EACH 8-dimensional
with Octonionic structure.
\vspace{12pt}

They are ALL isomorphic by the Spin(8) Triality Automorphism,
which can be represented by rotating or interchanging
the 3 arms of the Dynkin diagram of Spin(8).
\vspace{12pt}

The Triality isomorphism between spacetime
and fermion particles and fermion antiparticles
constitutes a
SUBTLE SUPERSYMMETRY between fermions and spacetime.
\vspace{12pt}

Ultraviolet finiteness of the D4-D5-E6 physics model
may be seen by considering, prior to dimensional reduction,
the generalized supersymmetry relationship between
the 28 gauge bosons
\vspace{12pt}

and
the 8 (first-generation) fermion particles and antiparticles.
\vspace{12pt}

In the 8-dimensional spacetime,
the dimension of each of the 28 gauge bosons in
the Lagrangian is 1,
\vspace{12pt}

and
the dimension of each of the 8 fermion particles is 7/2,
\vspace{12pt}

so that
the total dimension 28x1 = 28 of the gauge bosons
is equal to
the total dimension 8x(7/2) = 28 of the fermion particles.
\vspace{12pt}

After dimensional reduction of spacetime to 4 dimensions,
the 8 fermions get a 3-generation structure
\vspace{12pt}

and
the 28 gauge bosons are decomposed to produce
the Standard Model of
U(1) electromagnetism, SU(2) weak force, and SU(3) color force,
\vspace{12pt}

plus a Spin(5) = Sp(2) gauge field
that can produce gravity by the MacDowell-Mansouri mechanism.
\vspace{12pt}

Note that both Gravity and the Standard Model forces
are required for the cancellations that produce
the ultraviolet finiteness that is useful in
the Sakharov Zero Point Fluctuation model of gravity.
\vspace{12pt}

\newpage

\section{HyperDiamond Lattices.}

n-dimensional HyperDiamond structures $nHD$ are
constructed from $D_{n}$ lattices.
\vspace{12pt}

An n-dimensional HyperDiamond structures $nHD$ is a lattice
if and only if $n$ is even.
\vspace{12pt}

If $n$ is odd, the $nHD$ structure is
only a "packing", not a "lattice", because a
nearest neighbor link
from an origin vertex to a destination vertex
cannot be extended in the same direction
to get another
nearest neighbor link.
\vspace{12pt}

n-dimensional HyperDiamond structures $nHD$ are
constructed from $D_{n}$ lattices.
\vspace{12pt}

The lattices of type $D_{n}$ are n-dimensional Checkerboard
lattices, that is,
the alternate vertices of a ${\bf{Z}}^{n}$ hypercubic lattice.
\vspace{12pt}

A general reference on lattices is Conway and Sloane
\cite{CON}.
\vspace{12pt}

For the n-dimensional HyperDiamond lattice construction
from $D_{n}$,
Conway and Sloane use an
n-dimensional glue vector with $n$ coordinates
of length $0.5$:

$$[1] = (0.5, ..., 0.5)$$

\vspace{12pt}

Consider the 3-dimensional structure $3HD$.
\vspace{12pt}

Start with $D_{3}$, the fcc close packing in 3-space.
\vspace{12pt}

Make a second $D_{3}$ shifted by the glue vector
$(0.5, 0.5, 0.5)$.
\vspace{12pt}

Then form the union $D_{3} \cup  ([1] + D_{3})$.
\vspace{12pt}

That is the 3-dimensional crystal structure that is
made up of tetrahedral bonds.
\vspace{12pt}

The crystal structure of H2O Water Ice has such
tetrahedral bonds.
The bonds from Oxygen to Oxygen
are each half covalent and half Hydrogen.
\vspace{12pt}

Carbon Diamonds have tetrahedral bonds from Carbon to Carbon
that are purely covalent Carbon-Carbon.
\vspace{12pt}

HyperDiamond lattices were so named by David Finkelstein
because of the 3-dimensional Diamond crystal structure.
\vspace{12pt}

\subsection{From E8 to 4HD.}

When you construct an 8-dimensional HyperDiamond 8HD lattice,
you get
$$D8 \cup  ([1] + D8) = E8$$

The E8 lattice is in a sense fundamentally 4-dimensional,
and
the E8 HyperDiamond lattice naturally reduces to
the 4-dimensional HyperDiamond lattice.
\vspace{12pt}

To get from E8 to 4HD, reduce each of the D8 lattices
in the  $E8 = 8HD = D8 \cup  ([1] + D8)$ lattice
to D4 lattices.
\vspace{12pt}

We then get a 4-dimensional HyperDiamond
             $4HD = D4 \cup  ([1] + D4)$ lattice.
\vspace{12pt}

To see this,
start with $E8 = 8HD = D8 \cup  ([1] + D8)$.
\vspace{12pt}

We can write:

$$D8 = ( (  D4  ,0,0,0,0) \cup  (0,0,0,0,  D4  ) )
                        \cup
     (1,0,0,0,1,0,0,0) +  (   D4  ,  D4  ) $$

The third term is the diagonal term of
an orthogonal decomposition of D8,
and
the first two terms are orthogonal to each other:
\vspace{12pt}

associative 4-dimensional Physical Spacetime
\vspace{12pt}

and
coassociative 4-dimensional Internal Symmetry space.

Now,
we see that the orthogonal decomposition of
8-dimensional spacetime
into
4-dimensional associative Physical Spacetime
plus
4-dimensional Internal Symmetry space
gives
a decomposition of D8 into D4 + D4.
\vspace{12pt}

Since $E8 = D8 \cup  ([1] + D8)$,
and since
$$[1] = (0.5,0.5,0.5,0.5,0.5,0.5,0.5,0.5)$$

can be decomposed by
\vspace{12pt}

$$[1] = (0.5,0.5,0.5,0.5,0,0,0,0) + (0,0,0,0,0.5,0.5,0.5,0.5)$$
\vspace{12pt}

we have
\vspace{12pt}

$$E8 = D8 \cup  ([1] + D8) =$$

$$   =            (D4,0,0,0,0) + (0,0,0,0,D4)) $$
$$\cup$$
$$ (((0.5,0.5,0.5,0.5,0,0,0,0) + (0,0,0,0,0.5,0.5,0.5,0.5)) +$$
$$ ((D4,0,0,0,0) + (0,0,0,0,D4))) =$$
$$   = ((D4,0,0,0,0) \cup ((0.5,0.5,0.5,0.5,0,0,0,0)
+ (D4,0,0,0,0))) +$$
$$ ((0,0,0,0,D4) \cup ((0,0,0,0,0.5,0.5,0.5,0.5)
+ (0,0,0,0,D4))) $$

Since 4HD is $D4 \cup ([1] + D4)$,
\vspace{12pt}

$E8 = 8HD = 4HDa + 4HDca$
\vspace{12pt}

where $4HDa$ is
the 4-dimensional associative Physical Spacetime
\vspace{12pt}

and
$4HDca$ is
the 4-dimensional coassociative Internal Symmetry space.
\vspace{12pt}

\newpage

\subsection{4-dimensional HyperDiamond Lattice.}

The 4-dimensional HyperDiamond lattice HyperDiamond is
\newline
HyperDiamond = $D_{4} \cup  ([1] + D_{4})$.
\vspace{12pt}

The 4-dimensional HyperDiamond HyperDiamond
= $D_{4} \cup  ([1] + D_{4})$
is the ${\bf{Z}}^{4}$ hypercubic lattice with null edges.
\vspace{12pt}

It is the lattice that Michael Gibbs \cite{GIB} used
in his 1994 Georgia Tech Ph.D.
thesis advised by David Finkelstein.
\vspace{12pt}

The 8 nearest neighbors to the origin in the
4-dimensional HyperDiamond HyperDiamond lattice can be written
in octonion coordinates as:
\vspace{12pt}

\begin{equation}
\begin{array} {c}
(  1 + i + j + k) / 2  \\
(  1 + i - j - k) / 2  \\
(  1 - i + j - k) / 2  \\
(  1 - i - j + k) / 2  \\
(- 1 - i + j + k) / 2  \\
(- 1 + i - j + k) / 2  \\
(- 1 + i + j - k) / 2  \\
(- 1 - i - j - k) / 2  \\
\end{array}
\end{equation}

\vspace{12pt}

Here is an explicit construction of the 4-dimensional
HyperDiamond HyperDiamond lattice nearest neighbors
to the origin.
\vspace{12pt}

\newpage

Start with 24 vertices of a 24-CELL $D_{4}$ with squared norm 2:
\vspace{12pt}

\begin{equation}
\begin{array}{cccc}
+1   &    +1   &     0    &    0 \\
+1   &     0   &    +1    &    0 \\
+1   &     0   &     0    &   +1 \\
+1   &    -1   &     0    &    0 \\
+1   &     0   &    -1    &    0 \\
+1   &     0   &     0    &   -1 \\
-1   &    +1   &     0    &    0 \\
-1   &     0   &    +1    &    0 \\
-1   &     0   &     0    &   +1 \\
-1   &    -1   &     0    &    0 \\
-1   &     0   &    -1    &    0 \\
-1   &     0   &     0    &   -1 \\
 0   &    +1   &    +1    &    0 \\
 0   &    +1   &     0    &   +1 \\
 0   &    +1   &    -1    &    0 \\
 0   &    +1   &     0    &   -1 \\
 0   &    -1   &    +1    &    0 \\
 0   &    -1   &     0    &   +1 \\
 0   &    -1   &    -1    &    0 \\
 0   &    -1   &     0    &   -1 \\
 0   &     0   &    +1    &   +1 \\
 0   &     0   &    +1    &   -1 \\
 0   &     0   &    -1    &   +1 \\
 0   &     0   &    -1    &   -1 \\
\end{array}
\end{equation}

\vspace{12pt}

\newpage

Shift the 24 vertices by the glue vector
to get 24 more vertices $[1] + D_{4}$:
\vspace{12pt}

\begin{equation}
\begin{array}{cccc}
+1.5   &    +1.5   &     0.5   &     0.5 \\
+1.5   &     0.5   &    +1.5   &     0.5 \\
+1.5   &     0.5   &     0.5   &    +1.5 \\
+1.5   &    -0.5   &     0.5   &     0.5 \\
+1.5   &     0.5   &    -0.5   &     0.5 \\
+1.5   &     0.5   &     0.5   &    -0.5 \\
-0.5   &    +1.5   &     0.5   &     0.5 \\
-0.5   &     0.5   &    +1.5   &     0.5 \\
-0.5   &     0.5   &     0.5   &    +1.5 \\
-0.5   &    -0.5   &     0.5   &     0.5 \\
-0.5   &     0.5   &    -0.5   &     0.5 \\
-0.5   &     0.5   &     0.5   &    -0.5 \\
 0.5   &    +1.5   &    +1.5   &     0.5 \\
 0.5   &    +1.5   &     0.5   &    +1.5 \\
 0.5   &    +1.5   &    -0.5   &     0.5 \\
 0.5   &    +1.5   &     0.5   &    -0.5 \\
 0.5   &    -0.5   &    +1.5   &     0.5 \\
 0.5   &    -0.5   &     0.5   &    +1.5 \\
 0.5   &    -0.5   &    -0.5   &     0.5 \\
 0.5   &    -0.5   &     0.5   &    -0.5 \\
 0.5   &     0.5   &    +1.5   &    +1.5 \\
 0.5   &     0.5   &    +1.5   &    -0.5 \\
 0.5   &     0.5   &    -0.5   &    +1.5 \\
 0.5   &     0.5   &    -0.5   &    -0.5 \\
\end{array}
\end{equation}

\vspace{12pt}

\newpage

Of the 48 vertices of $D_{4} \cup ([1] + D_{4})$,
these 6 are nearest neighbors to the origin:
\vspace{12pt}

\begin{equation}
\begin{array}{cccc}
-0.5   &    -0.5   &     0.5   &     0.5 \\
-0.5   &     0.5   &    -0.5   &     0.5 \\
-0.5   &     0.5   &     0.5   &    -0.5 \\
 0.5   &    -0.5   &    -0.5   &     0.5 \\
 0.5   &    -0.5   &     0.5   &    -0.5 \\
 0.5   &     0.5   &    -0.5   &    -0.5 \\
\end{array}
\end{equation}

\vspace{12pt}

Two more nearest neighbors, also of squared norm 1,
of the origin
\vspace{12pt}

\begin{equation}
\begin{array}{cccc}

 0.5   &     0.5   &     0.5   &     0.5 \\
-0.5   &    -0.5   &    -0.5   &    -0.5 \\
\end{array}
\end{equation}

\vspace{12pt}

come from adding the glue vector to the origin
\vspace{12pt}

\begin{equation}
\begin{array}{cccc}
  0   &     0   &     0   &     0 \\
\end{array}
\end{equation}

\vspace{12pt}

and to the squared norm 4 point
\vspace{12pt}

\begin{equation}
\begin{array}{cccc}
  -1   &     -1   &     -1   &     -1 \\
\end{array}
\end{equation}

\vspace{12pt}

\newpage

\section{Spacetime and Internal Symmetry Space.}

The 4-dimensional spacetime of
this HyperDiamond Feynman Checkerboard
physics model is a HyperDiamond lattice that comes
from 4 dimensions of the
8-dimensional $E_{8}$ lattice spacetime.  If the basis of
the $E_{8}$ lattice is $\{1, i, j, k, E, I, J, K \}$,
then the basis of the 4-dimensional spacetime is the
associative part with basis $\{1, i, j, k \}$.
\vspace{12pt}

Therefore, the 4-dimensional spacetime lattice is
called the associative spacetime and denoted by $4HD_{a}$.
\vspace{12pt}

The 1-time and 3-space dimensions of the $4HD_{a}$ spacetime
can be represented by the 4 future lightcone links and the
4 past lightcone links as in the following pair of
"Square Diagrams"
of the 4 lines connecting the future ends
of the 4 future lightcone links
and
of the 4 lines connecting the past ends
of the 4 past lightcone links:
\vspace{12pt}

\begin{picture}(140,140)

\put(65,15){\line(1,0){100}}
\put(165,15){\line(0,1){100}}
\put(165,115){\line(-1,0){100}}
\put(65,115){\line(0,-1){100}}
\put(5,0){{\bf $T$ }$1+i+j+k$}
\put(165,0){{\bf $X$ }$1+i-j-k$}
\put(165,120){{\bf $Y$ }$1-i+j-k$}
\put(5,120){{\bf $Z$ }$1-i-j+k$}

\end{picture}

\begin{picture}(160,160)

\put(65,15){\line(1,0){100}}
\put(165,15){\line(0,1){100}}
\put(165,115){\line(-1,0){100}}
\put(65,115){\line(0,-1){100}}
\put(5,0){{\bf $-Y$ }$-1+i-j+k$}
\put(165,0){{\bf $-Z$ }$-1+i+j-k$}
\put(165,120){{\bf $-T$ }$-1-i-j-k$}
\put(5,120){{\bf $-X$ }$-1-i+j+k$}

\end{picture}

\vspace{12pt}

The 8 links $\{ {\bf{T,X,Y,Z,-T,-X,-Y,-Z}} \}$ correspond
to the 8 root vectors of the $Spin(5)$ de Sitter gravitation
gauge group, which has an 8-element Weyl group
$S_{2}^{2} \times S_{2}$.
\newline
The symmetry group of the 4 links of the future
lightcone is $S_{4}$,
the Weyl group of the 15-dimensional Conformal group
$SU(4)$ = $Spin(6)$.
\newline
10 of the 15 dimensions make up the de Sitter
$Spin(5)$ subgroup, and
\newline
the other 5 fix the "symmetry-breaking direction"
and scale of the
Higgs mechanism.
\vspace{12pt}

For more on this, see \cite{SMI7} and WWW URLs
\vspace{12pt}

http://galaxy.cau.edu/tsmith/cnfGrHg.html
\vspace{12pt}

http://www.innerx.net/personal/tsmith/cnfGrHg.html
\vspace{12pt}

\newpage

The Internal Symmetry Space of
this HyperDiamond Feynman Checkerboard
physics model is a HyperDiamond lattice that
comes from 4 dimensions
of the 8-dimensional $E_{8}$ lattice spacetime.
If the basis of
the $E_{8}$ lattice is $\{1, i, j, k, E, I, J, K \}$,
then the basis of the 4-dimensional
Internal Symmetry Space is the
coassociative part with basis $\{E, I, J, K \}$.
\vspace{12pt}

Therefore, the 4-dimensional spacetime lattice is
called the associative spacetime and denoted by $4HD_{ca}$.
\vspace{12pt}

Physically, the $4HD_{ca}$
Internal Symmetry Space should be thought of
as a space "inside" each vertex of the
$4HD_{a}$ HyperDiamond Feynman
Checkerboard spacetime, sort of like a Kaluza-Klein structure.
\vspace{12pt}

The 4 dimensions of the $4HD_{ca}$ Internal Symmetry Space are:
\newline
electric charge;
\newline
red color charge;
\newline
green color charge; and
\newline
blue color charge.
\vspace{12pt}

Each vertex of the $4HD_{ca}$ lattice has 8 nearest neighbors,
connected by lightcone links.  They have the algebraic structure
of the 8-element quaternion group $<2,2,2>$. \cite{COX2}
\vspace{12pt}

Each vertex of the $4HD_{ca}$ lattice has
24 next-to-nearest neighbors,
connected by two lightcone links.
They have the algebraic structure
of the 24-element binary tetrahedral group $<3,3,2>$ that is
associated with the 24-cell and the $D_{4}$ lattice. \cite{COX2}
\vspace{12pt}

\newpage

As with the time as space dimensions of the $4HD_{a}$ spacetime,
the E-electric and RGB-color dimensions of the $4HD_{ca}$
Internal Symmetry Space can be represented by
the 4 future lightcone links and the
4 past lightcone links.
\vspace{12pt}

However, in the $4HD_{ca}$ Internal Symmetry Space the
E Electric Charge should be treated as independent of
the RGB Color Charges.
\vspace{12pt}

As a result the following pair of "Square Diagrams"
look more like "Triangle plus Point Diagrams".

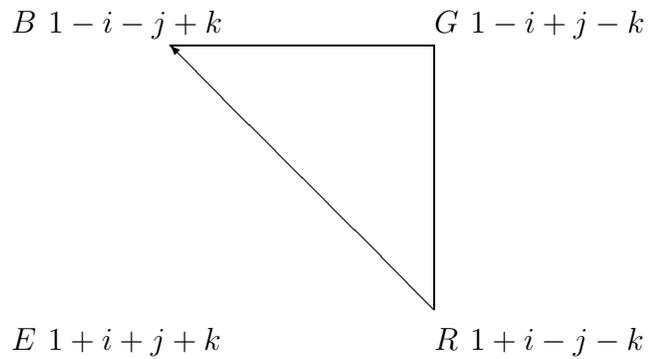
\begin{picture}(200,200)

\put(165,15){\line(0,1){100}}
\put(165,115){\line(-1,0){100}}
\put(5,0){{\bf $E$ }$1+i+j+k$}
\put(165,0){{\bf $R$ }$1+i-j-k$}
\put(165,120){{\bf $G$ }$1-i+j-k$}
\put(5,120){{\bf $B$ }$1-i-j+k$}
\put(165,15){\vector(-1,1){100}}

\end{picture}

\begin{picture}(200,200)

\put(65,15){\line(1,0){100}}
\put(65,115){\line(0,-1){100}}
\put(5,0){{\bf $-G$ }$-1+i-j+k$}
\put(165,0){{\bf $-B$ }$-1+i+j-k$}
\put(165,120){{\bf $-E$ }$-1-i-j-k$}
\put(5,120){{\bf $-R$ }$-1-i+j+k$}
\put(165,15){\line(-1,1){100}}

\end{picture}

The 2+6 links $\{ {\bf{E,-E;R,G,B,-R,-G,-B}} \}$ correspond to:
\vspace{12pt}

the 2 root vectors of the weak force $SU(2)$,
which has a 2-element Weyl group $S_{2}$; and
\vspace{12pt}

the 6 root vectors of the color force $SU(3)$,
which has a 6-element Weyl group $S_{3}$.
\vspace{12pt}

\vspace{12pt}

\vspace{12pt}

\vspace{12pt}

In calculations, it is sometimes convenient to use
the volumes of compact manifolds that represent spacetime,
internal symmetry space, and fermion representation space.
\vspace{12pt}

The compact manifold that represents 8-dim spacetime
is ${\bf R}P^1 \times S^7$, the Shilov boundary of
the bounded complex
homogeneous domain that corresponds to
$Spin(10) / (Spin(8) \times U(1))$.
\vspace{12pt}

The compact manifold that represents 4-dim spacetime
is ${\bf R}P^1 \times S^3$, the Shilov boundary of
the bounded complex
homogeneous domain that corresponds to
$Spin(6) / (Spin(4) \times U(1))$.
\vspace{12pt}

The compact manifold that represents
4-dim internal symmetry space
is ${\bf R}P^1 \times S^3$, the Shilov boundary
of the bounded complex
homogeneous domain that corresponds to
$Spin(6) / (Spin(4) \times U(1))$.
\vspace{12pt}

The compact manifold that represents
the 8-dim fermion representation space
is ${\bf R}P^1 \times S^7$, the Shilov boundary
of the bounded complex
homogeneous domain that corresponds to
$Spin(10) / (Spin(8) \times U(1))$.
\vspace{12pt}

The manifolds  ${\bf R}P^1 \times S^3$  and
${\bf R}P^1 \times S^7$ are
homeomorphic to  $S^1 \times S^3$  and   $S^1 \times S^7$,
which are untwisted trivial sphere bundles over $S^1$.
The corresponding twisted sphere bundles are
the generalized Klein bottles
$Klein(1,3) Bottle$  and  $Klein(1,7) Bottle$.
\vspace{12pt}

\subsection{Global Internal Symmetry Space and SpaceTime.}

Matti Pitkanen has suggested that the global structure
of 4-dimensional Spacetime and Internal Symmetry Space
should be given by 8-dimensional SU(3),
which decomposes into CP2 base
and U(2) fibre, both of which are 4-dimensional,
by SU(3) / U(2) = CP2.
\vspace{12pt}

Associative 4-dimensional Spacetime,
with Minkowski signature, is topologically
U(2) = SU(2)xU(1) = S3 x S1
\vspace{12pt}

which is consistent with the D4-D5-E6 model
Minkowski Spacetime of RP1 x S3.
\vspace{12pt}

If you do a Wick rotation to Euclidean signature,
you get a Spacetime that is naturally globally S4.
\vspace{12pt}

Coassociative 4-dimensional Internal Symmetry Space is CP2.
\vspace{12pt}

\newpage

\section{Feynman Checkerboards.}

The 2-dimensional Feynman Checkerboard \cite{FEY1, FEY2}
is a notably successful and useful representation of
the Dirac equation in 2-dimensional spacetime.
\vspace{12pt}

To build a Feynman 2-dimensional Checkerboard,
\newline
start with a 2-dimensional Diamond Checkerboard with
\newline
two future lightcone links and two past lightcone
links at each vertex.
\vspace{12pt}

The future lightcone then looks like
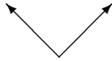
\vspace{12pt}

\begin{picture}(50,50)

\put(25,10){\vector(1,1){20}}
\put(25,10){\vector(-1,1){20}}

\end{picture}

\vspace{12pt}

If the 2-dimensional Feynman Checkerboard is
\newline
coordinatized by the complex plane $\bf{C}$:
\newline
the real axis $1$ is identified with the time axis $t$;
\newline
the imaginary axis $i$ is identified with the space axis
$x$; and
 \newline
the two future lightcone links are
$(1/\sqrt{2})(1 + i)$ and $(1/\sqrt{2})(1 - i)$.
\vspace{12pt}

In cylindrical coordinates $t,r$ with $r^{2} = x^{2}$,
\newline
the Euclidian metric is $t^{2} + r^{2}$ = $t^{2} + x^{2}$ and
\newline
the Wick-Rotated Minkowski metric with speed of light $c$ is
\newline
 $(ct)^{2} - r^{2}$ = $(ct)^{2} - x^{2}$.
\vspace{12pt}

For the future lightcone links to lI on
\newline
the 2-dimensional Minkowski lightcone, $c = 1$.
\vspace{12pt}

Either link is taken into the other link
by complex multiplication by $ \pm  i$.
\vspace{12pt}

Now, consider a path in the Feynman Checkerboard.
\newline
At a given vertex in the path, denote the
future lightcone link in
\newline
the same direction as the past path link by $1$, and
\newline
the future lightcone link in
the (only possible) changed direction by $ i$.
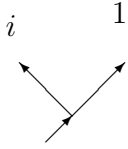
\vspace{12pt}

\begin{picture}(50,50)

\put(15,0){\vector(1,1){10}}
\put(25,10){\vector(1,1){20}}
\put(25,10){\vector(-1,1){20}}
\put(40,45){$1$}
\put(0,40){$ i$}

\end{picture}

\vspace{12pt}

The Feynman Checkerboard rule is that
\newline
if the future step at a vertex point of a given path
is in a different direction
\newline
from the immediately preceding step from the past,
\newline
then the path at the point of change gets a weight
of $ -i m \epsilon$,
\newline
where $m$ is the mass
(only massive particles can change directions),and
\newline
$\epsilon$ is the length of a path segment.
\vspace{12pt}

Here I have used the Gersch \cite{GER} convention
\newline
of weighting each turn by $ -im \epsilon$
\newline
rather than the Feynman \cite{FEY1, FEY2} convention
\newline
of weighting by $ +im \epsilon$, because Gersch's
convention gives a better nonrelativistic limit
in the isomorphic 2-dimensional Ising model \cite{GER}.
\vspace{12pt}

HOW SHOULD THIS BE GENERALIZED
\newline
TO HIGHER DIMENSIONS?
\vspace{12pt}

The 2-dim future light-cone is the 0-sphere $S^{2-2} = S^{0} =
\{ i, 1 \}$ ,
\vspace{12pt}

with $1$ representing a path step to the future in
\newline
the same direction as the path step from the past, and
\vspace{12pt}

$ i$ representing a path step to the future in a
\newline
(only 1 in the 2-dimensional Feynman Checkerboard lattice)
\newline
different direction from the path step from the past.
 \vspace{12pt}

The 2-dimensional Feynman Checkerboard lattice spacetime can be
\newline
represented by the complex numbers $\bf{C}$,
\newline
with $1,i$ representing the two future lightcone directions and
\newline
$-1,-i$ representing the two past lightcone directions.
\vspace{12pt}

Consider a given path in
\newline
the Feynman Checkerboard lattice 2-dimensional spacetime.
\vspace{12pt}

At any given vertex on the path in the lattice 2-dimensional
spacetime,
\newline
the future lightcone direction representing the
continuation of the path
\newline
in the same direction can be represented by $1$, and
\newline
the future lightcone direction representing
the (only 1 possible)
\newline
change of direction can be represented by $ i$ since either
\newline
of the 2 future lightcone directions can be
taken into the other
\newline
by multiplication by $ \pm  i$,
\newline
$+$ for a left turn and $-$ for a right turn.
\vspace{12pt}

If the path does change direction at the vertex,
\newline
then the path at the point of change gets a weight of
$ -im \epsilon$,
\newline
where $i$ is the complex imaginary,
\newline
$m$ is the mass
(only massive particles can change directions),and
\newline
$\epsilon$ is the timelike length of a path segment,
where the 2-dimensional speed of light is taken to be $1$.
\vspace{12pt}

Here I have used the Gersch \cite{GER} convention
\newline
of weighting each turn by $ -im \epsilon$
\newline
rather than the Feynman \cite{FEY1, FEY2} convention
\newline
of weighting by $ +im \epsilon$, because Gersch's
convention gives a better nonrelativistic limit
in the isomorphic 2-dimensional Ising model \cite{GER}.
\vspace{12pt}

For a given path, let
\newline
$C$ be the total number of direction changes, and
\newline
$c$ be the $c$th change of direction, and
\newline
$ i$ be the complex imaginary  representing
the $c$th change of direction.
\vspace{12pt}

$C$ can be no greater than the timelike
Checkerboard distance $D$
\newline
between the initial and final points.
\vspace{12pt}

The total weight for the given path is then
\vspace{12pt}

\begin{equation}
\prod_{0 \leq c \leq C} -i m \epsilon =
(m \epsilon)^{C} (\prod_{0 \leq c \leq C} -i) =
(-im \epsilon )^{C}
\end{equation}

\vspace{12pt}

The product is a vector in the direction
$ \pm  1$ or $ \pm  i$.
\vspace{12pt}

Let $N(C)$ be the number of paths with
$C$ changes in direction.
\vspace{12pt}

The propagator amplitude for the particle to go from
the initial vertex to the final vertex is the sum over all
paths of the weights, that is the path integral sum
over all weighted paths:
\vspace{12pt}

\begin{equation}
\sum_{0 \leq C \leq D} N(C) (-im \epsilon )^{C}
\end{equation}

\vspace{12pt}

The propagator phase is the angle between
\newline
the amplitude vector in the complex plane and the complex real axis.
\vspace{12pt}

Conventional attempts to generalize the
Feynman Checkerboard from
\newline
2-dimensional spacetime to $k$-dimensional spacetime
are based on
\newline
the fact that the 2-dimensional future light-cone
directions are
\newline
the 0-sphere $S^{2-2} = S^{0} = \{ i,1 \}$.
\vspace{12pt}

The $k$-dimensional continuous spacetime lightcone
directions are
\newline
the $(k-2)$-sphere $S^{k-2}$.
\vspace{12pt}

In 4-dimensional continuous spacetime, the lightcone
directions
are $S^{2}$.
\vspace{12pt}

Instead of looking for a 4-dimensional lattice
spacetime, Feynman
\newline
and other generalizers went from discrete $S^{0}$
to continuous $S^{2}$
\newline
for lightcone directions, and then tried to
construct a weighting
\newline
using changes of directions as rotations
in the continuous $S^{2}$, and
\newline
never (as far as I know) got any generalization that worked.
\vspace{12pt}

\newpage

The HyperDiamond HyperDiamond generalization has
discrete lightcone directions.
\vspace{12pt}

If the 4-dimensional Feynman Checkerboard is coordinatized by
\newline
the quaternions $\bf{Q}$:
\newline
the real axis $1$ is identified with the time axis $t$;
\newline
the imaginary axes $i,j,k$ are identified with the space
axes $x,y,z$; and
\newline
the four future lightcone links are
\newline
$(1/2)(1+i+j+k)$,
\newline
$(1/2)(1+i-j-k)$,
\newline
$(1/2)(1-i+j-k)$, and
\newline
$(1/2)(1-i-j+k)$.
\vspace{12pt}

In cylindrical coordinates $t,r$
\newline
with $r^{2} = x^{2}+y^{2}+z^{2}$,
\newline
the Euclidian metric is $t^{2} + r^{2}$ = $t^{2} +
x^{2}+y^{2}+z^{2}$ and
\newline
the Wick-Rotated Minkowski metric with speed of light $c$ is
\newline
 $(ct)^{2} - r^{2}$ = $(ct)^{2} - x^{2} -y^{2} -z^{2}$.
\vspace{12pt}

For the future lightcone links to lI on
\newline
the 4-dimensional Minkowski lightcone, $c = \sqrt{3}$.
\vspace{12pt}

Any future lightcone link is taken into any
other future lightcone
\newline
link by quaternion multiplication by
$ \pm  i$,  $ \pm  j$, or  $ \pm  k$.
\vspace{12pt}

For a given vertex on a given path,
\newline
continuation in the same
direction can be represented by the link $1$, and
\newline
changing direction can be represented by the
\newline
imaginary quaternion $ \pm  i, \pm  j, \pm  k$ corresponding to
\newline
the link transformation that makes the change of direction.
\vspace{12pt}

Therefore, at a vertex where a path changes direction,
\newline
a path can be weighted by quaternion imaginaries
\newline
just as it
is weighted by the complex imaginary in the 2-dimensional case.
\vspace{12pt}

If the path does change direction at a vertex, then
\newline
the path at the point of change gets a weight of
$-im \epsilon$, $-jm \epsilon$, or $-km \epsilon$
\newline
where $i,j,k$ is the quaternion imaginary representing
the change of direction,
\newline
$m$ is the mass (only massive particles can change directions),
and
\newline
$\sqrt{3} \epsilon$ is the timelike length of a path segment,
\newline
where the 4-dimensional speed of light is taken to be
$\sqrt{3}$.
\vspace{12pt}

For a given path,
\newline
let $C$ be the total number of direction changes,
\newline
$c$ be the $c$th change of direction, and
\newline
$e_{c}$ be the quaternion imaginary $i,j,k$ representing
the $c$th change of direction.
\vspace{12pt}

$C$ can be no greater than the timelike Checkerboard distance
$D$
\newline
between the initial and final points.
\vspace{12pt}

The total weight for the given path is then
\vspace{12pt}

\begin{equation}
\prod_{0 \leq c \leq C} -e_{c} m \sqrt{3} \epsilon =
(m \sqrt{3} \epsilon)^{C} (\prod_{0 \leq c \leq C} -e_{c})
\end{equation}

\vspace{12pt}

Note that since the quaternions are not commutative,
\newline
the product must be taken in the correct order.
\vspace{12pt}

The product is a vector in the direction $ \pm  1$,
$ \pm  i$, $ \pm  j$, or $ \pm  k$.
and
\vspace{12pt}

Let $N(C)$ be the number of paths with $C$ changes in direction.
\vspace{12pt}

The propagator amplitude for the particle to go
\newline
from the initial vertex to the final vertex is
\newline
the sum over all paths of the weights,
\newline
that is the path integral sum over all weighted paths:
\vspace{12pt}

\begin{equation}
\sum_{0 \leq C \leq D} N(C) (m \sqrt{3} \epsilon)^{C}
(\prod_{0 \leq c \leq C} -e_{c})
\end{equation}

\vspace{12pt}

The propagator phase is the angle between
\newline
the amplitude vector in quaternionic 4-space and
\newline
the quaternionic real axis.
\vspace{12pt}

The plane in quaternionic 4-space defined by
\newline
the amplitude vector and the quaternionic real axis
\newline
can be regarded as the complex plane of the propagator phase.
\vspace{12pt}

\newpage

\section{Charge = Amplitude to Emit Gauge Boson.}

\subsection{What Factors Determine Charge?}

In the HyperDiamond Feynman Checkerboard model
the charge of a particle is the amplitude for a particle
to emit a gauge boson of the relevant force.
Neutral particles do not emit gauge bosons.
\vspace{12pt}

Force strengths are probalilities, or squares of amplitudes
for emission of gauge bosons, or squares of charges,
so that calculation of charges is equivalent to calculation
of force strengths.
\vspace{12pt}

Three factors determine the probability for emission of
a gauge boson from an origin spacetime vertex to a target vertex:
\vspace{12pt}

the part of the Internal Symmetry Space
of the target spacetime vertex that is available for the gauge boson
to go to from the origin vertex;
\vspace{12pt}

the volume of the spacetime link that is available for the gauge
boson to go through from the origin vertex to the
target vertex; and
\vspace{12pt}

an effective mass factor for forces
(such as the Weak force and Gravity)
that, in the low-energy ranges of our experiments,
are carried effectively by gauge bosons that are not
massless high-energy.
\vspace{12pt}

In the $D_{4}-D_{5}-E_{6}$ Lagrangian continuum version
of this physics model, force strength probabilities are
calculated
in terms of relative volumes of bounded
complex homogeneous domains and
their Shilov boundaries.
\vspace{12pt}

The relationship between
the $D_{4}-D_{5}-E_{6}$ Lagrangian continuum approach
and
the HyperDiamond Feynman Checkerboard discrete approach
is that:
\vspace{12pt}

the bounded complex homogeneous domains correspond to
\newline

harmonic functions of generalized Laplacians
\newline

that determine heat equations, or diffusion equations;
\vspace{12pt}

while the amplitude to emit gauge bosons in the
HyperDiamond Feynman Checkerboard is a process that
is similar to diffusion, and
therefore also corresponds to a generalized Laplacian.
\vspace{12pt}

Details of the $D_{4}-D_{5}-E_{6}$ Lagrangian continuum approach
can be found on the World Wide Web at URLs
\vspace{12pt}

http://xxx.lanl.gov/abs/hep-ph/9501252
\vspace{12pt}

http://galaxy.cau.edu/tsmith/d4d5e6hist.html
\vspace{12pt}

For the discrete HyperDiamond Feynman Checkerboard
approach of this paper, the only free charge parameter
is the charge of the $Spin(5)$ gravitons in the
MacDowell-Mansouri formalism of Gravity.  Note that
these $Spin(5)$ gravitons are NOT the ordinary spin-2
gravitons of the low-energy region in which we live.
The charge of the $Spin(5)$ gravitons is taken to be unity, 1,
so that its force strength is also unity, 1.
All other force strengths are determined as ratios
with respect to the $Spin(5)$ gravitons and each other.
\vspace{12pt}

The four forces of the HyperDiamond Feynman Checkerboard
model are Gravity, the Color force, the Weak force,
and Electromagnetism.
\vspace{12pt}

The charge of each force is the amplitude for
one of its gauge bosons to be emitted from a given origin
vertex of the spacetime HyperDiamond lattice
and go to a neighboring target vertex.
\vspace{12pt}

The force strength of each force is the square of the
charge amplitude, or the probablility for
one of its gauge bosons to be emitted from a given origin
vertex of the spacetime HyperDiamond lattice
and go to a neighboring target vertex.
\vspace{12pt}

The HyperDiamond Feynman Checkerboard model calculations
are actually done for force strengths, or probablilities,
because it is easier to calculate probablilities.
\vspace{12pt}

The force strength probability for a gauge boson to
be emitted from an origin spacetime HyperDiamond vertex
and go to a target vertex is the product of three things:
\vspace{12pt}

the volume $Vol(M_{IS_{force}})$ of the
target Internal Symmetry Space,
that is, the part of the Internal Symmetry Space
of the target spacetime vertex that is
available for the gauge boson
to go to from the origin vertex;
\vspace{12pt}

the volume ${Vol(Q_{force})} \over {{Vol(D_{force})}^{ \left( 1
\over m_{force} \right) }}$ of the spacetime link to
the target spacetime vertex from the origin vertex; and
\vspace{12pt}

an effective mass factor $1 \over \mu_{force}^2$ for forces
(such as the Weak force and Gravity)
that, in the low-energy ranges of our experiments,
are carried effectively by gauge bosons that are not
massless high-energy $SU(2)$ or $Spin(5)$ gauge bosons,
but are either massive Weak bosons due to the Higgs mechanism
or effective spin-2 gravitons.  For other forces, the
effective mass factor is taken to be unity, 1.
\vspace{12pt}

Therefore, the force strength of a given force is

\begin{equation}
\alpha_{force} = \left(1 \over \mu_{force}^2 \right)
\left( Vol(M_{IS_{force}})
\right)
\left( {Vol(Q_{force})} \over {{Vol(D_{force})}^{ \left( 1
\over m_{force} \right) }} \right)
\end{equation}

\vspace{12pt}

The symbols have the following meanings:
\vspace{12pt}

$\alpha_{force}$ represents the force strength;
\vspace{12pt}

$\mu_{force}$ represents the effective mass;
\vspace{12pt}

$M_{IS_{force}}$ represents the part of the target
Internal Symmetry Space that is available for the gauge
boson to go to;
\vspace{12pt}

$Vol(M_{IS_{force}})$ stands for volume of $M_{IS_{force}}$;
\vspace{12pt}

$Q_{force}$ represents the link from the origin
to the target that is available for the gauge
boson to go through;
\vspace{12pt}

$Vol(Q_{force})$ stands for volume of $Q_{force}$;
\vspace{12pt}

$D_{force}$ represents the complex bounded homogeneous domain
of which $Q_{force}$ is the Shilov boundary;
\vspace{12pt}

$m_{force}$ is the dimensionality of $Q_{force}$,
which is 4 for Gravity and the Color force,
2 for the Weak force (which therefore is considered to
have two copies of $Q_{W}$
for each spacetime HyperDiamond link),
and 1 for Electromagnetism (which therefore is considered to
have four copies of $Q_{E}$
for each spacetime HyperDiamond link)
\vspace{12pt}

${Vol(D_{force})}^{ \left( 1 \over m_{force} \right) }$
stands for
a dimensional normalization factor (to reconcile the dimensionality
of the Internal Symmetry Space of the target vertex
with the dimensionality of the link from the origin to the
target vertex).
\vspace{12pt}

\vspace{12pt}

The force strength formula is stated in terms
of continuum structures,
such as volumes of manifolds, Shilov Boundaries, etc.,
rather than in discrete terms.
We recognize that a discrete version
of the calculations would be the fundamentally correct way
to calculate in the discrete HyperDiamond Feynman Checkerboard
model, but it is easier for us to look up relevant manifold
volumes than to write and execute computer code to do the
discrete calculations.
\vspace{12pt}

However, since the HyperDiamond Feynman Checkerboard lattice
spacing is Planck length and therefore much smaller
than the relevant distances for any experiments that
we want to describe, we think that the continuum calculations
are good approximations of the fundamental discrete HyperDiamond
Feynman Checkerboard calculations.
\vspace{12pt}

The geometric volumes needed for the calculations,
mostly taken from Hua \cite{HUA}, are
\vspace{12pt}

\begin{equation}
\begin{array}{||c||c|c||c|c||c|c||}
\hline
Force & M & Vol(M) & Q
& Vol(Q) & D & Vol(D)  \\
\hline
& & & & & & \\
gravity & S^4 & 8\pi^{2}/3
& {\bf R}P^1 \times S^4  & 8\pi^{3}/3
& IV_{5} & \pi^{5}/2^{4} 5! \\
\hline
& & & & & &\\
color & {\bf C}P^2 & 8\pi^{2}/3
& S^5 & 4\pi^{3}
& B^6 \: (ball) & \pi^{3}/6 \\
\hline
& & & & & & \\
weak & {S^2} \times {S^2} & 2 \times {4 \pi}
& {\bf R}P^1 \times S^2 & 4 \pi^2
& IV_{3} & \pi^{3} / 24 \\
\hline
& & & & & & \\
e-mag  & T^4  & 4 \times {2\pi}
& -  & -
& -  & - \\
\hline
\end{array}
\end{equation}

Using these numbers, the results of the
calculations are the relative force strengths
at the characteristic energy level of the
generalized Bohr radius of each force:

\begin{equation}
\begin{array}{|c|c|c|c|c|}
\hline
Gauge \: Group & Force & Characteristic
& Geometric & Total \\
& & Energy & Force & Force \\
& & & Strength & Strength \\
\hline
& & & & \\
Spin(5) & gravity & \approx 10^{19} GeV
& 1 & G_{G}m_{proton}^{2} \\
& & & & \approx 5 \times 10^{-39} \\
\hline
& & & & \\
SU(3) & color & \approx 245 MeV & 0.6286
& 0.6286 \\
\hline
& & & & \\
SU(2) & weak & \approx 100 GeV & 0.2535
& G_{W}m_{proton}^{2} \approx  \\
& & & & \approx 1.05 \times 10^{-5} \\
\hline
& & & & \\
U(1) & e-mag  & \approx 4 KV
& 1/137.03608  & 1/137.03608 \\
\hline
\end{array}
\end{equation}

\vspace{12pt}

The force strengths are given at the characteristic
energy levels of their forces, because the force
strengths run with changing energy levels.
\vspace{12pt}

The effect is particularly pronounced with the color
force.

\vspace{12pt}

The color force strength was calculated
at various energies according to renormalization group
equations, with the following results:
\vspace{12pt}

\begin{equation}
\begin{array}{|c|c|}
\hline
Energy \: Level & Color \: Force \: Strength \\
\hline
&  \\
245 MeV & 0.6286 \\
& \\
5.3 GeV & 0.166 \\
& \\
34 GeV & 0.121  \\
& \\
91 GeV & 0.106 \\
& \\
\hline
\end{array}
\end{equation}

\vspace{12pt}

Shifman \cite{SHI} in a paper at
\vspace{12pt}

http://xxx.lan.gov/abs/hep-ph/9501222

\vspace{12pt}

has noted that Standard Model global fits at the $Z$ peak,
about $91 \; GeV$, give a color force strength of about
0.125 with $\Lambda_{QCD} \approx 500 \; MeV$,
\newline
whereas low energy results and lattice calculations
give  a color force strength at the $Z$ peak of about
0.11 with $\Lambda_{QCD} \approx 200 \; MeV$.

\vspace{12pt}

In the remainder of this Chapter 8,
we discuss further the concepts of Target Internal Symmetry
Space, Link to Target, and Effective Mass Factors,
and then we discuss in more detail each of the four forces.
\vspace{12pt}

In this HyperDiamond Feynman Checkerboard model,
all force strengths are represented as ratios with
respect to the geometric force strength of Gravity
(that is, the force strength of Gravity without using
the Effective Mass factor).

\subsubsection{Volume of Target Internal Symmetry Space.}

$M_{IS_{force}}$ represents the part of the target
Internal Symmetry Space that is available for the gauge
boson to go to; and
\vspace{12pt}

$Vol(M_{IS_{force}})$ stands for volume of $M_{IS_{force}}$.
\vspace{12pt}

What part of the target Internal Symmetry Space is
available for the gauge boson to go to?
\vspace{12pt}

Each vertex of the spacetime HyperDiamond lattice
is not just a point, but also contains its own
Internal Symmetry Space (also a HyperDiamond lattice),
so the amplitude for a gauge boson to go from one vertex
to another depends
\vspace{12pt}

not only on the spacetime link between the vertices
\vspace{12pt}

but also on the degree of connection the gauge boson
has with the Internal Symmetry Spaces of the vertices.
\vspace{12pt}

If the gauge boson can connect any vertex in the
origin Internal Symmetry Space with any vertex in
the destination Internal Symmetry Space,
then the gauge boson has full connectivity between
the Internal Symmetry Spaces.
\vspace{12pt}

However, if the gauge boson can connect a vertex in the
origin Internal Symmetry Space only with some,
but not any, of the vertices in
the destination Internal Symmetry Space,
then the gauge boson has only partial connectivity between
the Internal Symmetry Spaces, and
has a lower amplitude to be emitted from the origin
spacetime vertex to the destination spacetime vertex.
\vspace{12pt}

The amount of connectivity between the Internal Symmetry Spaces
is the geometric measure of the charge of a force,
and therefore of its force strength.
\vspace{12pt}

To represent the gauge boson connectivity
between Internal Symmetry Spaces,
it is useful to label the basis of the Internal Symmetry Space
by the degrees of freedom of the forces:
\vspace{12pt}

electric charges $\{ +1, -1 \}$ ; and
\vspace{12pt}

color (red, green, blue) charges
$\{ +r, -r; +g, -g; +b, -b \}$ .
\vspace{12pt}

so that the total basis of the Internal Symmetry Space is
\vspace{12pt}

$\{ +1, -1, +r, -r; +g, -g; +b, -b \}$

\vspace{12pt}

This connectivity can be measured by comparing
\vspace{12pt}

the full target Internal Symmetry Space HyperDiamond lattice
\vspace{12pt}

with the subspace of the target
Internal Symmetry Space HyperDiamond lattice
that is the image of a given point of
the origin Internal Symmetry Space under all the
transformations of the Internal Symmetry Space
of the gauge group of the force.
\vspace{12pt}

The $M_{IS_{force}}$ target Internal Symmetry Space manifolds,
each irreducible component of which has dimension $m_{force}$,
for the four forces are:
\vspace{12pt}

\begin{equation}
\begin{array}{|c|c|c|c|}
\hline
Gauge \: Group & Symmetric \: Space & m_{force}
& M_{IS_{force}}  \\
\hline
& & & \\
Spin(5) & Spin(5) \over Spin(4)  & 4 & S^4\\
& & & \\
SU(3) & SU(3) \over {SU(2) \times U(1)}
& 4  & {\bf C}P^2 \\
& & & \\
SU(2) & SU(2) \over U(1)  & 2 & S^2 \times S^2 \\
& & & \\
U(1) & U(1)  & 1 & S^1 \times S^1 \times S^1
\times S^1 \\
& & & \\
\hline
\end{array}
\end{equation}

\vspace{12pt}

\subsubsection{Volume of Link to Target.}

If, as in the case of the Electromagnetic $U(1)$ photon,
there is only one gauge boson that can go through the link
from an origin spacetime HyperDiamond vertex to a target vertex,
then the link to the target vertex is like a one-lane road.
\vspace{12pt}

For the Weak $SU(2)$ force, the Color $SU(3)$ force,
and $Spin(5)$ Gravity, there are, respectively, 3, 8, and 10
gauge bosons that can go through the link
from an origin spacetime HyperDiamond vertex to a target vertex,
so that the link to the target vertex is
like a multi-lane highway.
\vspace{12pt}

The volume of the link to the target vertex
is not measured just by the number of gauge bosons,
but by the volume of the minimal manifold $Q_{force}$
that can carry two things:
\vspace{12pt}

the gauge bosons, with gauge group $G_{force}$; and
\vspace{12pt}

the $U(1)$ phase of the propagator from origin to target.
\vspace{12pt}

The first step in constructing $Q_{force}$ is to
find a manifold whose local isotropy symmetry group
is $G_{force} \times U(1)$ so that it can carry
both $G_{force}$ and $U(1)$.
\vspace{12pt}

To do that, look for the smallest Hermitian symmetric space
of the form $K / (G_{force} \times U(1))$.
\vspace{12pt}

Having found that Hermitian symmetric space,
then go to its corrresponding
complex bounded homogeneous domain, $D_{force}$.
\vspace{12pt}

Then take $Q_{force}$ to be the Shilov boundary of $D_{force}$.
\vspace{12pt}

$Q_{force}$ is then the minimal manifold that can
carry both $G_{force}$ and $U(1)$, and so is the
manifold that should represent the link
from origin to target vertex.
\vspace{12pt}

The $Q_{force}$, Hermitian symmetric space,
and $D_{force}$ manifolds for the four forces are:
\vspace{12pt}

\begin{equation}
\begin{array}{|c|c|c|c|c|}
\hline
Gauge & Hermitian & Type & m_{force}
& Q_{force}  \\
Group & Symmetric & of & & \\
& Space & D_{force} & & \\
\hline
& & & & \\
Spin(5) & Spin(7) \over {Spin(5) \times U(1)}
& IV_{5} &4 & {\bf R}P^1 \times S^4 \\
& & & & \\
SU(3) & SU(4) \over {SU(3) \times U(1)}
& B^6 \: (ball) &4 & S^5 \\
& & & & \\
SU(2) & Spin(5) \over {SU(2) \times U(1)}
& IV_{3} & 2 & {\bf R}P^1 \times S^2 \\
& & & & \\
U(1) & -  & - & 1  & - \\
& & & & \\
\hline
\end{array}
\end{equation}

\vspace{12pt}

The geometric volumes of
the target Internal Symmetry Space $M_{IS_{force}}$,
the link volume $Q_{force}$,
and the bounded complex domains $D_{force}$ of which
the link volume is the Shilov boundary,
mostly taken from Hua \cite{HUA}, are:
\vspace{12pt}

\begin{equation}
\begin{array}{||c||c|c||c|c||c|c||}
\hline
Force & M & Vol(M) & Q
& Vol(Q) & D & Vol(D)  \\
\hline
& & & & & & \\
gravity & S^4 & 8\pi^{2}/3
& {\bf R}P^1 \times S^4  & 8\pi^{3}/3
& IV_{5} & \pi^{5}/2^{4} 5! \\
\hline
& & & & & &\\
color & {\bf C}P^2 & 8\pi^{2}/3
& S^5 & 4\pi^{3}
& B^6 \: (ball) & \pi^{3}/6 \\
\hline
& & & & & & \\
weak & {S^2} \times {S^2} & 2 \times {4 \pi}
& {\bf R}P^1 \times S^2 & 4 \pi^2
& IV_{3} & \pi^{3} / 24 \\
\hline
& & & & & & \\
e-mag  & T^4  & 4 \times {2\pi}
& -  & -
& -  & - \\
\hline
\end{array}
\end{equation}

\vspace{12pt}

The geometric part of the force strength is
formed from the product of the volumes of
the target Internal Symmetry Space $M_{IS_{force}}$
and the link volume $Q_{force}$.
\vspace{12pt}

To take the product properly,
we must take into account how
the target Internal Symmetry Space $M_{IS_{force}}$,
which lives at the target vertex,
and the link volume $Q_{force}$,
which lives on the link connecting the origin vertex
to the target vertex and so can be regarded as
containing both the origin vertex and the target vertex,
fit together.
\vspace{12pt}

The dimension $m_{force}$ of each irreducible component of
the target Internal Symmetry Space $M_{IS_{force}}$
is less than the dimension of the link volume $Q_{force}$,
\vspace{12pt}

which in turn is less than the dimension of
the bounded complex domain $D_{force}$ of which
the link volume $Q_{force}$ is the Shilov boundary.
\vspace{12pt}

Since the link volume $Q_{force}$ is a Shilov boundary,
it can be regarded as a shell whose Shilov interior is
the bounded complex domain $D_{force}$.
\vspace{12pt}

If we were to merely take the product of the
volume of the target Internal Symmetry Space $M_{IS_{force}}$
with the volume of the link volume $Q_{force}$,
we would be overcounting the contribution of
the link volume $Q_{force}$ because of its dimension
is higher than the dimension of each irreducible component of
the target Internal Symmetry Space $M_{IS_{force}}$.
\vspace{12pt}

To get rid of the overcounting,
we should make the link volume $Q_{force}$ compatible with
each irreducible component of
the target Internal Symmetry Space $M_{IS_{force}}$.
\vspace{12pt}

Since the target Internal Symmetry Space $M_{IS_{force}}$
has fundamentally a 4-dimensional HyperDiamond structure,
each irreducible component is fundamentally made up of
$m_{force}$-dimensional hypercubic cells,
\vspace{12pt}

where the irreducible component hypercubic cells are
\newline

4-dimensional for Gravity and the Color force,
\newline

2-dimensional for the Weak force,
\newline

and 1-dimensional for Electromagnetism.
\vspace{12pt}

Since $D_{force}$ is the Shilov interior of $Q_{force}$,
$Q_{force}$ would be compatible with $M_{IS_{force}}$
if $D_{force}$ were mapped 1-1 onto
an $m_{force}$-dimensional hypercubic cell
in an irreducible component of
the target Internal Symmetry Space $M_{IS_{force}}$.
\vspace{12pt}

To do this, first construct a hypercube of the same dimension
$m_{force}$ as an irreducible component of
the target Internal Symmetry Space $M_{IS_{force}}$
and the same volume as $D_{force}$.
\vspace{12pt}

The edge length of such a hypercube is
the $m_{force}$-th root of the volume of
the bounded complex domain $D_{force}$.
\vspace{12pt}

Since the hypercubes in the fundamental HyperDiamond structures
of the target Internal Symmetry Space $M_{IS_{force}}$
and its irreducible components are unit hypercubes
with edge length 1,
\vspace{12pt}

we must, to make the link volume $Q_{force}$ compatible with
each irreducible component of
the target Internal Symmetry Space $M_{IS_{force}}$,
\vspace{12pt}

divide the volume of the link volume $Q_{force}$
by the $m_{force}$-th root of the volume of
the bounded complex domain $D_{force}$
\vspace{12pt}

to make $D_{force}$ the right size to fit
the hypercubes in the fundamental HyperDiamond structures
of the target Internal Symmetry Space $M_{IS_{force}}$
and its irreducible components.
\vspace{12pt}

In other words,
to reconcile the dimensionality of the link volume $Q_{force}$
to the dimensionality of the target Internal Symmetry Space,
divide $Q_{force}$ by
\vspace{12pt}

${Vol(D_{force})}^{ \left( 1 \over m_{force} \right) }$.
\vspace{12pt}

The resulting volume
\vspace{12pt}

${Vol(Q_{force})} \over {{Vol(D_{force})}^{ \left( 1 \over m_{force}
\right) }}$
\vspace{12pt}

is then the correctly normalized volume of the spacetime link that is
available for the gauge boson to go through to get to
the target spacetime vertex from the origin vertex,
\vspace{12pt}

which correctly normalized volume should be used in
multiplying by the volume of
he target Internal Symmetry Space $M_{IS_{force}}$
to get the geometric part of the force strength.
\vspace{12pt}

\subsubsection{Effective Mass Factors.}

In the low-energy ranges of our experiments,
the Weak force and Gravity are carried effectively
by gauge bosons that are not
massless high-energy $SU(2)$ or $Spin(5)$ gauge bosons,
but are either massive Weak bosons due to the Higgs mechanism
or effective spin-2 gravitons.
\vspace{12pt}

$\mu_{force}$ represents the effective mass for
the Weak force and Gravity in the low-energy range.
\vspace{12pt}

In the low-energy range, the forece strengths of
the Weak force and Gravity have an effective mass factor
$1 \over \mu_{force}^2$.
\vspace{12pt}

For other forces, the Color force and Electromagnetism,
the effective mass factor is taken to be unity, 1.
\vspace{12pt}

\subsection{Gravity.}

For Gravity, $Spin(5)$ gravitons can carry all of the
Internal Symmetry Space charges
\vspace{12pt}

$\{ +1, -1; +r, -r; +g, -g; +b, -b \}$
\vspace{12pt}

$Spin(5)$ gravitons act transitively on the
4-dimensional manifold $S^{4} = Spin(5) / Spin(4)$,
so that they can take any  given point in
the origin Internal Symmetry Space into any point in the
target Internal Symmetry Space.
\vspace{12pt}

The full connectivity of the Gravity of $Spin(5)$ gravitons
is geometrically represented by $M_{IS_{G}}$
as the 4-dimensional sphere $S^{4} = Spin(5) / Spin(4)$
whose volume is $8\pi^{2}/3$.
\vspace{12pt}

The link manifold to the target vertex is
\vspace{12pt}

$Q_{G} = {\bf R}P^1 \times S^4 = ShilovBdy(D_{G})$
with volume $8\pi^{3}/3$
\vspace{12pt}

The bounded complex homogeneous domain $D_{G}$
\newline

is of type $IV_{5}$  with volume $\pi^{5}/2^{4} 5!$.
\vspace{12pt}

It corresponds to the Hermitian Symmetric Space
$$Spin(7)/(Spin(5) \times U(1)$$
\vspace{12pt}

For $Spin(5)$ Gravity, $m_{G}$ is 4.
\vspace{12pt}

For $Spin(5)$ Gravity, $\mu_{G} = m_{Planck}$, so that
\vspace{12pt}

${1 \over \mu_{G}^2} = {1 \over {m_{Planck}^{2}}}$.

\vspace{12pt}

Therefore, the force strength of Gravity is:
\vspace{12pt}

$\alpha_{G} = \left(1 \over \mu_{G}^2 \right)
\left( Vol(M_{IS_{G}})
\right)
\left( {Vol(Q_{G})} \over
{{Vol(D_{G})}^{ \left( 1 \over m_{G} \right) }} \right)$

\vspace{12pt}

which is $G_{G}m_{proton}^{2} \approx 5 \times 10^{-39}$
\vspace{12pt}

at the characteristic energy level of $\approx 10^{19} GeV$,
the Planck energy.
\vspace{12pt}

The only factor different from 1 in the force strength of
Gravity is the Effective Mass factor,
because force strengths in the
HyperDiamond Feynman Checkerboard model
are represented as ratios with respect to the strongest
geometric force, Gravity, so that the geometric force strength
factors for Gravity are cancelled to unity by taking the ratio
with themselves.
\vspace{12pt}

For an estimate of the Planck mass calculated in
the spirit of the HyperDiamond Feynman Checkerboard model, see
\vspace{12pt}

http://galaxy.cau.edu/tsmith/Planck.html

\vspace{12pt}

http://www.innerx.net/personal/tsmith/Planck.html

\vspace{12pt}

The action of Gravity on Spinors is given by
the Papapetrou Equations.

\vspace{12pt}

\subsection{Color Force.}

For the Color force, $SU(3)$ gluons carry the
Internal Symmetry Space color charges
\vspace{12pt}

$\{ +r, -r; +g, -g; +b, -b \}$

\vspace{12pt}

$SU(3)$ gluons act transitively on the
4-dimensional manifold
${\bf C}P^2 = SU(3) / (SU(2) \times U(1))$,
so that they can take any  given point in
the origin Internal Symmetry Space into any point in the
target Internal Symmetry Space.
\vspace{12pt}

The full connectivity of the Color force of $SU(3)$ gluons
is geometrically represented by $M_{IS_{C}}$
as ${\bf C}P^{2} = SU(3) / (SU(2) \times U(1))$
whose volume is $8\pi^{2}/3$.
\vspace{12pt}

The link manifold to the target vertex is
\vspace{12pt}

$Q_{C} = S^{5} = ShilovBdy(B^{6})$
with volume $4\pi^{3}$
\vspace{12pt}

The bounded complex homogeneous domain
$D_{C}$ is of type $B^{6} (ball)$
with volume $\pi^{3}/6$
\vspace{12pt}

For the $SU(3)$ Color force, $m_{C}$ is 4.
\vspace{12pt}

For the $SU(3)$ Color force, $\mu_{C} = 1 $, so that
\vspace{12pt}

for the $SU(3)$ Color force, ${1 \over \mu_{C}^2} = 1 $
\vspace{12pt}

Therefore, the force strength of the Color force,
which is sometimes conventionally denoted by
$\alpha_{S}$ (for strong)
as well as by $\alpha_{C}$, is:
\vspace{12pt}

$\alpha_{C} = \alpha_{S}
= \left(1 \over \mu_{C}^2 \right) \left( Vol(M_{IS_{C}})
\right)
\left( {Vol(Q_{C})} \over
{{Vol(D_{C})}^{ \left( 1 \over m_{C} \right) }} \right)$

\vspace{12pt}

which, when divided by the geometric force strength of Gravity,
is $0.6286$
\vspace{12pt}

at the characteristic energy level of $\approx 245 MeV$.
\vspace{12pt}

Force strength constants and particle masses
are not really "constant" when you measure them,
as the result of your measurement will depend on
the energy at which you measure them.
Measurements at one energy level can be related to
measurements at another by renormalization equations.
The lightest experimentally observable particle
based on the color force is the pion,
which is a quark-antiquark pair made up of the lightest
quarks, the up and down quarks.
A quark-antiquark pair is the carrier of the strong force,
and mathematically resembles a bivector gluon,
which is the carrier of the color force.
The charactereistic energy level of pions is
the square root of the sum of the squares of
the masses of the two charged and one neutral pion.
It is about 245 MeV (to more accuracy 241.4 MeV).
The gluon-carried color force strength is renormalized
to higher energies from about 245 MeV in the conventional way.

The Color force strength was calculated
at various energies according to renormalization group
equations, as shown at
\vspace{12pt}

http://galaxy.cau.edu/tsmith/cweRen.html

\vspace{12pt}

http://www.innerx.net/personal/tsmith/cweRen.html

\vspace{12pt}

with the following results:

\begin{equation}
\begin{array}{|c|c|}
\hline
Energy \: Level & Color \: Force \: Strength \\
\hline
&  \\
245 MeV & 0.6286 \\
& \\
5.3 GeV & 0.166 \\
& \\
34 GeV & 0.121  \\
& \\
91 GeV & 0.106 \\
& \\
\hline
\end{array}
\end{equation}

\vspace{12pt}

Shifman \cite{SHI}, in a paper at
\vspace{12pt}

http://xxx.lan.gov/abs/hep-ph/9501222

\vspace{12pt}

has noted that Standard Model global fits at the $Z_{0}$ peak,
about $91 \; GeV$, give a color force strength of about
0.125 with $\Lambda_{QCD} \approx 500 \; MeV$,
\vspace{12pt}

whereas low energy results and lattice calculations
give  a color force strength at the $Z_{0}$ peak of about
0.11 with $\Lambda_{QCD} \approx 200 \; MeV$.
\vspace{12pt}

The low energy results and lattice calculations are
closer to the tree level HyperDiamond Feynman Checkerboard model
value at $91 \; GeV$ of 0.106.
\vspace{12pt}

Also, the HyperDiamond Feynman Checkerboard model has
$\Lambda_{QCD} \approx 245 \; MeV$
\vspace{12pt}

For the pion mass, upon which the $\Lambda_{QCD}$
calculation depends, see
\vspace{12pt}

http://galaxy.cau.edu/tsmith/SnGdnPion.html

\vspace{12pt}

http://www.innerx.net/personal/tsmith/SnGdnPion.html

\vspace{12pt}

\subsection{Weak Force.}

For the Weak force, $SU(2)$ Weak bosons carry the
Internal Symmetry Space electric charges
\vspace{12pt}

$\{ +1, -1 \}$

\vspace{12pt}

$SU(2)$ Weak bosons act transitively on the
2-dimensional manifold $S^{2} = SU(2) / U(1)$,
so that two copies of
$S^{2}$, in the form of $S^{2} \times S^{2}$,
 are required so that
they can take any given point in
the origin Internal Symmetry Space into any point in the
target Internal Symmetry Space.
\vspace{12pt}

Each of the two connectivity components of the Weak force of
$SU(2)$ Weak bosons is geometrically represented by
$M_{IS_{W}}$
as $S^{2} = SU(2) / U(1)$
whose volume is $4\pi$.
\vspace{12pt}

The total manifold $M_{IS_{W}}$ is $S^{2} \times S^{2}$,
with volume $2 \times 4\pi$
\vspace{12pt}

The link manifold to the target vertex is
\vspace{12pt}

$Q_{W} = {\bf R}P^{1} \times S^{2} = ShilovBdy(D_{W})$
with volume $4\pi^{3}$.
\vspace{12pt}

The bounded complex homogeneous domain
$D_{W}$ is of type $IV_{3}$
with volume $4\pi^{2}$.
\vspace{12pt}

It corresponds to the  Hermitian Symmetric Space
\newline

$Spin(5)/(SU(2) \times U(1)$.
\vspace{12pt}

For the $SU(2)$ Weak force, $m_{W}$ is 2.
\vspace{12pt}

For the $SU(2)$ Weak force, due to the Higgs mechanism,
\newline

$\mu_{W} = \sqrt{m_{W+}^{2} + m_{W-}^{2} + m_{W_{0}}^{2}}$,
\newline

so that
\vspace{12pt}

${1 \over \mu_{W}^2}
= {1 \over {m_{W+}^{2} +m_{W-}^{2} + m_{W_{0}}^{2}}}$.

\vspace{12pt}

Therefore, the force strength of the Weak force is:
\vspace{12pt}

$\alpha_{W} = \left(1 \over \mu_{W}^2 \right) \left( Vol(M_{IS_{W}})
\right)
\left( {Vol(Q_{W})} \over
{{Vol(D_{W})}^{ \left( 1 \over m_{W} \right) }} \right)$

\vspace{12pt}

which, when divided by the geometric force strength
\newline
of Gravity, is
$G_{W}m_{proton}^{2} \approx 1.05 \times 10^{-5}$
\vspace{12pt}

at the characteristic energy level of $\approx 100 GeV$.
\vspace{12pt}

The geometric component of the Weak force strength,
that is, everything but the effective mass factor,
has the value $0.2535$ when divided by the geometric
force strength of Gravity,
\vspace{12pt}

\subsection{Electromagnetism.}

For Electromagnetixm, $U(1)$ photons carry no
Internal Symmetry Space charges.
\vspace{12pt}

$U(1)$ Weak bosons act transitively on the
1-dimensional manifold $S^{1} = U(1)$,
so that four copies of $S^{1}$,
in the form of the 4-torus $T^{4}$,
are required so that they can take any given point in
the origin Internal Symmetry Space into any point in the
target Internal Symmetry Space.
\vspace{12pt}

Each of the four connectivity components
of the Electromagnetism of
$U(1)$ photons is geometrically represented by $M_{IS_{E}}$
as $S^{1} = U(1)$
whose volume is $2\pi$.
\vspace{12pt}

The total manifold $M_{IS_{E}}$ is
$T^{4} = S^{1} \times S^{1} \times S^{1} \times S^{1}$,
with volume $4 \times 2\pi$
\vspace{12pt}

The link manifold to the target vertex is trivial for the
abelian neutral U(1) photons of Electromagnetism,
so we take $Q_{E}$ and $D_{E}$ to be equal to unity.
\vspace{12pt}

For $U(1)$ Electromagnmetism, $m_{E}$ is 1.
\vspace{12pt}

For $U(1)$ Electromagnmetism, $\mu_{W} = 1 $, so that
\vspace{12pt}

${1 \over \mu_{E}^2} = 1$.
\vspace{12pt}

Therefore, the force strength of Electromagnetism is:
\vspace{12pt}

$\alpha_{E} = \left(1 \over \mu_{E}^2 \right)
\left( Vol(M_{IS_{E}})
\right)
\left( {Vol(Q_{E})} \over
{{Vol(D_{E})}^{ \left( 1 \over m_{E} \right) }} \right)$

\vspace{12pt}

which, when divided by the geometric force strength of Gravity,
is $\alpha_{E} = 1/137.03608$
\vspace{12pt}

at the characteristic energy level of $\approx 4 KeV$.
\vspace{12pt}

\vspace{12pt}

\newpage

\section{Mass = Amplitude to Change Direction.}

In the HyperDiamond Feynman Checkerboard model
the mass parameter $m$ is the amplitude for a particle
to change its spacetime direction.  Massless particles
do not change direction, but continue on the same
lightcone path.
\vspace{12pt}

In the $D_{4}-D_{5}-E_{6}$ Lagrangian continuum version
of this physics model, particle masses are calculated in terms
of relative volumes of bounded complex homogeneous domains and
their Shilov boundaries.
\vspace{12pt}

The relationship between
the $D_{4}-D_{5}-E_{6}$ Lagrangian continuum approach
and
the HyperDiamond Feynman Checkerboard discrete approach
is that:
\vspace{12pt}

the bounded complex homogeneous domains correspond
\newline

to harmonic functions of generalized Laplacians
\newline

that determine heat equations, or diffusion equations;
\vspace{12pt}

while the amplitude to change directions in the
HyperDiamond Feynman Checkerboard is a diffusion process
in the HyperDiamond spacetime, also corresponding to
a generalized Laplacian.
\vspace{12pt}

Details of the $D_{4}-D_{5}-E_{6}$ Lagrangian continuum approach
can be found on the World Wide Web at URLs
\vspace{12pt}

http://xxx.lanl.gov/abs/hep-ph/9501252
\vspace{12pt}

http://galaxy.cau.edu/tsmith/d4d5e6hist.html
\vspace{12pt}

For the discrete HyperDiamond Feynman Checkerboard
approach of this paper, the only free mass parameter
is the mass of the Higgs scalar.  All other particle
masses are determined as ratios with respect to the
Higgs scalar and each other.
\vspace{12pt}

The Higgs mass is set equal to 259.031 GeV in the
HyperDiamond Feynman Checkerboard model, a figure
chosen so that the ratios will give an electron
mass of 0.5110 MeV.
\vspace{12pt}

Effectively, in the HyperDiamond Feynman Checkerboard model,
\newline
the electron mass is fixed at 0.5110 MeV
\newline

and all other masses are determined from it
\newline
by the ratios calculated in the model.
\vspace{12pt}

\subsection{Higgs Scalar, Gauge Bosons, and Fermions.}

Recall from Chapter 4 that 16x16 = 256-dimensional $DCl(0,8)$
has structure

\vspace{12pt}

\[
\begin{array}{ccccccccccccccccccccc}
&&{\bf1}&&{\bf8}&&{\bf28}&&56&&70&&56&&28&&8&&1&&\\
\end{array}
\]

\vspace{12pt}

that gives three types of particles for which mass ratios
can be calculated in the HyperDiamond Feynman Checkerboard
model:
\vspace{12pt}

the Higgs Scalar;
\vspace{12pt}

the 28 bivector gauge bosons;
\vspace{12pt}

and the 8 + 8 = 16 half-spinor fermions.
\vspace{12pt}

\subsection{Higgs Scalar Mass.}

There is only one Higgs scalar,
and we have chosen its mass to be 259.031 GeV.
\vspace{12pt}

\subsection{Gauge Boson Masses.}

After dimensional reduction to 4-dimensional spacetime,
the 28 $Spin(0,8)$ gauge bosons split into two groups:

12 Standard Model gauge bosons, which are
\vspace{12pt}

8 $SU(3)$ gluons,
\vspace{12pt}

3 $SU(2)$ weak bosons, and
\vspace{12pt}

1 $U(1)$ photon;
\vspace{12pt}

and
\vspace{12pt}

16 $U(4)$ gauge bosons,
\vspace{12pt}

which reduce to 15 $SU(4) = Spin(6)$ gauge bosons
plus one $U(1)$ phase for particle propagator amplitudes
(this phase is what makes the sum-over-histories quantum
theory interferences work).
\vspace{12pt}

The 15  $SU(4) = Spin(6)$ gauge bosons further reduce
to 10 $Spin(5)$ deSitter gravitons that give physical
gravity by the MacDowell-Mansouri mechanism as described in
\vspace{12pt}

http://xxx.lanl.gov/abs/hep-ph/9501252
\vspace{12pt}

and 4 conformal generators and 1 scale generator.
\vspace{12pt}

The 4 conformal generators couple to the Higgs scalar
so that it becomes the mass-giver by the Higgs mechanism
as described in
\vspace{12pt}

http://xxx.lanl.gov/abs/hep-ph/9501252
\vspace{12pt}

and the 1 scale generator
represents the scale that we chose
by setting the Higgs scalar mass at 259.031 GeV.
\vspace{12pt}

All these gauge bosons are massless at very high energies,
but at energies comparable to the Higgs mass and below,
the Higgs scalar couples to the $SU(2)$ weak bosons
to give them mass.
\vspace{12pt}

\subsubsection{Weak Boson Masses.}

Denote the 3 $SU(2)$ massless high-energy weak bosons
by $W_{+}$, $W_{-}$, and $W_{0}$.
\vspace{12pt}

The triplet $ \{ W_{+}, W_{-}, W_{0} \} $
couples directly with the Higgs scalar, so that the
total mass of the triplet $ \{ W_{+}$, $W_{-}$, $W_{0} \} $
is equal to the mass of the Higgs scalar,  259.031 GeV.
\vspace{12pt}

What are individual masses of members
of the triplet $\{ W_{+}, W_{-}, W_{0} \}$?
\vspace{12pt}

The entire triplet $ \{ W_{+}, W_{-}, W_{0} \} $
can be represented by the 3-sphere $S^{3}$.
\vspace{12pt}

The Hopf fibration of $S^{3}$ as
$S^{1} \rightarrow  S^{3} \rightarrow  S^{2}$
gives a decomposition of the $W$ bosons
into the neutral $W_{0}$ corresponding to $S^{1}$ and
the charged pair $W_{+}$ and $W_{-}$ corresponding
to $S^{2}$.

\vspace{12pt}

The mass ratio of the sum of the masses of
$W_{+}$ and $W_{-}$ to
the mass of $W_{0}$
should be the volume ratio of
the $S^{2}$ in $S^{3}$ to
the $S^{1}$ in ${S3}$.

\vspace{12pt}

The unit sphere $S^{3} \subset R^{4}$ is
normalized by $1 / 2$.

\vspace{12pt}

The unit sphere $S^{2} \subset R^{3}$ is
normalized by $1 / \sqrt{3}$.

\vspace{12pt}

The unit sphere $S^{1} \subset R^{2}$ is
normalized by $1 / \sqrt{2}$.

\vspace{12pt}

The ratio of the sum of the $W_{+}$ and $W_{-}$ masses to
the $W_{0}$ mass should then be
$(2  / \sqrt{3}) V(S^{2}) / (2 / \sqrt{2}) V(S^{1}) =
1.632993$.
\vspace{12pt}

Since the total mass of the triplet
$ \{ W_{+}, W_{-}, W_{0} \} $
is 259.031 GeV, and the charged wek bosons have
equal mass, we have
\vspace{12pt}

$m_{W_{+}} = m_{W_{-}} = 80.326 \; GeV$, and
$m_{W_{0}} = 98.379 \; GeV$.

\vspace{12pt}

\subsubsection{Parity Violation, Effective Masses, and
Weinberg Angle.}

The charged $W_{\pm}$ neutrino-electron interchange
\newline
must be symmetric with the electron-neutrino interchange,
\newline
so that the absence of right-handed neutrino particles requires
\newline
that the charged $W_{\pm}$ $SU(2)$
weak bosons act only on left-handed electrons.
\vspace{12pt}

Each gauge boson must act consistently
on the entire Dirac fermion particle sector,
so that the charged $W_{\pm}$ $SU(2)$ weak bosons
act only on left-handed fermions of all types.
\vspace{12pt}

The neutral $W_{0}$ weak boson does not interchange Weyl
neutrinos with Dirac fermions, and so is not restricted
to left-handed fermions,
but also has a component that acts on both types of fermions,
both left-handed and right-handed, conserving parity.
\vspace{12pt}

However, the neutral $W_{0}$ weak bosons are related to
the charged $W_{\pm}$ weak bosons by custodial $SU(2)$
symmetry, so that the left-handed component of the
neutral $W_{0}$ must be equal to the left-handed (entire)
component of the charged $W_{\pm}$.
\vspace{12pt}

Since the mass of the $W_{0}$ is greater than the mass
of the $W_{\pm}$, there remains for the $W_{0}$ a component
acting on both types of fermions.
\vspace{12pt}

Therefore the full $W_{0}$ neutral weak boson interaction
is proportional to
$(m_{W_{\pm}}^{2} / m_{W_{0}}^{2})$
acting on left-handed fermions
and
\vspace{12pt}

$(1 - (m_{W_{\pm}}^{2} / m_{W_{0}}^{2}))$ acting
on both types of fermions.
\vspace{12pt}

If $(1 - (m_{W_{\pm}}2 / m_{W_{0}}^{2}))$ is defined to be
$\sin{\theta_{w}}^{2}$ and denoted by $\xi$, and
\vspace{12pt}

if the strength of the $W_{\pm}$ charged weak force
(and of the custodial $SU(2)$ symmetry) is denoted by $T$,
\vspace{12pt}

then the $W_{0}$ neutral weak interaction can be written as:
\vspace{12pt}

$W_{0L} \sim T + \xi$ and $W_{0LR} \sim \xi$.
\vspace{12pt}

Since the $W_{0}$ acts as $W_{0L}$ with respect to the
parity violating $SU(2)$ weak force and
\vspace{12pt}

as $W_{0LR}$ with respect to the parity conserving $U(1)$
electromagnetic force of the $U(1)$ subgroup of $SU(2)$,
\vspace{12pt}

the $W_{0}$ mass $m_{W_{0}}$ has two components:
\vspace{12pt}

the parity violating $SU(2)$ part $m_{W_{0L}}$ that is
equal to $m_{W_{\pm}}$ ; and
\vspace{12pt}

the parity conserving part $m_{W_{0LR}}$ that acts like a
heavy photon.
\vspace{12pt}

As $m_{W_{0}}$ = 98.379 GeV = $m_{W_{0L}} + m_{W_{0LR}}$, and
\vspace{12pt}

as $m_{W_{0L}} = m_{W_{\pm}} = 80.326 \; GeV$,
\vspace{12pt}

we have $m_{W_{0LR}} = 18.053 \; GeV$.

\vspace{12pt}

Denote by $\tilde{\alpha_{E}} = \tilde{e}^{2}$ the force
strength of the weak parity conserving $U(1)$
electromagnetic type force that acts through the
$U(1)$ subgroup of $SU(2)$.

\vspace{12pt}

The electromagnetic force strength
$\alpha_{E} = e^{2} = 1 / 137.03608$ was calculated
in Chapter 8 using
the volume $V(S^{1})$ of an $S^{1} \subset R^{2}$,
normalized by $1 / \sqrt{2}$.

\vspace{12pt}

The $\tilde{\alpha_{E}}$ force is part of the $SU(2)$ weak
force whose strength $\alpha_{w} = w^{2}$ was calculated
in Chapter 8 using the volume $V(S^{2})$ of an
$S^{2} \subset  R^{3}$,
normalized by $1  / \sqrt{3}$.

\vspace{12pt}

Also, the electromagnetic force strength $\alpha_{E} = e^{2}$
was calculated in Chapter 8 using a
4-dimensional spacetime with global structure of
the 4-torus $T^{4}$ made up of four $S^{1}$ 1-spheres,

\vspace{12pt}

while the $SU(2)$ weak force strength
$\alpha_{w} = w^{2}$ was calculated in Chapter 8
using two 2-spheres $S^{2} \times S^{2}$,
each of which contains one 1-sphere of
the $\tilde{\alpha_{E}}$ force.

\vspace{12pt}

Therefore
$\tilde{\alpha_{E}} = \alpha_{E} (\sqrt{2} /
\sqrt{3})(2 / 4) = \alpha_{E} / \sqrt{6}$,
\vspace{12pt}

 $\tilde{e}  = e / \sqrt[4]{6} = e / 1.565$ , and

\vspace{12pt}

the mass $m_{W_{0LR}}$ must be reduced to an effective value

\vspace{12pt}

$m_{W_{0LR}eff} = m_{W_{0LR}} / 1.565$
= 18.053/1.565 = 11.536 GeV

\vspace{12pt}

for the $\tilde{\alpha_{E}}$ force to act like
an electromagnetic force in the 4-dimensional
spacetime HyperDiamond Feynman Checkerboard model:

\vspace{12pt}

$\tilde{e} m_{W_{0LR}} = e (1/5.65) m_{W_{0LR}} = e m_{Z_{0}}$,

\vspace{12pt}

where the physical effective neutral weak boson is
denoted by $Z_{0}$.

\vspace{12pt}

Therefore, the correct HyperDiamond Feynman
Checkerboard values for
weak boson masses and the Weinberg angle $\theta_{w}$ are:

\vspace{12pt}

$m_{W_{+}} = m_{W_{-}} = 80.326 \; GeV$;

\vspace{12pt}

$m_{Z_{0}} = 80.326 +11.536 = 91.862 \; GeV$; and

\vspace{12pt}

$\sin{\theta_{w}}^{2} = 1 - (m_{W_{\pm}} /
m_{Z_{0}})^{2} = 1 - ( 6452.2663 / 8438.6270 ) = 0.235$.

\vspace{12pt}

Radiative corrections are not taken into account here,
and may change these tree-level
HyperDiamond Feynman Checkerboard values somewhat.

\vspace{12pt}

\subsection{Fermion Masses.}

First generation fermion particles are represented
by octonions as follows:
\vspace{12pt}

 \begin{equation}
\begin{array}{|c|c|} \hline
Octonion  & First \: Generaton \\ \hline
basis \: element & Fermion \: Particle\\ \hline
1 & e-neutrino   \\ \hline
i & red \: up \: quark \\ \hline
j & green \: up \: quark \\ \hline
k & blue \: up \: quark  \\ \hline
E & electron \\ \hline
I & red \: down \: quark  \\ \hline
J & green \: down \: quark  \\ \hline
K & blue \: down \: quark  \\ \hline
\end{array}
\end{equation}

\vspace{12pt}

First generation fermion antiparticles are represented
by octonions in a similiar way.
\vspace{12pt}

Second generation fermion particles and antiparticles
are represented by pairs of octonions.
\vspace{12pt}

Third generation fermion particles and antiparticles
are represented by triples of octonions.
\vspace{12pt}

In the HyperDiamond Feynman Checkerboard model,
there are no higher generations of fermions than the Third.
\vspace{12pt}

This can be seen algebraically as a consequence of the
fact that the Lie algebra series $E_{6}$, $E_{7}$, and $E_{8}$,
has only 3 algebras, which in turn is a consequence of
non-associativity of octonions, as described in
\vspace{12pt}

http://galaxy.cau.edu/tsmith/E678.html
\vspace{12pt}

http://www.innerx.net/personal/tsmith/E678.html
\vspace{12pt}

or geometrically as a consequence of the fact that,
\vspace{12pt}

if you reduce the original 8-dimensional spacetime
\newline

into associative 4-dimensional physical spacetime
\newline

and coassociative 4-dimensional Internal Symmetry Space,
\vspace{12pt}

then, if you look in the original 8-dimensional spacetime
at a fermion (First-generation represented by a single octonion)
propagating from one vertex to another,
there are only 4 possibilities for the same propagation
after dimensional reduction:
\vspace{12pt}

1 - the origin and target vertices are both
in the associative 4-dimensional physical spacetime,
in which case the propagation is unchanged, and the
fermion remains a FIRST generation fermion represented
by a single octonion;
\vspace{12pt}

2 - the origin vertex is in the associative spacetime,
and the target vertex in in the Internal Symmetry Space,
in which case there must be a new link from
the original target vertex in the Internal Symmetry Space
to a new target vertex in the associative spacetime,
and a second octonion can be introduced at the original
target vertex in connection with the new link,
so that the fermion can be regarded after dimensional reduction
as a pair of octonions, and therefore as
a SECOND generation fermion;
\vspace{12pt}

3 - the target vertex is in the associative spacetime,
and the origin vertex in in the Internal Symmetry Space,
in which case there must be a new link to
the original origin vertex in the Internal Symmetry Space
from a new origin vertex in the associative spacetime,
so that a second octonion can be introduced at the original
origin vertex in connection with the new link,
so that the fermion can be regarded after dimensional reduction
as a pair of octonions, and
therefore as a SECOND generation fermion;
and
\vspace{12pt}

4 - both the origin vertex and the target vertex
are in the Internal Symmetry Space,
in which case there must be a new link to
the original origin vertex in the Internal Symmetry Space
from a new origin vertex in the associative spacetime,
and a second new link from the original target vertex
in the Internal Symmetry Space to a new target vertex
in the associative spacetime,
so that a second octonion can be introduced at the original
origin vertex in connection with the first new link,
and a third octonion can be introduced at the original
target vertex in connection with the second new link,
so that the fermion can be regarded after dimensional reduction
as a triple of octonions,
and therefore as a THIRD generation fermion.
\vspace{12pt}

As there are no more possibilities,
there are no more generations.
\vspace{12pt}

\newpage

\subsubsection{Renormalization.}

Particle masses and force strength constants
are not really "constant" when you measure them,
as the result of your measurement will depend on
the energy at which you measure them.
Measurements at one energy level can be related to
measurements at another by renormalization equations.
\vspace{12pt}

The particle masses calculated in the $D_{4}-D_{5}-E_{6}$ model
are, with respect to renormalization, each defined at
the energy level of the calculated particle mass.
\vspace{12pt}

For leptons, such as the electron, muon, and tauon,
which carry no color charge,
you can renormalize conventionally from that energy level
to "translate" the result to another energy level,
because those particles are not "confined" and
so can be experimentally observed as "free particles"
("free" means "not strongly bound to other particles,
except for virtual particles of the active
vacuum of spacetime").
\vspace{12pt}

For quarks, which are confined and
cannot be experimentally observed as free particles,
the situation is more complicated.
In the $D_{4}-D_{5}-E_{6}$ model, the calculated quark masses
are considered to be constituent masses.
\vspace{12pt}

In hep-ph/9802425,
Di Qing, Xiang-Song Chen, and Fan Wang,
of Nanjing University, present a qualitative QCD analysis
and a quantitative model calculation
to show that the constituent quark model
[after mixing a small amount (15
remains a good approximation
even taking into account the nucleon spin structure
revealed in polarized deep inelastic scattering.
\vspace{12pt}

The effectiveness of
the NonRelativistic model of light-quark hadrons
is explained by, and affords experimental Support for,
the Quantum Theory of David Bohm (see quant-ph/9806023).
\vspace{12pt}

Consitituent particles are Pre-Quantum particles in the
\newline

sense that their properties are calculated without
\newline

using sum-over-histories Many-Worlds quantum theory.
\newline

("Classical" is a commonly-used synonym for "Pre-Quantum".)
\vspace{12pt}

Since experiments are quantum sum-over-histories processes,
experimentally observed particles are Quantum particles.
\vspace{12pt}

The lightest experimentally observable particle
containing quarks is the pion,
which is a quark-antiquark pair made up of the lightest
quarks, the up and down quarks.
\vspace{12pt}

A quark-antiquark pair is the carrier of the strong force,
and mathematically resembles a bivector gluon,
which is the carrier of the color force.
\vspace{12pt}

The charactereistic energy level of pions is
the square root of the sum of the squares of
the masses of the two charged and one neutral pion.
It is about 245 MeV (to more accuracy 241.4 MeV).
\vspace{12pt}

The gluon-carried color force strength is renormalized
to higher energies from about 245 MeV in the conventional way.
\vspace{12pt}

What about quarks, as opposed to gluons?
\vspace{12pt}

Gluons are represented by quark-antiquark pairs,
but a quark is a single quark.
\vspace{12pt}

The lightest particle containing a quark that is not
coupled to an antiquark is the proton,
which is a stable (except with respect to quantum gravity)
3-quark color neutral particle.
\vspace{12pt}

The characteristic energy level of the proton
is about 1 GeV (to more accuracy 938.27 MeV).
\vspace{12pt}

Quark masses are renormalized to higher energies
from about 1 GeV (or from their calculated mass,
below which they do not exist except virtually)
in the conventional way.
\vspace{12pt}

What about the 3 quarks (up, down, and strange)
that have constituent masses less than 1 GeV?
\vspace{12pt}

Below 1 GeV, they can only exist
(if not bound to an antiquark) within a proton,
so their masses are "flat", or do not "run",
in the energy range below 1 GeV.
\vspace{12pt}

Since the 3 quarks, up, down, and strange,
are the only ones lighter than a proton,
they can be used as the basis for a useful
low-energy theory, Chiral Perturbation Theory,
that uses the group $SU(3) \times SU(3)$,
or, if based only on the lighter up and down quarks
that uses the group $SU(2) \times SU(2)$.
\vspace{12pt}

A useful theory at high energies,
much above 1 GeV, is Perturbative QCD,
that treats the quarks and gluons as free,
which they are asymptotically
as energies become very high.
\vspace{12pt}

\subsubsection{Perturbative QCD, Chiral PT, Lattice GT.}

To do calculations in theories such as
Perturbative QCD and Chiral Perturbation Theory,
you need to use effective quark masses that
are called current masses.  Current quark masses are different
from the Pre-Quantum constituent quark masses of our model.
\vspace{12pt}

The current mass of a quark is defined in our model as
the difference between
the constituent mass of the quark
and
the density of the lowest-energy sea of virtual gluons,
quarks, and antiquarks, or 312.75 MeV.
\vspace{12pt}

Since the virtual sea is a quantum phenomenon,
the current quarks
of Perturbative QCD and Chiral Perturbation Theory are,
in my view, Quantum particles.
\vspace{12pt}

The relation between current masses and constituent masses
may be explained, at least in part,
by the Quantum Theory of David Bohm.
\vspace{12pt}

Therefore, our model is unconventional in that:
\vspace{12pt}

a current quark is viewed as a composite combination
of a fundamental constituent quark
and Quantum virtual sea gluon, quarks, and antiquarks
(compare the conventional picture of, for example,
hep-ph/9708262, in which current quarks are Pre-Quantum
and constituent quarks are Quantum composites); and
\vspace{12pt}

the input current quarks of
Perturbative QCD and Chiral Perturbation Theory
are Quantum, and not Pre-Quantum, so that we view
Perturbative QCD and Chiral Perturbation Theory as
effectively "second-order" Quantum theories
(rather than fundamental theories)
that are most useful in describing phenomena
at high and low energy levels, respectively.
\vspace{12pt}

Therefore,
in Perturbative QCD and Chiral Perturbation Theory,
the up and down quarks roughly massless.
One result is that the current masses
can then be used as input for
the $SU(3) \times SU(3)$ Chiral Perturbation Theory
that, although it is only approximate because the
constituent mass of the strange quark is about 312 MeV,
rather than nearly zero,
can be useful in calculating meson properties.
\vspace{12pt}

WHAT ARE THE REGIONS OF VALIDITY OF PERTURBATIVE QCD
and CHIRAL PERTURBATION THEORY?
\vspace{12pt}

Perturbative QCD is useful at high energies.
If Perturbative QCD is valid at energies above 4.5 GeV,
then Yndurian in hep-ph/9708300 has shown that
lower bounds for current quark masses are:
\vspace{12pt}

$m_{s}$       at least 150 MeV
(compare $D_{4}-D_{5}-E_{6}$ 312   MeV)
\vspace{12pt}

$m_{u} + m_{d}$    at least  10 MeV
(compare $D_{4}-D_{5}-E_{6}$   6.5 MeV)
\vspace{12pt}

$m_{u} - m_{d}$    at least   2 MeV
(compare $D_{4}-D_{5}-E_{6}$   2.1 MeV)
\vspace{12pt}

Yndurian bases his estimates on positivity,
and uses $\bar{MS}$ masses defined at $1 GeV^2$.
\vspace{12pt}

Chiral Perturbation Theory is useful at low energies.
Lellouch, Rafael, and Taron in hep-ph/9707523 have shown
roughly similar lower bounds using Chiral Perturbation Theory.
\vspace{12pt}

Lattice Gauge Theory calculations of Gough et al
in hep-ph/9610223
give light quark current $\bar{MS}$ masses at $1 GeV^2$ as:
\vspace{12pt}

$m_{s}$       59    to  101   MeV
\vspace{12pt}

$m_{u} + m_{d}$     4.6  to    7.8 MeV
\vspace{12pt}

Clearly,
the strange quark current masses of Lattice Gauge Theory
are a lot lighter than those calculated by
Perturbative QCD and Chiral Perturbation Theory,
as well as the $D_{4}-D_{5}-E_{6}$l.
\vspace{12pt}

Further, recent observation of the decay of K+ to
pion+ and a muon-antimuon pair gives a branching ratio
with respect to the decay of K+ to pion+ and e+ e-
that is 0.167, which is about 2 sigma below the
Chiral Perturbation Theory prediction of 0.236.
\vspace{12pt}

These facts indicate that Perturbative QCD,
Chiral Perturbation Theory, and Lattice Gauge Theory
are approximations to fundamental theory,
each useful in some energy regions,
but not fully understood in all energy regions.
\vspace{12pt}

Conventional Lattice Gauge Theory for fermions,
such as quarks, has some fundamental problems:
\vspace{12pt}

The conventional lattice Dirac operator is afflicted
with the Fermion Doubling Problem, in which
nearest-neighbor lattice sites are occupied
with (for 4-dim spacetime) $2^4 = 16$ times too many
fermions;.
\vspace{12pt}

The conventional solutions to the Fermion Doubling
Problem are to add non-local terms that violate
Chiral Symmetry on the lattice.
If you are trying to do Chiral Perturbation Theory
on the lattice, that
seems to be a bad idea.
\vspace{12pt}

To solve the Fermion Doubling Problem without
violating Chiral Symmetry, Bo Feng, Jianming Li,
and Xingchang Song have proposed to modify the
conventional lattice Dirac operator by adding a
non-local term that (like an earlier approach
of Drell et. al., Phys. Rev. D 14 (1976) 1627) couples
all lattice sites along a given direction instead
of coupling only nearest-neighbor sites. Their modified
lattice Dirac operator not only preserved Chiral Symmetry,
it also gives the conventional
D'Alembertian operator, and they are able to construct
the Weinberg-Salam Electro-Weak model on
a lattice.
\vspace{12pt}

Conventional Lattice Gauge Theory is formulated
somewhat differently from  another approach
to formulating physics models on lattices:
Feynman Checkerboards,
\vspace{12pt}

\newpage

\subsubsection{Truth Quark Mass.}

In the HyperDiamond Feynman Checkerboard model,
the Higgs scalar couples directly with
the particle-antiparticle pair
made of femions with the most charge in the highest generation.
\vspace{12pt}

That means Third Generation fermions
made of triples of octonions,
\vspace{12pt}

carrying both color and electric charge, and therefore quarks
rather than leptons,
\vspace{12pt}

and carrying electric charge of magnitude
$2/3$ rather than $1/3$,
and therefore:
\vspace{12pt}

the Higgs scalar couples most strongly with the Truth quark,
whose tree-level constituent mass of 129.5155 GeV is
somewhat lower than, but close to,
the Higgs scalar mass of about 146 GeV.
\vspace{12pt}

The HyperDiamond Feynman Checkerboard model value of
about 130 GeV is substantially different from
the roughly 175 GeV figure advocated by FermiLab.
\vspace{12pt}

I think that the FermiLab figure is incorrect.
\vspace{12pt}

The Fermilab figure is based on analysis of
semileptonic events.  We think that the Fermilab
semileptonic analysis does not handle background correctly,
and ignores signals in the data that are in rough
agreement with our tree level constituent mass
of about 130 GeV.
\vspace{12pt}

Further, I think that dileptonic events are more
reliable for Truth quark mass determination,
even though there are fewer of them than semileptonic events.
\vspace{12pt}

I disagree with the Fermilab D0 analysis of dileptonic events,
which Fermilab says are in the range of 168.3 GeV.
\vspace{12pt}

My analysis of those dileptonic events
gives a Truth quark mass of about 136.7 GeV,
in rough agreement with
the D4-D5-E6 model tree level Truth quark constituent mass
of about 130 GeV.
\vspace{12pt}

More details about these issues,
including gif images of Fermilab data histograms
and other relevant experimental results,
can be found on the World Wide Web at URLs
\vspace{12pt}

http://galaxy.cau.edu/tsmith/TCZ.html
\vspace{12pt}

http://www.innerx.net/personal/tsmith/TCZ.html
\vspace{12pt}

I consider the mass of the Truth quark to be
a good test of our theory, as our theory can
be falsified if we turn out to be wrong in our
interpretation of experimental results.

\subsubsection{Other Fermion Masses.}

In the HyperDiamond Feynman Checkerboard model,
the masses of the other fermions are calculated from
the mass of the Truth quark, with the following results for
individual tree-level lepton masses and quark constituent masses:

$m_{e}$ = 0.5110 MeV;
\newline
$m_{\nu_{e}}$ = $m_{\nu_{\mu}}$ = $m_{\nu_{\tau}}$ = 0;
\newline
$m_{d}$ = $m_{u}$ = 312.8 MeV  (constituent quark mass);
\newline
$m_{\mu}$ = 104.8 MeV;
\newline
$m_{s}$ =  625 MeV  (constituent quark mass);
\newline
$m_{c}$ =  2.09 GeV  (constituent quark mass);
\newline
$m_{\tau}$ =  1.88 GeV;
\newline
$m_{b}$ =  5.63 GeV  (constituent quark mass);

These results when added up give a total mass of
\newline
first generation fermions:
\begin{equation}
\Sigma_{f_{3}} = 7.508 \; GeV
\end{equation}
\vspace{12pt}

\vspace{12pt}

Here is how the individual fermion mass calculations
are done:

\vspace{12pt}

The Weyl fermion neutrino has at tree level
only the left-handed state,
whereas the Dirac fermion electron and quarks can have
both left-handed and right-handed states,
so that the total number of states corresponding
to each of the half-spinor $Spin(0,8)$ representations is 15.

\vspace{12pt}

In all generations, neutrinos are massless at tree level.
However, even though massless at tree level,
neutrinos are spinors and therefore are acted
upon by Gravity as shown by the Papapetrou Equations.
\vspace{12pt}

Further,
in Quantum Field Theory at Finite Temperature,
the gravitational equivalence principle may be violated,
causing mixing among neutrinos of different generations.

\vspace{12pt}

In the HyperDiamond Feynman Checkerboard model,
the first generation
fermions correspond to octonions ${\bf O}$, while second
generation fermions correspond to pairs of octonions
${\bf O} \times {\bf O}$ and third generation fermions
correspond to triples of octonions
${\bf O} \times {\bf O} \times {\bf O}$.

\vspace{12pt}

To calculate the fermion masses in the model,
the volume of a compact manifold representing the
spinor fermions $S_{8+}$ is used.
It is the parallelizable manifold $S^7\times RP^1$.

\vspace{12pt}

 Also, since gravitation is coupled to mass,
the infinitesimal generators of the MacDowell-Mansouri
gravitation group, $Spin(0,5)$, are relevant.

\vspace{12pt}

The calculated quark masses are constituent masses, not
current masses.

\vspace{12pt}

In the HyperDiamond Feynman Checkerboard model,
fermion masses are
calculated as a product of four factors:
\begin{equation}
V(Q_{fermion}) \times N(Graviton) \times N(octonion) \times Sym
\end{equation}

$V(Q_{fermion})$ is the volume of the part of the half-spinor
\newline
fermion particle manifold $S^7 \times RP^1$ that is
\newline
related to the fermion particle by photon,
weak boson, and gluon interactions.

\vspace{12pt}

$N(Graviton)$ is the number of types of
$Spin(0,5)$ graviton related to
the fermion.  The 10 gravitons correspond to the 10 infinitesimal
generators of $Spin(0,5)$ = $Sp(2)$.
\newline
2 of them are in the Cartan subalgebra.
\newline
6 of them carry color charge, and may therefore be considered as
corresponding to quarks.
\newline
The remaining 2 carry no color charge, but
may carry electric charge and so may be
considered as corresponding
to electrons.
\newline
One graviton takes the electron into itself,
and the other can only take the first-generation electron
into the massless electron
neutrino.
\newline
Therefore only one graviton should correspond to the mass
of the first-generation electron.
\newline
The graviton number ratio of the down quark to the
first-generation electron is therefore 6/1 = 6.

\vspace{12pt}

$N(octonion)$ is an octonion number factor relating up-type quark
masses to down-type quark masses in each generation.

\vspace{12pt}

$Sym$ is an internal symmetry factor, relating  2nd and 3rd
generation massive leptons to first generation fermions.
\newline
It is not used in first-generation calculations.

\vspace{12pt}

The ratio of the down quark constituent mass to the electron mass
is then calculated as follows:
\newline
Consider the electron, e.
\newline
By photon, weak boson, and gluon interactions,
e can only be taken into 1, the massless neutrino.
\newline
The electron and neutrino, or their antiparticles,
cannot be combined to produce any of the
massive up or down quarks.
\newline
The neutrino, being massless at tree level,
does not add anything to the mass formula for the electron.
\newline
Since the electron cannot be related to any other massive Dirac
fermion, its volume $V(Q_{electron})$ is taken to be 1.

\vspace{12pt}

Next consider a red down quark $I$.
\vspace{12pt}

By gluon interactions, $I$ can be taken into
$J$ and $K$,
the blue and green down quarks.
\vspace{12pt}

By also using weak boson interactions, it can be taken into
$i$, $j$, and $k$, the red, blue, and green up quarks.
\vspace{12pt}

Given the up and down quarks, pions can be formed
from quark-antiquark pairs, and the pions can decay to produce
electrons and neutrinos.
\vspace{12pt}

Therefore the red down quark (similarly, any down quark)
is related to any part of $S^7\times {\bf R}P^1$,
the compact manifold corresponding to
\vspace{12pt}

$$\{ 1, i, j, k, I, J, K, E \}$$
\vspace{12pt}

and therefore a down quark should have a spinor manifold
volume factor $V(Q_{down quark}$ of the volume of
$S^7\times {\bf R}P^1$.
\vspace{12pt}

The ratio of the down quark spinor manifold volume factor to
the electron spinor manifold volume factor is just
\vspace{12pt}

\begin{equation}
V(Q_{down quark}) / V(Q_{electron}) =
V(S^7\times {\bf R}P^1)/1 = \pi ^{5} / 3.
\end{equation}

\vspace{12pt}

Since the first generation graviton factor is 6,
\vspace{12pt}

\begin{equation}
md/me = 6V(S^7 \times {\bf R}P^1) = 2 {\pi}^5 = 612.03937
\end{equation}

\vspace{12pt}

As the up quarks correspond to $i$, $j$, and $k$,
which are the octonion transforms under
$E$ of $I$, $J$, and $K$ of
the down quarks, the up quarks and down quarks
have the same constituent mass $m_{u} = m_{d}$.

\vspace{12pt}

Antiparticles have the same mass as the corresponding
particles.

\vspace{12pt}

Since the model only gives ratios of massses,
the mass scale is
fixed by assuming that the electron mass $m_{e}$ = 0.5110 MeV.

\vspace{12pt}

Then, the constituent mass of the down quark is
$m_{d}$ = 312.75 MeV, and
\newline
the constituent mass for the up quark is
$m_{u}$ = 312.75 MeV.

\vspace{12pt}

As the proton mass is taken to be the sum of the constituent
masses of its constituent quarks
\begin{equation}
m_{proton} = m_{u} + m_{u} + m_{d} = 938.25 \; MeV
\end{equation}
The $D_{4}-D_{5}-E_{6}$ model calculation is close to
the experimental value of 938.27 MeV.

\vspace{12pt}

The third generation fermion particles correspond to triples of
octonions.  There are $8^3$ = 512 such triples.

\vspace{12pt}

The triple $\{ 1,1,1 \}$ corresponds to the tau-neutrino.

\vspace{12pt}

The other 7 triples involving only $1$ and $E$ correspond
to the tauon:
$$\{ E, E, E \},
\{ E, E, 1 \},
\{ E, 1, E \},
\{ 1, E, E \},
\{ 1, 1, E \},
\{ 1, E, 1 \},
\{ E, 1, 1 \} $$,

The symmetry of the 7 tauon triples is the same as the
symmetry of the 3 down quarks, the 3 up quarks, and the electron,
so the tauon mass should be the same as the sum of the masses of
the first generation massive fermion particles.

\vspace{12pt}

Therefore the tauon mass 1.87704 GeV.

\vspace{12pt}

The calculated Tauon mass of 1.88 GeV is a sum
of first generation fermion masses, all of which are
valid at the energy level of about 1 GeV.
\vspace{12pt}

However, as the Tauon mass is about 2 GeV,
the effective Tauon mass should be renormalized
from the energy level of 1 GeV (where the mass is 1.88 GeV)
to the energy level of 2 GeV.
\vspace{12pt}

Such a renormalization should reduce the mass.
If the renormalization reduction were about 5 percent,
the effective Tauon mass at 2 GeV would be about 1.78 GeV.
\vspace{12pt}

The 1996 Particle Data Group Review of Particle Physics gives
a Tauon mass of 1.777 GeV.
\vspace{12pt}

\vspace{12pt}

Note that all triples corresponding to the
tau and the tau-neutrino are colorless.

\vspace{12pt}

The beauty quark corresponds to 21 triples.
\newline
They are triples of the same form as the 7 tauon triples,
but for $1$ and $I$,  $1$ and $J$, and  $1$ and$ K$,
which correspond to the red, green, and blue beauty quarks,
respectively.

\vspace{12pt}

The seven triples of the red beauty quark correspond
to the seven triples of the tauon,
except that the beauty quark interacts with 6 $Spin(0,5)$
gravitons while the tauon interacts with only two.

\vspace{12pt}

The beauty quark constituent mass should be
the tauon mass times the
third generation graviton factor 6/2 = 3, so the B-quark mass is
\newline
$m_{b}$ = 5.63111 GeV.

\vspace{12pt}

The calculated Beauty Quark mass of 5.63 GeV is a
consitituent mass, that is,
it corresponds to the conventional pole mass plus 312.8 MeV.
\vspace{12pt}

Therefore, the calculated Beauty Quark mass of 5.63 GeV
corresponds to a conventional pole mass of 5.32 GeV.
\vspace{12pt}

The 1996 Particle Data Group Review of Particle Physics gives
a lattice gauge theory Beauty Quark pole mass as 5.0 GeV.
\vspace{12pt}

The pole mass can be converted to an MSbar mass
if the color force strength constant $alpha_{s}$ is known.
The conventional value of $alpha_{s}$ at about 5 GeV is about 0.22.
Using $alpha_{s(5GeV)} = 0.22$,
a pole mass of 5.0 GeV gives an MSbar 1-loop mass of 4.6 GeV,
and an MSbar 1,2-loop mass of 4.3, evaluated at about 5 GeV.
\vspace{12pt}

If the MSbar mass is run from 5 GeV up to 90 GeV,
the MSbar mass decreases by about 1.3 GeV,
giving an expected MSbar mass of about 3.0 GeV at 90 GeV.
DELPHI at LEP has observed the Beauty Quark
and found a 90 GeV MSbar mass of about 2.67 GeV,
with error bars +/- 0.25 (stat) +/- 0.34 (frag) +/- 0.27 (theo).
\vspace{12pt}

Note that the D4-D5-E6 model calculated mass of 5.63 GeV
corresponds to a pole mass of 5.32 GeV,
which is somewhat higher than the conventional value of 5.0 GeV.
\vspace{12pt}

However,
the D4-D5-E6 model calculated value of
the color force strength constant $alpha_{s}$ at
about 5 GeV is about 0.166,
\vspace{12pt}

while
the conventional value of
the color force strength constant $alpha_{s}$ at
about 5 GeV is about 0.216,
\vspace{12pt}

and
the D4-D5-E6 model calculated value of
the color force strength constant $alpha_{s}$ at
about 90 GeV is about 0.106,
\vspace{12pt}

while
the conventional value of
the color force strength constant $alpha_{s}$ at
about 90 GeV is about 0.118.
\vspace{12pt}

The D4-D5-E6 model calculations gives
a Beauty Quark pole mass (5.3 GeV)
that is about 6 percent higher
than the conventional Beauty Quark pole mass (5.0 GeV),
\vspace{12pt}

and
a color force strength $alpha_{s}$ at 5 GeV (0.166) such that
1 + $alpha_{s}$ = 1.166 is about 4 percent lower
than the conventional value of 1 + $alpha_{s}$ = 1.216 at 5 GeV.

\vspace{12pt}

Note particularly that triples of the type $\{ 1, I, J \}$,
$\{ I, J, K \}$, etc.,
do not correspond to the beauty quark, but to the truth quark.

\vspace{12pt}

The truth quark corresponds to the remaining 483 triples, so the
constituent mass of the red truth quark is 161/7 = 23 times the
red beauty quark mass, and the red T-quark mass is
\begin{equation}
m_{t} = 129.5155 \; GeV
\end{equation}

The blue and green truth quarks are defined similarly.

\vspace{12pt}

The tree level T-quark constituent mass
gives a Truth quark-antiquark
mass of 259.031 GeV.

\vspace{12pt}

The tree level T-quark constituent mass rounds off to 130 GeV.

\vspace{12pt}

These results when added up give a total mass of
\newline
third generation fermions:
\begin{equation}
\Sigma_{f_{3}} = 1,629 \; GeV
\end{equation}
\vspace{12pt}

\newpage

The second generation fermion calculations are:

The second generation fermion particles correspond
to pairs of octonions.
\newline
There are 82 = 64 such pairs.
\newline
The pair $\{ 1,1 \}$ corresponds to the $\mu$-neutrino.
\newline
the pairs $\{ 1, E \}$, $\{ E, 1 \}$, and
$\{ E, E \}$ correspond to the muon.
\newline
Compare the symmetries of the muon pairs to the symmetries
of the first generation fermion particles.
\newline
The pair $\{ E, E \}$ should correspond
to the $E$ electron.
\newline
The other two muon pairs have a symmetry group S2,
which is 1/3 the size of the color symmetry group S3
which gives the up and down quarks their mass of 312.75 MeV.

\vspace{12pt}

Therefore the mass of the muon should be the sum of
the $\{ E, E \}$ electron mass and
the $\{ 1, E \}$, $\{ E, 1 \}$ symmetry mass,
which is 1/3 of the up or down quark mass.

\vspace{12pt}

Therefore,  $m_{\mu}$ = 104.76 MeV.
\newline
\vspace{12pt}

Note that all pairs corresponding to
the muon and the $\mu$-neutrino are colorless.

\vspace{12pt}

The red, blue and green strange quark each corresponds
to the 3 pairs involving $1$ and $I$, $J$, or $K$.

\vspace{12pt}

The red strange quark is defined as the three pairs
$1$ and $I$, because $I$ is the red down quark.
\newline
Its mass should be the sum of two parts:
the $\{ I, I \}$ red down quark mass, 312.75 MeV, and
the product of the symmetry part of the muon mass, 104.25 MeV,
times the graviton factor.

\vspace{12pt}

Unlike the first generation situation,
massive second and third generation leptons can be taken,
by both of the colorless gravitons that
may carry electric charge, into massive particles.

\vspace{12pt}

Therefore the graviton factor for the second and
\newline
third generations is 6/2 = 3.

\vspace{12pt}

Therefore the symmetry part of the muon mass times
the graviton factor 3 is 312.75 MeV, and
the red strange quark constituent mass is
$$m_{s} = 312.75 \; MeV + 312.75 \; MeV = 625.5 \; MeV$$

The blue strange quarks correspond to the
three pairs involving $J$,
the green strange quarks correspond to the
three pairs involving $K$,
and their masses are determined similarly.

\vspace{12pt}

The charm quark corresponds to the other 51 pairs.
Therefore, the mass of the red charm quark should
be the sum of two parts:

\vspace{12pt}

the $\{ i, i \}$, red up quark mass, 312.75 MeV; and

\vspace{12pt}

the product of the symmetry part of the strange quark
mass, 312.75 MeV, and

\vspace{12pt}

the charm to strange octonion number factor 51/9,
which product is 1,772.25 MeV.

\vspace{12pt}

Therefore the red charm quark constituent mass is
$$m_{c} = 312.75 \; MeV + 1,772.25 \; MeV = 2.085 \; GeV$$

The blue and green charm quarks are defined similarly,
and their masses are calculated similarly.
\vspace{12pt}

The calculated Charm Quark mass of 2.09 GeV is a
constituent mass, that is,
it corresponds to the conventional pole mass plus 312.8 MeV.
\vspace{12pt}

Therefore, the calculated Charm Quark mass of 2.09 GeV
corresponds to a conventional pole mass of 1.78 GeV.
\vspace{12pt}

The 1996 Particle Data Group Review of Particle Physics gives
a range for the Charm Quark pole mass from 1.2 to 1.9 GeV.
\vspace{12pt}

The pole mass can be converted to an MSbar mass
if the color force strength constant $alpha_{s}$ is known.
\vspace{12pt}

The conventional value of $alpha_{s}$ at about 2 GeV is about 0.39,
which is somewhat lower than the D4-D5-E6 model value.
\vspace{12pt}

Using $alpha_{s(2GeV)} = 0.39$,
a pole mass of 1.9 GeV gives an MSbar 1-loop mass of 1.6 GeV,
evaluated at about 2 GeV.
\vspace{12pt}

These results when added up give a total mass of
second generation fermions:
\begin{equation}
\Sigma_{f_{2}} = 32.9 \; GeV
\end{equation}

\vspace{12pt}

\newpage

\subsection{K-M Parameters.}

The following formulas use the above masses to
calculate Kobayashi-Maskawa parameters:

\begin{equation}
phase \; angle \; \epsilon = \pi / 2
\end{equation}

\begin{equation}
\sin{\alpha} = [m_{e}+3m_{d}+3m_{u}] /
\sqrt{ [m_{e}^{2}+3m_{d}^{2}+3m_{u}^{2}] +
[m_{\mu}^{2}+3m_{s}^{2}+3m_{c}^{2}] }
\end{equation}

\begin{equation}
\sin{\beta} = [m_{e}+3m_{d}+3m_{u}] /
\sqrt{ [m_{e}^{2}+3m_{d}^{2}+3m_{u}^{2}] +
[m_{\tau}^{2}+3m_{b}^{2}+3m_{t}^{2}] }
\end{equation}

\begin{equation}
\sin{\tilde{\gamma}} = [m_{\mu}+3m_{s}+3m_{c}] /
\sqrt{ [m_{\tau}^{2}+3m_{b}^{2}+3m_{t}^{2}] +
[m_{\mu}^{2}+3m_{s}^{2}+3m_{c}^{2}] }
\end{equation}

\begin{equation}
\sin{\gamma} = \sin{\tilde{\gamma}}
\sqrt{\Sigma_{f_{2}} / \Sigma_{f_{1}}}
\end{equation}

\vspace{12pt}

The resulting Kobayashi-Maskawa parameters are:

\begin{equation}
\begin{array}{|c|c|c|c|}
\hline
& d & s & b
  \\
\hline
u & 0.975 & 0.222 & -0.00461 i  \\
c & -0.222 -0.000191 i & 0.974 -0.0000434 i & 0.0423  \\
t & 0.00941 -0.00449 i & -0.0413 -0.00102 i & 0.999  \\
\hline
\end{array}
\end{equation}

For Z0 neutral weak boson processes, which are suppressed
by the GIM mechanism of cancellation of virtual subprocesses,
the matrix can be labelled either by
\vspace{12pt}

(u c t) input  and  (u'c't')  output,
\vspace{12pt}

or by
\vspace{12pt}

(d s b) input  and  (d's'b')  output:
\vspace{12pt}

Since neutrinos of all three generations are massless,
the lepton sector has no K-M mixing.
\vspace{12pt}

\newpage

\section{Protons, Pions, and Physical Gravitons.}

In his 1994 Georgia Tech Ph. D. thesis under David Finkelstein,
{\it{Spacetime as a Quantum Graph}}, Michael Gibbs \cite{GIB}
describes some 4-dimensional HyperDiamond lattice structures,
that he considers likely candidates to represent physical
pcrticles.
\vspace{12pt}

The terminology used by Michael Gibbs in his thesis \cite{GIB}
is useful with respect to the model he constructs.  Since his
model is substantially different from my HyperDiamond Feynman
Checkerboard in some respects, I use a different terminology
here.  However, I want to make it clear that I have borrowed
these particular structures from his thesis.
\vspace{12pt}

Three useful HyperDiamond structures are:
\vspace{12pt}

3-link Rotating Propagator, useful for building a
proton out of 3 quarks;
\vspace{12pt}

2-link Exchange Propagator, useful for building a
pion out of a quark and an antiquark; and
\vspace{12pt}

4-link Propagator, useful for building a physical spin-2
physical graviton out of $Spin(5)$ Gauge bosons..
\vspace{12pt}

In the 2-dimensional Feynman Checkerboard, there is
only one massive particle, the electron.
\vspace{12pt}

What about the $D_{4}-D_{5}-E_{6}$ model, or any other
model that has different particles with different masses?
\vspace{12pt}

In the context of Feynman Checkerboards, mass is just
the amplitude for a particle to have a change of direction
in its path.
\vspace{12pt}

More massive particles will change direction more often.
\vspace{12pt}

In the $D_{4}-D_{5}-E_{6}$ model, the HyperDiamond Feynman Checkerboard
fundamental path segment length $\epsilon$ of any particle
the Planck length $L_{PL}$.
\vspace{12pt}

However, in the sum over paths for a particle of mass $m$,
\newline
it is a useful approximation to consider the path segment
length to be the Compton wavelength $L_{m}$ of the mass $m$,
$$L_{m} =  h/mc$$
\vspace{12pt}

That is because the distances between
direction changes in the vast
bulk of the paths will be at least $L_{m}$, and
\newline
those distances will be approximately
integral multiples of $L_{m}$,
\newline
so that $L_{m}$ can be used as
the effective path segment length.
\vspace{12pt}

This is an important approximation
because the Planck length $L_{PL}$
is about $10^{-33}$ cm,
\newline
while the effective length $L_{100GeV}$ for a
particle of mass $100$ $GeV$ is about $10^{-16}$ cm.
\vspace{12pt}

In this section, the HyperDiamond lattice
is given quaternionic coordinates.
\newline
The orgin $0$ designates the beginning of the path.
\newline
The 4 future lightcone links
from the origin are given the coordinates
\newline
$1+i+j+k$, $1+i-j-k$, $1-i+j-k$, $1-i-j+k$
\vspace{12pt}

The path of a "Particle" at rest in space,
moving 7 steps in time,
is denoted by
\vspace{12pt}

\[ \begin{array}{|c|}
Particle \\
0 \\
1 \\
2 \\
3 \\
4 \\
5 \\
6 \\
7
\end{array} \]

Note that since the HyperDiamond speed of light is
$\sqrt{3}$, the
path length is $7 \sqrt{3}$.
\vspace{12pt}

\newpage

The path of a "Particle" moving along a lightcone path in
the $1+i+j+k$ direction for 7 steps with no change of direction
is
\vspace{12pt}

\[ \begin{array}{|c|}
Particle \\
0 \\
 1+i+j+k \\
2+2i+2j+2k \\
3+3i+3j+3k \\
4+4i+4j+4k \\
5+5i+5j+5k \\
6+6i+6j+6k \\
7+7i+7j+7k
\end{array} \]

\vspace{12pt}

At each step in either path, the future lightcone can
be represented by a "Square Diagram"
of lines connecting the future
ends of the 4 future lightcone links leading from the vertex at
which the step begins.

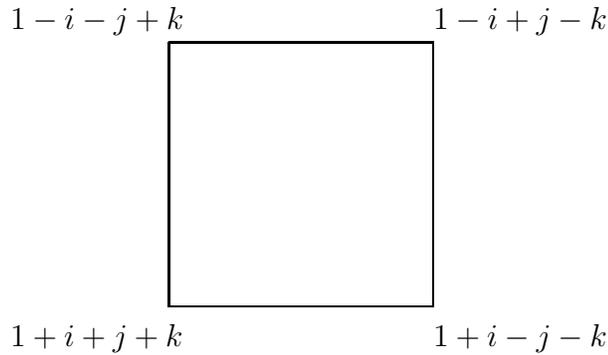
\begin{picture}(200,200)

\put(65,15){\line(1,0){100}}
\put(165,15){\line(0,1){100}}
\put(165,115){\line(-1,0){100}}
\put(65,115){\line(0,-1){100}}
\put(5,0){$1+i+j+k$}
\put(165,0){$1+i-j-k$}
\put(165,120){$1-i+j-k$}
\put(5,120){$1-i-j+k$}

\end{picture}

\vspace{12pt}

In the following subsections,
protons, pions, and physical gravitons
will be represented by multiparticle paths.
\newline
The multiple particles representing protons, pions, and physical
gravitons will be shown on sequences of such Square Diagrams,
as well as by a sequence of coordinates.
\newline
The coordinate sequences will be given only for a representative
sequence of timelike steps, with no space movement, because
\newline
the notation for a timelike sequence is clearer and
\newline
it is easy to transform a sequence of timelike steps into
a sequence of lightcone link steps, as shown above.
\vspace{12pt}

Only in the case of gravitons will it be useful to explicitly
discuss a path that moves in space as well as time.
\vspace{12pt}

\newpage

\subsection{3-Quark Protons.}

Since particle masses can only be observed experimentally
for particles that can exist in a free state
("free" means "not strongly bound to other particles,
except for virtual particles of the active vacuum of spacetime"),
and
since quarks do not exist in free states,
the quark masses that we calculate are interpreted
as constituent masses (not current masses).
\vspace{12pt}

The relation between current masses and constituent masses
may be explained, at least in part,
by the Quantum Theory of David Bohm.
\vspace{12pt}

In hep-ph/9802425,
Di Qing, Xiang-Song Chen, and Fan Wang,
of Nanjing University, present a qualitative QCD analysis
and a quantitative model calculation
to show that the constituent quark model
[after mixing a small amount (15
remains a good approximation
even taking into account the nucleon spin structure
revealed in polarized deep inelastic scattering.
\vspace{12pt}

The effectiveness of
the NonRelativistic model of light-quark hadrons
is explained by, and affords experimental Support for,
the Quantum Theory of David Bohm (see quant-ph/9806023).
\vspace{12pt}

Consitituent particles are Pre-Quantum particles
\newline

in the sense that their properties are calculated without
\newline
using sum-over-histories Many-Worlds quantum theory.
\newline

("Classical" is a commonly-used synonym for "Pre-Quantum".)
\vspace{12pt}

Since experiments are quantum sum-over-histories processes,
experimentally observed particles are Quantum particles.
\vspace{12pt}

Consider the experimentally observed proton.
A proton is a Quantum particle containing 3 constituent quarks:
two up quarks and one down quark;
one Red, one Green, and one Blue.
The 3 Pre-Quantum constituent quarks are called "valence" quarks.
They are bound to each other by $SU(3)$ QCD.
The constituent quarks "feel" the effects of QCD
by "sharing" virtual gluons and virtual quark-antiquark pairs
that come from the vacuum in sum-over-histories quantum theory.
\vspace{12pt}

Since the 3 valence constituent quarks within the proton
are constantly surrounded by the shared virtual gluons
and virtual quark-antiquark pairs,
the 3 valence constituent quarks can be said to
"swim" in a "sea" of virtual gluons and quark-antiquark pairs,
which are called "sea" gluons, quarks, and antiquarks.
\vspace{12pt}

In the model,
the proton is the most stable bound state of 3 quarks,
so that the virtual sea within the proton is at the
lowest energy level that is experimentally observable.
\vspace{12pt}

The virtual sea gluons are massless $SU(3)$ gauge bosons.
Since the lightest quarks are up and down quarks,
the virtual sea quark-antiquark pairs that most often
appear from the vacuum are up or down pairs,
each of which have the same constituent mass, 312.75 MeV.
If you stay below the threshold energy of the strange quark,
whose constituent mass is about 625 MeV,
the low energy sea within the proton contains only
the lightest (up and down) sea quarks and antiquarks,
so that the Quantum proton lowest-energy background sea
has a density of 312.75 MeV.
(In the model, "density" is mass/energy per unit volume,
where the unit volume is Planck-length in size.)
\vspace{12pt}

Experiments that observe the proton as a whole
do not "see" the proton's internal virtual sea, because
the paths of the virtual sea gluon, quarks, and antiquarks
begin and end within the proton itself.
Therefore, the experimentally observed mass
of the proton is the sum of the 3 valence quarks,
3 x 32.75 MeV, or 938.25 MeV
which is very close to the experimental value
of about 938.27 MeV.
\vspace{12pt}

To study the internal structure of hadrons, mesons, etc.,
you should use sum-over-histories quantum theory
of the $SU(3)$ color force $SU(3)$
.
Since that is computationally very difficult
\vspace{12pt}

(For instance, in my view,
the internal structure of a proton looks like
a nonperturbative QCD soliton.
\vspace{12pt}

See WWW URLs
http://galaxy.cau.edu/tsmith/SolProton.html
\vspace{12pt}

http://www.innerx.net/personal/tsmith/SolProton.html )
\vspace{12pt}

you can use approximate theories that correspond
to your experimental energy range.
\vspace{12pt}

For high energy experiments,
such as Deep Inelastic Scattering,
you can use Perturbative QCD.
For low energies,
you can use Chiral Perturbation Theory.
Renormalization equations are conventionally used
to relate experimental observations that are
made at different energy levels.
\vspace{12pt}

The HyperDiamond structure used to approximate the proton is
the 3-link Rotating Propagator, in which 3 quarks orbit their
center somewhat like the 3 balls of an Argentine bola.
\vspace{12pt}

\newpage

Here is a coordinate sequence representation of the approximate
HyperDiamond Feynman Checkerboard path of a proton:

\vspace{12pt}

\[ \begin{array}{|c|c|c|}
R-Quark & G-Quark & B-Quark \\
1+i-j-k & 1-i+j-k & 1-i-j+k \\
2-i+j-k & 2-i-j+k & 2+i-j-k \\
3-i-j+k & 3+i-j-k & 2-i+j-k \\
4+i-j-k & 4-i+j-k & 4-i-j+k \\
\end{array} \]

\vspace{12pt}

The following page contains a Square Diagram representation of
the approximate HyperDiamond Feynman Checkerboard
path of a proton at times 1,2, 3, and 4:
\vspace{12pt}

\newpage

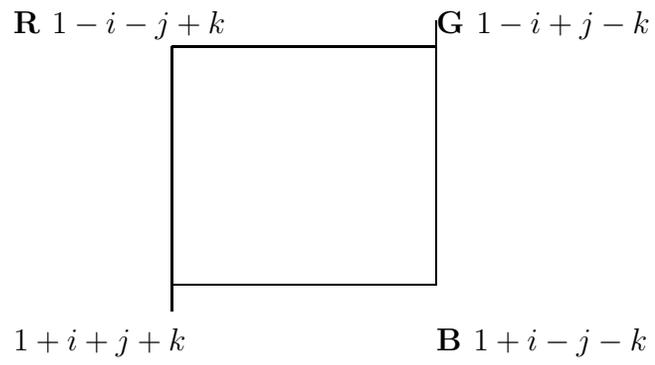
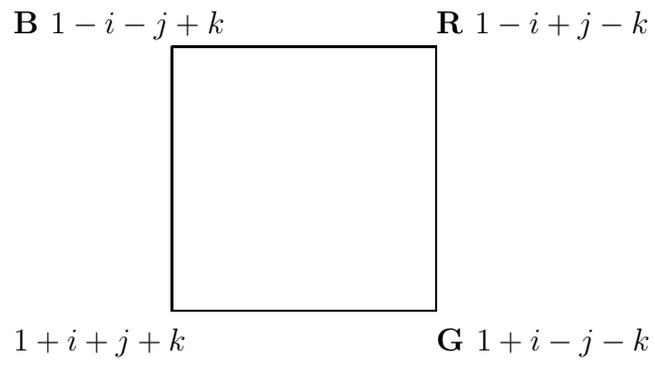
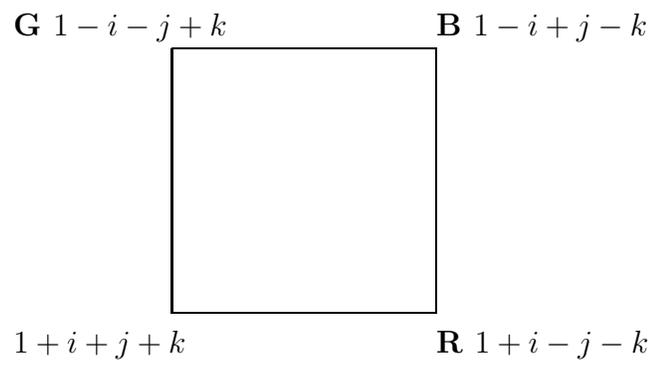
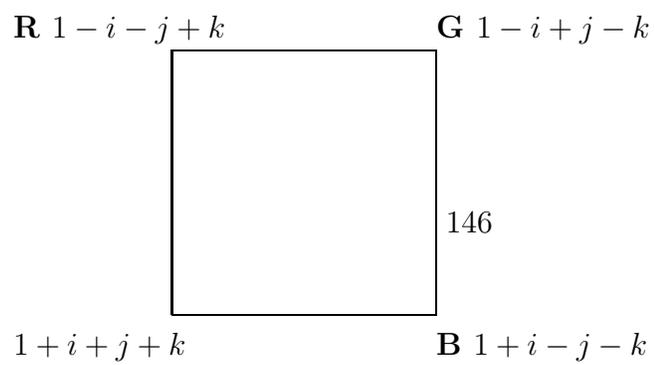
\begin{picture}(200,200)(0,410)

\put(65,595){\line(1,0){100}}
\put(165,595){\line(0,1){100}}
\put(165,685){\line(-1,0){100}}
\put(65,685){\line(0,-1){100}}
\put(5,570){$1+i+j+k$}
\put(165,570){{\bf B }$1+i-j-k$}
\put(165,690){{\bf G }$1-i+j-k$}
\put(5,690){{\bf R }$1-i-j+k$}

\put(65,395){\line(1,0){100}}
\put(165,395){\line(0,1){100}}
\put(165,495){\line(-1,0){100}}
\put(65,495){\line(0,-1){100}}
\put(5,380){$1+i+j+k$}
\put(165,380){{\bf G }$1+i-j-k$}
\put(165,500){{\bf R }$1-i+j-k$}
\put(5,500){{\bf B }$1-i-j+k$}

\put(65,205){\line(1,0){100}}
\put(165,205){\line(0,1){100}}
\put(165,305){\line(-1,0){100}}
\put(65,305){\line(0,-1){100}}
\put(5,190){$1+i+j+k$}
\put(165,190){{\bf R }$1+i-j-k$}
\put(165,310){{\bf B }$1-i+j-k$}
\put(5,310){{\bf G }$1-i-j+k$}

\put(65,15){\line(1,0){100}}
\put(165,15){\line(0,1){100}}
\put(165,115){\line(-1,0){100}}
\put(65,115){\line(0,-1){100}}
\put(5,0){$1+i+j+k$}
\put(165,0){{\bf B }$1+i-j-k$}
\put(165,120){{\bf G }$1-i+j-k$}
\put(5,120){{\bf R }$1-i-j+k$}

\end{picture}

\newpage

In the HyperDiamond Feynman Checkerboard model, where
the proton is represented by two up quarks and one down quark,
and quark masses are constituent masses:
\vspace{12pt}

the spins of the quarks should be in the lowest energy state,
with one spin anti-parallel to the other two,
so that the spin of the proton is

$$+ 1/2 + 1/2 - 1/2 = + 1/2$$

\vspace{12pt}

the color charge of the proton is

$$+ red + blue + green = 0$$

so that the pion is color-neutral;
\vspace{12pt}

the electric charge of the proton is

$$+ 2/3 + 2/3 - (- 1/3) = +1$$

\vspace{12pt}

the theoretical tree-level mass is

$$3 x 312.75 MeV = 938.25 MeV$$

while the experimental mass is 938.27 MeV;
\vspace{12pt}

the proton is stable with respect to decay
by the color, weak, and electromagnetic forces,
while decay by the gravitational force
is so slow that it cannot be observed with present technology.
\vspace{12pt}

\vspace{12pt}

\vspace{12pt}

\subsubsection{Neutron Mass.}

WHAT ABOUT THE NEUTRON MASS?
\vspace{12pt}

According to the 1986 CODATA Bulletin No. 63,
the experimental value of the neutron mass is 939.56563(28) Mev,
and the experimental value of the proton is 938.27231(28) Mev.
\vspace{12pt}

The neutron-proton mass difference 1.3 Mev is due to
the fact that the proton consists
of two up quarks and one down quark,
while the neutron consists of one up quark and two down quarks.
\vspace{12pt}

The magnitude of the electromagnetic energy difference
$m_{N} - m_{P}$ is about 1 Mev,
\vspace{12pt}

but the sign is wrong:

$$m_{N} - m_{P} = -1 Mev$$

\vspace{12pt}

and
the proton's electromagnetic mass is greater
than the neutron's.
\vspace{12pt}

The difference in energy between the bound states,
neutron and proton,
is not due to a difference between
the Pre-Quantum constituent masses
of the up quark and the down quark,
calculated in the theory to be equal.
\vspace{12pt}

It is due to the difference between the Quantum color force
interactions of the up and down constituent valence quarks
with the gluons and virtual sea quarks
in the neutron and the proton.
\vspace{12pt}

An up valence quark, constituent mass 313 Mev,
does not often swap places with a 2.09 Gev charm sea quark,
but a 313 Mev down valence quark can more often swap places
with a 625 Mev strange sea quark.
\vspace{12pt}

Therefore the Quantum color force constituent mass
of the down valence quark is heavier by about
\vspace{12pt}

$$(m_{s} - m_{d})  (m_{d} / m_{s})^2   a(w)  V_{12}
=  312  \times  0.25   \times  0.253  \times  0.22   Mev
=   4.3 Mev$$

\vspace{12pt}

(where $a(w) = 0.253$ is the geometric part
of the weak force strength
and
$V_{12} = 0.22$ is the K-M parameter mixing generations 1 and 2)
\vspace{12pt}

so that the Quantum color force constituent mass $Qm_{d}$
of the down quark is
           $$Qm_{d} = 312.75 + 4.3 = 317.05 MeV$$
\vspace{12pt}

Similarly,
the up quark Quantum color force mass increase is about
\vspace{12pt}

$$(m_{c} - m_{u})  (m_{u} / m_{c})^2   a(w)  V_{12}
=  1777  \times  0.022   \times  0.253  \times  0.22   Mev
=   2.2 Mev$$

\vspace{12pt}

so that the Quantum color force constituent mass $Qm_{u}$
of the up quark is
           $$Qm_{u} = 312.75 + 2.2 = 314.95 MeV$$

\vspace{12pt}

The Quantum color force Neutron-Proton mass difference is
\vspace{12pt}

$$m_{N} - m_{P} = Qm_{d} - Qm_{u}  =  317.05 Mev - 314.95 Mev
= 2.1 Mev$$

\vspace{12pt}

Since the electromagnetic Neutron-Proton mass difference is

roughly $m_{N} - m_{P} = -1 MeV$
\vspace{12pt}

the total theoretical Neutron-Proton mass difference is

$$m_{N} - m_{P}  =  2.1 Mev - 1 Mev = 1.1 Mev$$

\vspace{12pt}

an estimate that is fairly close
to the experimental value of 1.3 Mev.
\vspace{12pt}

Note that in the equation
\vspace{12pt}

$(ms - md)  (md/ms)^2   a(w)  |Vds|  =   4.3 Mev$
\vspace{12pt}

$Vds$ is a mixing of down and strange by a neutral Z0,
compared to the more conventional Vus mixing by charged W.
\vspace{12pt}

Although real neutral Z0 processes are suppressed
by the GIM mechanism,
\vspace{12pt}

which is a cancellation of virtual processes,
the process of the equation is strictly a virtual process.
\vspace{12pt}

Note also that the K-M mixing parameter $|Vds|$ is linear.
\vspace{12pt}

Mixing (such as between a down quark and a strange quark)
is a two-step process,
that goes approximately as the square of $|Vds|$:
\vspace{12pt}

First
the down quark changes to a virtual strange quark,
producing one factor of $|Vds|$.
\vspace{12pt}

Then, second,
the virtual strange quark changes back to a down quark,
producing a second factor of $|Vsd|$, which is approximately
equal to $|Vds|$.
\vspace{12pt}

Only the first step (one factor of $|Vds|$) appears in the
Quantum mass formula used to determine the neutron mass.
\vspace{12pt}

If you measure the mass of a neutron,
that measurement includes a sum over a lot of histories
of the valence quarks inside the neutron.
\vspace{12pt}

In some of those histories, in my view,
you will "see" some of the two valence down quarks
in a virtual transition state that is at a time
after the first action, or change from down to strange,
\vspace{12pt}

and
before the second action, or change back.
\vspace{12pt}

Therefore, you should take into account
those histories in the sum in which you see a strange valence quark,
\vspace{12pt}

and you get the linear factor $|Vds|$ in the above equation.
\vspace{12pt}

Note that
if there were no second generation fermions,
or
if the second generation quarks had equal masses,
\vspace{12pt}

then
the proton would be heavier than the neutron
(due to the electromagnetic difference)
\vspace{12pt}

and
the hydrogen atom would decay into a neutron,
and
there would be no stable atoms in our world.
\vspace{12pt}

\subsection{Quark-AntiQuark Pions.}

In this HyperDiamond Feynman Checkerboard version of the
$D_{4}-D_{5}-E_{6}$
model, pions are made up of first generation
valence Quark-AntiQuark
pairs.
\newline
The pion bound state of valence Quark-AntiQuark pairs
has a soliton structure that would,
projected onto a 2-dimensional spacetime,
be a Sine-Gordon breather.
See WWW URLs
\vspace{12pt}

http://galaxy.cau.edu/tsmith/SnGdnPion.html
\vspace{12pt}

http://www.innerx.net/personal/tsmith/SnGdnPion.html
\vspace{12pt}

Marin, Eilbeck, and Russell, in their paper
\vspace{12pt}

Localized Moving Breathers in a 2-D Hexagonal Lattice,
\vspace{12pt}

show "...that highly localized in-plane breathers can
propagate in specific directions with minimal lateral
spreading ...
This one-dimensional behavior in a two-dimensional lattice
was called quasi-one-dimensional (QOD) ..."
\vspace{12pt}

(In their paper, dimensionality refers to spatial dimensionality.)
\vspace{12pt}

They use QOD behavior to describe
phenomena in muscovite mica crytstals.
\vspace{12pt}

The D4-D5-E6 model uses similar QOD behavior to
describe the pion.
\vspace{12pt}

Take the pion as the fundamental quark-antiquark structure
and
assume that the quark masses are constituent masses that
include the effects of the complicated structure
of sea quarks and binding gluons within the pion
at the energy level of about 313 MeV
that corresponds to the Compton wavelength of
the first generation quark constituent mass.
\vspace{12pt}

The sea quarks and binding gluons within the pion
constitute the dressing of the proton valence quarks
with the quark and gluon sea
in which the valence quarks swim within the pion.
\vspace{12pt}

The quark and gluon sea has characteristic energy
of roughly the constituent quark masses of 312.8 MeV
so that the valence quarks float freely within the pion sea.
\vspace{12pt}

To express the assumption of free-floating quarks
more mathematically, define the relationship between
the calculated constituent quark masses, denoted by $m_{q}$,
and
QCD Lagrangian current quark masses, denoted by $M_{q}$,
by

$$M_{q} = m_{q} - m_{u} = m_{q} - m_{d} = m_{q} - 312.8 MeV$$

\vspace{12pt}

This makes, for the QCD Lagrangian,
the up and down quarks roughly massless.
\vspace{12pt}

One result is that the current masses
can then be used as input for the
$SU(3) \times SU(3)$ chiral theory
that, although it is only approximate because
the constituent mass of the strange quark is about 312 MeV,
rather than nearly zero,
can be useful in calculating meson properties.
\vspace{12pt}

In hep-ph/9802425,
Di Qing, Xiang-Song Chen, and Fan Wang,
of Nanjing University, present a qualitative QCD analysis
and a quantitative model calculation
to show that the constituent quark model
[after mixing a small amount (15
remains a good approximation
even taking into account the nucleon spin structure
revealed in polarized deep inelastic scattering.
\vspace{12pt}

The effectiveness of
the NonRelativistic model of light-quark hadrons
is explained by, and affords experimental Support for,
the Quantum Theory of David Bohm (see quant-ph/9806023).
\vspace{12pt}

The HyperDiamond structure used to approximate the pion is the
2-link Exchange Propagator.
\vspace{12pt}

\newpage

Here is a coordinate sequence representation of the approximate
HyperDiamond Feynman Checkerboard path of a pion:
\vspace{12pt}

\[ \begin{array}{|c|c|}
Quark & AntiQuark \\

0 & 0 \\
1+i+j+k & 1-i+j-k \\
2 & 2 \\
3-i+j-k & 3+i+j+k \\
4 & 4 \\
5+i+j+k & 5-i+j-k \\
6 & 6 \\
\end{array} \]

\vspace{12pt}

The following page contains a Square Diagram representation of
the approximate HyperDiamond Feynman Checkerboard path of a pion
at times 1, 3, and 5 (at 0, 2, 4, and 6, both the quark and
the antiquark are at the origin):
\vspace{12pt}

\newpage

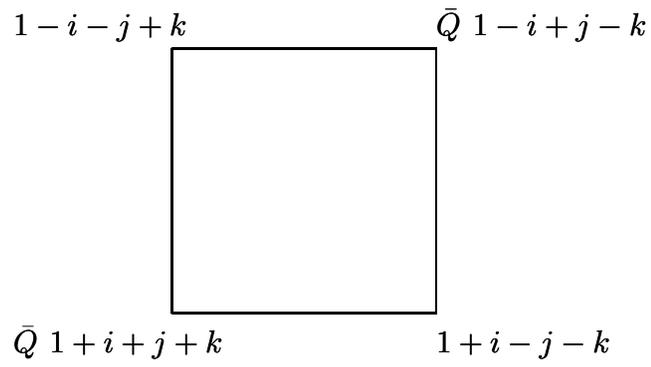
\begin{picture}(200,200)(0,410)

\put(65,585){\line(1,0){100}}
\put(165,585){\line(0,1){100}}
\put(165,685){\line(-1,0){100}}
\put(65,685){\line(0,-1){100}}
\put(5,570){{\bf $Q$ }$1+i+j+k$}
\put(165,570){$1+i-j-k$}
\put(165,690){{\bf $\bar{Q}$ }$1-i+j-k$}
\put(5,690){$1-i-j+k$}

\put(65,585){\line(1,0){100}}
\put(165,585){\line(0,1){100}}
\put(165,685){\line(-1,0){100}}
\put(65,685){\line(0,-1){100}}
\put(5,570){{\bf $\bar{Q}$ }$1+i+j+k$}
\put(165,570){$1+i-j-k$}
\put(165,690){{\bf $Q$ }$1-i+j-k$}
\put(5,690){$1-i-j+k$}

\put(65,585){\line(1,0){100}}
\put(165,585){\line(0,1){100}}
\put(165,685){\line(-1,0){100}}
\put(65,685){\line(0,-1){100}}
\put(5,570){{\bf $Q$ }$1+i+j+k$}
\put(165,570){$1+i-j-k$}
\put(165,690){{\bf $\bar{Q}$ }$1-i+j-k$}
\put(5,690){$1-i-j+k$}

\end{picture}

\vspace{12pt}

For more details about pions and other mesons, see WWW URLs
\vspace{12pt}

http://galaxy.cau.edu/tsmith/Sets2Quarks10.html
\vspace{12pt}

http://www.innerx.net/personal/tsmith/Sets2Quarks10.html
\vspace{12pt}

\newpage

\subsection{Spin-2 Physical Gravitons.}

In this HyperDiamond Feynman Checkerboard version
of the $D_{4}-D_{5}-E_{6}$ model,
spin-2 physical gravitons are made up of the 4 translation
spin-1 gauge bosons of the 10-dimensional
$Spin(5)$ de Sitter subgroup
of the 15-dimensional $Spin(6)$ Conformal group used to
construct Einstein-Hilbert gravity in the
$D_{4}-D_{5}-E_{6}$ model
described in URLs
\vspace{12pt}

http://xxx.lanl.gov/abs/hep-ph/9501252
\vspace{12pt}

http://xxx.lanl.gov/abs/quant-ph/9503009.
\vspace{12pt}

The action of Gravity on Spinors is given by
the Papapetrou Equations.
\vspace{12pt}

 The spin-2 physical gravitons are massless,
but they can have
energy up to and including the Planck mass.
\vspace{12pt}

Unlike the pions and protons,
which are made up of fermion quarks
that live on vertices of the HyperDiamond Lattice,
the gravitons are gauge bosons
that live on the links of the HyperDiamond Lattice.
\vspace{12pt}

The Planck energy spin-2 physical gravitons are really
fundamental structures with
HyperDiamond Feynman Checkerboard path length $L_{Planck}$.
\vspace{12pt}

\newpage

Here is a coordinate sequence representation of
\newline
the HyperDiamond Feynman Checkerboard path of a fundamental
\newline
Planck-mass spin-2 physical graviton,
\newline
where ${\bf T,  X, Y, Z }$ represent infinitesimal generators
\newline
of the $Spin(5)$ de Sitter group:
\vspace{12pt}

\[ \begin{array}{|c|c|c|c|}
{\bf T } & {\bf X } & {\bf Y } & {\bf Z } \\
1+i+j+k & 1+i-j-k & 1-i+j-k & 1-i-j+k \\
2+i+j+k & 2+i-j-k & 2-i+j-k & 2-i-j+k \\
3+i+j+k & 3+i-j-k & 3-i+j-k & 3-i-j+k \\
4+i+j+k & 4+i-j-k & 4-i+j-k & 4-i-j+k \\
\end{array} \]

\vspace{12pt}

\vspace{12pt}

The following is a Square Diagram representation of the
HyperDiamond Feynman Checkerboard path
of a fundamental Planck-mass
spin-2 physical graviton at times 1, 2, 3, and 4:
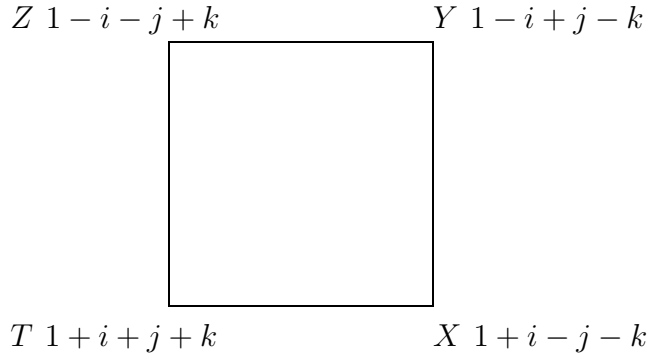
\vspace{12pt}

\begin{picture}(200,200)

\put(65,15){\line(1,0){100}}
\put(165,15){\line(0,1){100}}
\put(165,115){\line(-1,0){100}}
\put(65,115){\line(0,-1){100}}
\put(5,0){{\bf $T$ }$1+i+j+k$}
\put(165,0){{\bf $X$ }$1+i-j-k$}
\put(165,120){{\bf $Y$ }$1-i+j-k$}
\put(5,120){{\bf $Z$ }$1-i-j+k$}

\end{picture}

\vspace{12pt}

The representation above is for a timelike path
at rest in space.
\vspace{12pt}

\newpage

With respect to gravitons,
we can see something new and different
by letting the path move in space as well.
\vspace{12pt}

Let ${\bf T}$ and ${\bf Y}$
represent time and longitudinal space, and
\newline
${\bf X}$ and  ${\bf Z}$ represent transverse space.
\vspace{12pt}

Then, as discussed in Feynman's {\it Lectures on Gravitation},
pp. 41-42 \cite{FEY}, the Square Diagram representation shows
that our spin-2 physical graviton is indeed a spin-2 particle.
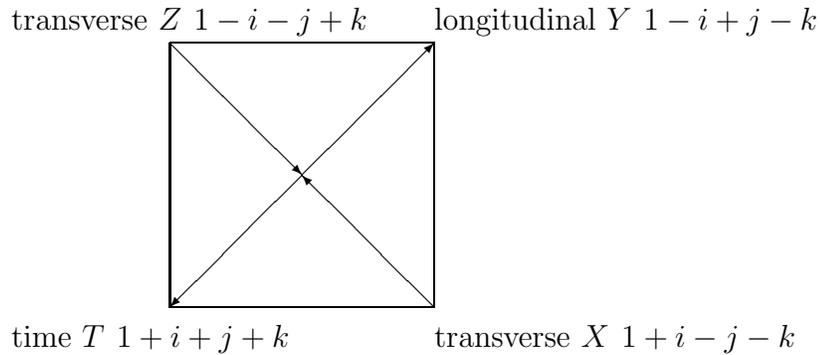
\vspace{12pt}

\begin{picture}(200,200)

\put(65,15){\line(1,0){100}}
\put(165,15){\line(0,1){100}}
\put(165,115){\line(-1,0){100}}
\put(65,115){\line(0,-1){100}}
\put(5,0){time {\bf $T$ }$1+i+j+k$}
\put(165,0){transverse {\bf $X$ }$1+i-j-k$}
\put(165,120){longitudinal {\bf $Y$ }$1-i+j-k$}
\put(5,120){transverse {\bf $Z$ }$1-i-j+k$}
\put(115,65){\vector(1,1){50}}
\put(115,65){\vector(-1,-1){50}}
\put(165,15){\vector(-1,1){50}}
\put(65,115){\vector(1,-1){50}}

\end{picture}

\vspace{12pt}

Spin-2 physical gravitons of energy less than the Planck mass
are more complicated composite gauge boson structures with
approximate HyperDiamond Feynman Checkerboard path length
$L_{graviton energy}$.
\vspace{12pt}

They can be deformed from a square shape, but retain
their spin-2 nature as described by Feynman \cite{FEY}.
\vspace{12pt}

\subsubsection{Planck Mass.}

An estimated calculation of the Planck mass is at WWW URLs
\vspace{12pt}

http://galaxy.cau.edu/tsmith/Planck.html
\vspace{12pt}

http://www.innerx.net/personal/tsmith/Planck.html
\vspace{12pt}

Here is a summary of a combinatorial calculation:
\vspace{12pt}

Consider an isolated single point,
or vertex in the lattice picture of spacetime.
In the HyperDiamond Feynman Checkerboard model,
fermions live on vertices,
and only first-generation fermions can live on a single vertex.
\vspace{12pt}

(The second-generation fermions live on
two vertices that act at our energy levels very much like one,
and
the third-generation fermions live on three vertices that act at
our energy levels very much like one.)
\vspace{12pt}

At a single spacetime vertex, a Planck-mass black hole
\newline

is the Many-Worlds quantum sum of all possible virtual
\newline

first-generation particle-antiparticle fermion pairs
\newline

permitted by the Pauli exclusion principle
to live on that vertex.
\vspace{12pt}

The Planck mass in 4-dimensional spacetime is
the sum of masses of all possible
virtual first-generation particle-antiparticle fermion pairs
permitted by the Pauli exclusion principle.
\vspace{12pt}

There are 8 fermion particles and 8 fermion antiparticles
for a total of 64 particle-antiparticle pairs.
A typical combination should have several quarks,
several antiquarks,
a few colorless quark-antiquark pairs
that would be equivalent to pions,
and some leptons and antileptons.
\vspace{12pt}

Due to the Pauli exclusion principle,
no fermion lepton or quark could be present at
the vertex more than twice
unless they are in the form of boson pions,
colorless first-generation quark-antiquark pairs
that are not subject to the Pauli exclusion principle.
\vspace{12pt}

Of the 64 particle-antiparticle pairs, 12 are pions.
\vspace{12pt}

A typical combination should have about 6 pions.
\vspace{12pt}

If all the pions are independent,
the typical combination should have a mass of
$0.14 \times 6 GeV = 0.84 GeV$.
\vspace{12pt}

However, just as the pion mass of 0.14 GeV is
less than the sum of the masses of a quark and an antiquark,
pairs of oppositely charged pions may
form a bound state of less mass
than the sum of two pion masses.
\vspace{12pt}

If such a bound state of oppositely charged pions
has a mass as small as 0.1 GeV,
\vspace{12pt}

and if the typical combination has one such pair
and 4 other pions,
then
the typical combination should have a mass
in the range of 0.66 GeV.
\vspace{12pt}

Summing over all $2^{64}$ combinations,
the total mass of a one-vertex universe should give:
\vspace{12pt}

$$m_{Planck} = 1.217 -1.550 \times 10^{19} GeV$$

There is also a quaternionic calculation of about
$1.3 \times 10^{19} GeV$ at WWW URLs
\vspace{12pt}

http://galaxy.cau.edu/tsmith/Planck.html
\vspace{12pt}

http://www.innerx.net/personal/tsmith/Planck.html
\vspace{12pt}

\newpage

\appendix

\section{Errata for Earlier Papers.}

Errata for Earlier Papers can be found at WWW URLs
\vspace{12pt}

http://galaxy.cau.edu/tsmith/Errata.html
\vspace{12pt}

http://www.innerx.net/personal/tsmith/Errata.html

\end{document}